\documentclass[12pt,aps,prd,onecolumn,nofootinbib,superscriptaddress,preprintnumbers,balancelastpage]{revtex4}
\pdfoutput=1
\usepackage{verbatim}
\usepackage{amssymb,amsmath,latexsym,graphics, graphicx,epsfig,multirow,comment,hyperref,appendix, slashed, verbatim} 
\usepackage{graphicx,subfigure, slashed}
\usepackage{epstopdf}
\usepackage{url}
\usepackage{appendix}
\usepackage{color}
\usepackage{caption}
\newcommand{\gsim}{\lower.7ex\hbox{$\;\stackrel{\textstyle>}{\sim}\;$}}
\newcommand{\lsim}{\lower.7ex\hbox{$\;\stackrel{\textstyle<}{\sim}\;$}}

\def\OO{{\cal O}}

\def\EE{{\cal E}}
\def\CC{{\cal C}}

\def\GG{{\cal G}}

\newcommand{\TeV}{\,\mathrm{TeV}}
\newcommand{\GeV}{\,\mathrm{GeV}}

\newcommand{\ifb}{\,\mathrm{fb}^{-1}}

\newcommand{\half}{{\frac{1}{2}  }}

\newcommand{\MET}{\slashed{E}_T}

\newcommand{\nkt}{{n_{k_T}}}
\newcommand{\nca}{{n_{\text{CA}}}}

\newcommand{\be}{\begin{eqnarray}}
\newcommand{\ee}{\end{eqnarray}}
\newcommand{\bea}{\begin{eqnarray}}
\newcommand{\eea}{\end{eqnarray}}

\newcommand{\bef}{\begin{figure}[htbp]\begin{center}}
\newcommand{\eef}{\end{center}\end{figure}}

\newcommand{\secref}[1]{Sec.~\ref{Sec: #1}}

\begin{document}

\begin{flushright}
\mbox{\normalsize SLAC-PUB-15698}
\end{flushright}
\vskip 80 pt

\title{Learning How to Count:  A High Multiplicity Search for the LHC }

\author{Sonia El Hedri}
\affiliation{
SLAC, Stanford University, Menlo Park, CA 94025 USA}

\author{Anson Hook}
\affiliation{
Institute for Advanced Studies, Princeton University, Princeton, NJ 08544 USA}

\author{Martin Jankowiak}
\affiliation{
Institut f\"{u}r Theoretische Physik, Universit\"{a}t Heidelberg, Germany}

\author{Jay G. Wacker}
\affiliation{
SLAC, Stanford University, Menlo Park, CA 94025 USA}


\begin{abstract}
\vskip 15 pt
\begin{center}
{\bf Abstract}
\end{center}
\vskip -8 pt
$\quad$ 
We introduce a search technique that is sensitive to a broad class of signals with large final state multiplicities.  Events are clustered into large radius jets and jet substructure techniques are used to count the number of subjets within each jet.  The search consists of a cut on the
total number of subjets in the event as well as the summed jet mass and missing energy.  Two different techniques for counting
subjets are described and expected sensitivities are presented for eight benchmark signals.  These signals exhibit diverse phenomenology, including 2-step cascade decays, direct three body decays, and multi-top final states.  We find improved sensitivity to these 
signals as compared to previous high multiplicity searches as well as a reduced reliance on missing energy requirements. 
One benefit of this approach is that it allows for natural data driven estimates of the QCD background.
\end{abstract}

\maketitle
\newpage
\tableofcontents

\setcounter{section}{0}
\renewcommand{\thesection}{\arabic{section}}
\renewcommand{\thesubsection}{\arabic{section}.\arabic{subsection}}
\renewcommand{\thesubsubsection}{\arabic{section}.\arabic{subsection}.\arabic{subsubsection}}
\renewcommand{\thetable}{\arabic{table}}
\makeatletter \renewcommand*{\p@subsection}{} \renewcommand*{\p@subsubsection}{}

\section{Introduction}

The search for physics beyond the Standard Model is a central focus of LHC research. 
The motivations for extensions of the Standard Model are multi-faceted, spanning such diverse physics
topics as the identity of dark matter, the radiative stability of the weak scale, the unification of forces, and
the origin of the baryon asymmetry of the Universe.  Apart from these specific theory inputs, which suggest
certain classes of models, there is the generic goal of thoroughly probing the weak scale for new physics---whatever that might be. 
In order to cover the vast range of possibilities for new physics that open up when the input from theory is loosened,
it is imperative for the LHC to carry out an extensive experimental program that is sensitive to the widest possible range
of new physics signatures.

Since it is not possible to perform model independent searches for new physics---doing so is limited
by both theoretical and experimental systematic uncertainties---it is necessary to design searches for new physics
that are targeted at specific experimental signatures.  Typically such searches are based on exploiting a single key handle that 
significantly reduces the dominant Standard Model backgrounds, namely those originating from QCD jet production. 
For instance, it is common to require hard electroweak particles such as leptons or photons or (large amounts of) missing energy.
Requiring b-tagged jets, massive jet resonances or extremely high energy events can provide alternative avenues for parametrically 
reducing the QCD background.  These considerations exist both because of triggering requirements necessary for permanently
storing events to tape and because realistic systematic uncertainties in any case limit the degree to which increased integrated luminosity leads
to increased sensitivity.

The past several years have seen increasing attention being paid to high multiplicities (i.e.~requiring that events have more than $\sim6$ final state jets)
as a way of parametrically reducing QCD contributions to searches for new physics.  Historically this approach was
motivated by searches for black holes at the LHC (see e.g.~ref.~\cite{Dimopoulos:2001hw}), but more recently high multiplicity searches have been advocated as an effective technique
for helping to probe scenarios with natural supersymmetry as well as those with baryonic R-parity violating
supersymmetry \cite{Bramante:2011xd,Papucci:2011wy,kumarsusy,Ruderman:2012jd,Bhattacherjee:2013gr,Curtin:2012rm}.  High multiplicity
final states also arise in theories of strong dynamics where new colored objects can produce four to eight top quarks through the
production of ``coloron'' vector resonances or colored technipions \cite{Hill:1991at,Dicus:1994sw}.  Finally, some models introduced to
explain the magnitude of the $t\bar{t}$ asymmetry measured at the Tevatron also predict large final state multiplicites \cite{Gross:2012bz}.

One of the challenges of using high multiplicity as a handle for reducing QCD backgrounds is that the background rate is 
intrinsically difficult to calculate.  The current state of the art for tree-level jet production is $2\rightarrow 7$,
with $2\rightarrow 6$ being the most that is typically feasible.  Significantly, these tree-level calculations have unquantified
uncertainties in their rates and distributions.  One of the computational challenges is that high multiplicity final states have enormous
configuration spaces; for instance the 10 jet final state has 28 dimensions, with the consequence that is unfeasible to densely populate the configuration
space with Monte Carlo events.  Many of the configuration space variables, such as the angular separations between jets, $\Delta R_{ij}$,
have not been studied in depth and historically have been unreliably calculated with tree-level Monte Carlo.

One way that high multiplicity backgrounds are estimated is by extrapolating from lower multiplicities.
There are well-known approximate empirical scaling relations connecting the $N$ jet production rate to the $N+1$ rate.
There has been some progress in deriving these from first principles (see e.g.~ref.~\cite{Gerwick:2012hq, Gerwick:2012fw}); nevertheless, this approach comes with large uncertainties in the
rates that will cause these searches to quickly become systematically limited.  Additionally, for large $N$ the
$p_T$ spectrum for the $N$th jet becomes increasingly soft and harder to measure accurately, leading to additional uncertainties.

Recently alternative approaches to gaining sensitivity to high multiplicity final states have been developed in the spirit
of the ``fat jet approach'' introduced\footnote{See the BOOST 2010 \& 2011 proceedings \cite{BOOST} for additional references.} in ref.~\cite{Seymour:1993mx, Butterworth:2008sd}.
In effect these proposals factorize the problem: instead of directly counting jets, the event is clustered into a fixed
number of large radius jets ($N=$ 4, 5, or 6) whose substructure is then further scrutinized.  Because the jet radius is large,
these $N$ ``fat'' jets will incorporate most of the radiation in the central region of the detector.  
The particles that would have formed multiple small radius jets in the traditional approach are clustered together into large radius jets.  
These fat jets will automatically have substructure, some of which will appear to originate from hard $1\rightarrow 2$ parton shower splittings.
While such splittings certainly occur in QCD, they occur relatively rarely with the result that requiring multiple fat jets to have multiple hard splittings
helps to separate signal events with high multiplicities from the dominant low multiplicity QCD background. 

The fat jet approach has an additional benefit in that it may be better suited to getting a good handle on 
systematic uncertainties in the QCD backgrounds.
Large radius jets, particularly well separated ones, have properties that are relatively independent from one another, since their dynamics are
largely driven by the parton shower, which is local in nature. For instance, the 
mass of one fat jet is not strongly correlated with the masses of other fat jets in the event.
By exploiting the approximate independence of jets, QCD backgrounds can be estimated with a data driven analysis.
Effectively the large configuration space we started out with has been factorized into a much smaller fat jet configuration space tied to 
$N$ (approximately identical) configuration spaces encoding the fat jets' substructure.  This factorization lends itself to the measurement of
appropriate jet templates, which can then be combined with fat jet distributions to arrive at background estimates.
This approach is of obvious practical importance to experimental searches, but it is also
useful for theoretical calculations of the backgrounds, since searching for obvservables that reduce QCD backgrounds by
six to ten orders of magnitude while acquiring enough statistics in Monte Carlo can be challenging to prohibitive. 

Specifically, this study builds on a recent paper \cite{Hook:2012fd} that proposed $M_J$, the sum over fat jet masses, as an effective observable
for separating high multiplicity signals from Standard Model backgrounds.
Jet mass is the simplest jet substructure observable, but it is also one of the coarsest.  This article advocates
a more refined use of jet substructure to probe high multiplicity events, in particular by counting the number of subjets inside each fat jet. 
This may seem similar to clustering events into small radius jets and counting the resulting number of jets, 
but the potential to systematically estimate QCD backgrounds with a data driven approach distinguishes it from the traditional approach. 

This article is organized as follows. 
\secref{Subjets} presents the two subjet counting techniques introduced in this paper.
\secref{MC} describes how the backgrounds were generated in Monte Carlo. 
\secref{DD} describes how the backgrounds were validated against ATLAS data and how the QCD backgrounds should be amenable to data driven estimates. 
\secref{Searches} introduces eight benchmark signals and presents the expected sensitivity of our search strategy. A comparison to previous high multiplicity searches is also made.
We conclude in \secref{Discussion} with some general discussion.

\section{Counting Subjets}
\label{Sec: Subjets}

This section describes the two subjet counting techniques implemented in this study, $n_{\rm{k_T}}$ and $\nca$.  The former
is a straightforward application of the exclusive $k_T$ algorithm \cite{Catani:1993hr}. The latter counts subjets by 
recursively inspecting the structure of the Cambridge/Aachen \cite{Wobisch:1998wt} clustering tree of the fat jet.
Both are implemented using {\tt FastJet} 3 \cite{Cacciari:2011ma}.
As we will see, searches incorporating these methods yield improved sensitivity to the high multiplicity  signals considered in \secref{Searches}.
The two algorithms result in very similar expected limits, with the difference that $\nca$ does better than $n_{\rm{k_T}}$
in cases where the QCD backgrounds are more important.

\subsection{Wide radius jets}
\label{Sec: fatjets}

Wide radius jets are now a standard technique in high energy physics searches.  
The basic structure of our high multiplicity search is built around such fat jets and is common to both subjet counting techniques.  First the event is clustered
into fat jets with $R_0=1.2$ using the anti-$k_T$ jet algorithm \cite{Cacciari:2008gp}.  Next the fat jets are trimmed  \cite{Krohn:2009th}
with the parameters  $r_{\text{cut}} = 0.3$ and $f_{\text{cut}} = 0.05$. While fat jets are particularly sensitive to pile-up, making some sort
of jet grooming necessary, 
it has been shown that jet substructure observables such as jet mass and N-subjettiness \cite{Thaler:2010tr} have a significantly reduced sensitivity to pile-up effects with this choice of parameters \cite{ATLAS:2012am}.
Next the leading fat jet is required to have $p_T > 100 \GeV$, while subleading fat jets are required to have $p_T > 50 \GeV$.  
Only those events with four or more such fat jets are considered.
Then the subjet count $n_i$ of each of the four leading fat jets is calculated using one of the two algorithms described below.  Finally
cuts are made on the observables 
\begin{eqnarray}
M_J \equiv \sum_{i=1}^4 m_i \qquad\qquad N_{J} \equiv \sum_{i=1}^4 n_i
\end{eqnarray}
and the missing transverse energy ($\MET$).

\subsection{Counting with $k_T$}
\label{Sec: kTMethod}

The exclusive $k_T$ algorithm is defined via two metrics, $d_{ij}$ and $d_{iB}$, and a dimensionful resolution parameter 
$d_{\rm{cut}}$ \cite{Salam:2009jx}.  The jet-jet metric $d_{ij}$ and the jet-beam metric $d_{i}$ are defined as:
\begin{eqnarray}
d_{ij} = \min \left[p_{Ti}^{2}, p_{Tj}^{2} \right] \Delta R_{ij}^2 \qquad \qquad d_{i} = p_{Ti}^{2}
\end{eqnarray}
Here, $p_{Ti}$ is the transverse momentum of protojet $i$ and  $\Delta R_{ij}^2\equiv\Delta y_{ij} ^2 + \Delta \phi_{ij}^2$.  The 
exclusive mode of the algorithm proceeds by sequentially clustering pairs of protojets, stopping once all the $d_{ij}$ and $d_{i}$
are above $d_{\rm{cut}}$. When the smallest metric is $d_{ij}$, $i$ and $j$ are combined. When the smallest metric is $d_{i}$,
protojet $i$ is set aside as a beam jet. The total number of subjets (in the exclusive sense) 
is then given by the number of protojets that are not beam jets and that remain once the clustering step
has terminated.\footnote{The beam jets, in the rare case they appear in the reclustering of our fat jets, are soft and at the periphery of the fat jet so the fact that they are discarded is just as we should like.}  
For our particular application we define $n_{\rm{k_T}}$ by taking
\begin{eqnarray}
\sqrt{d_{\rm{cut}}} =  f_{\rm{k_T}} p_{TJ},
\end{eqnarray}
 where $p_{TJ}$ is the total transverse momentum of the fat jet and $f_{\rm{k_T}}$ is a dimensionless parameter that we take to be given by
$$f_{\rm{k_T}}=0.065$$ 
throughout.  This value of $f_{\rm{k_T}}$ leads to good separation between signal and background,
although for the range of signals considered the separation depends only weakly on the particular value used.
Then for a given fat jet $n_{\rm{k_T}}$ is taken to be the number of subjets identified by exclusive $k_T$ with 
$$p_T\ge p_{T\text{ cut}}=40\GeV.$$
The dependence on the parameter $p_{\rm{Tcut}}$ is rather weak for the massive fat jets of interest, 
which contain softer subjets only infrequently.

A significant advantage of this definition of $n_{\rm{k_T}}$ is that a typical QCD jet will, due to
its asymmetric energy sharing (a hard core surrounded by soft radiation), have a small number of subjets since much of the soft radiation will
be clustered with the core (see Fig.~\ref{Fig: exampleNs}).  This is in contrast to a naive application of Cambridge-Aachen for 
reclustering, which can yield a large number of subjets even for a single-pronged QCD jet.  

\subsection{Counting with Cambridge-Aachen}
\label{Sec: CAMethod}

The $n_{\rm{k_T}}$ algorithm introduced in \secref{kTMethod} was entirely generic in its motivation and approach. The present method, denoted by $\nca$, aims instead to count the number
of hard partons {\it consistent with the decay of a massive particle}. The $\nca$ algorithm  is explicitly constructed to: 
\begin{itemize}
\item  identify massive substructure; and
\item `undercount' the number of subjets within a given fat jet if the energy sharing among the subjets is very asymmetric.  
\end{itemize}
The latter requirement is made because asymmetric sharing of energy between subjets is a telltale sign of subjets generated via the parton shower.
The method is in the spirit
of the various substructure algorithms that have emerged since the introduction of the BDRS procedure \cite{Butterworth:2008sd} and which make use of the information
encoded in the clustering tree of the jet. In particular, it is closely related to an intermediate step in the HEPTopTagger \cite{Plehn:2009rk,HEPTT}.

We determine the number of subjets by unclustering the fat jet down to the mass scale $m_{\rm{cut}}$, throwing out subjets with an asymmetric energy sharing as defined by $y_{\rm{cut}}$.  The number of identified subjets that then pass an additional $p_T$ cut yields $\nca$.  
In detail, the method is defined as follows:
\begin{enumerate}
\item Cluster a given fat jet using the Cambridge/Aachen algorithm.
\item To define $\nca$ inspect the clustering tree of the fat jet via the following recursive procedure.
\item Uncluster $j$ into $j_1$ and $j_2$ with $p_{\rm{T1}} > p_{\rm{T2}}$.
\item If $m_j < m_{\rm{cut}}$ or $dR(j_1,j_2) < R_{\rm{min}}$ consider $j$ a subjet and exit the recursion.
\item If $p_{\rm{T2}} < y_{\rm{cut}} \cdot (p_{\rm{T1}}+p_{\rm{T2}})$ throw out $j_2$.
\item Continue the recursion on $j_1$ and (if it is retained) $j_2$.
\item When the recursion is complete, count the number of identified subjets with $p_{\rm{T}} > p_{\rm{Tcut}}$; this number is $\nca$.
\end{enumerate}
So, for example, an idealized two-pronged jet initiated by the hadronic decay of an energetic 
$Z$ boson would yield (supposing that the decay angle is such as to yield a roughly symmetric energy sharing) 
a count $\nca = 1$ for $m_{\rm{cut}} > m_Z$ and $\nca = 2$ for $m_{\rm{cut}} < m_Z$.

Throughout this study we use the following parameters: 
\begin{eqnarray}
m_{\rm{cut}}=30 \GeV, \quad y_{\rm{cut}}=0.10, \quad R_{\rm{min}}=0.15, \quad p_{\rm{Tcut}}=30\GeV.
\end{eqnarray}
These values lead to good separation between signal and background, although for the range of signals considered the separation
provided by $\nca$ depends only weakly on the particular values used.

\subsection{Comparison of $\nkt$ and $\nca$}

This section has introduced two distinct subjet counting techniques, and it is interesting to ask how they are related.  A detailed comparison between the two algorithms is complicated by the fact that each is defined by several parameters.  For simplicity we restrict ourselves to the parameter choices made above. 
Qualitatively, the two algorithms have a number of similar features.

\begin{figure}[t]
\includegraphics[width=0.505\linewidth]{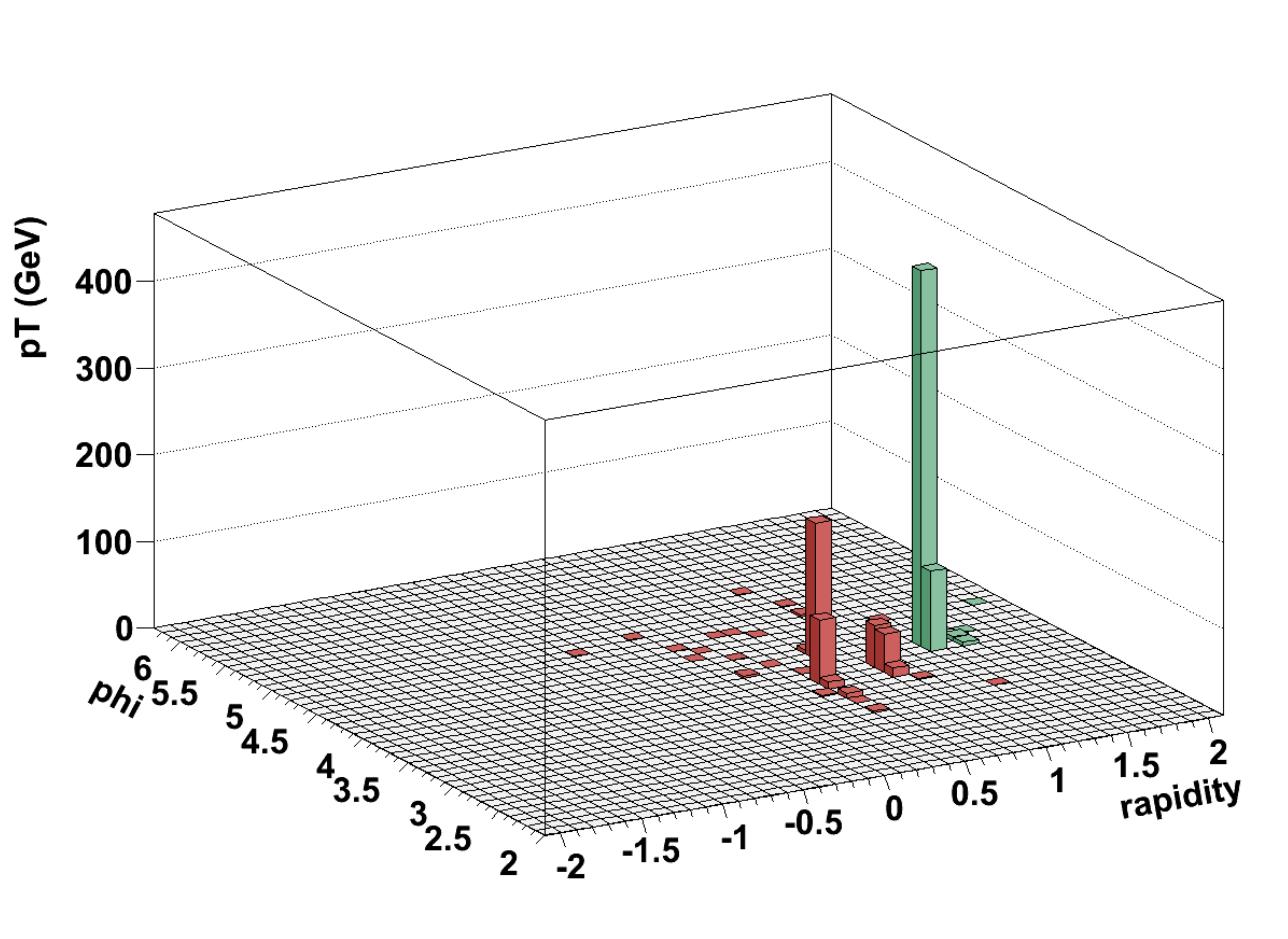}
\hspace{-4mm}
\includegraphics[width=0.505\linewidth]{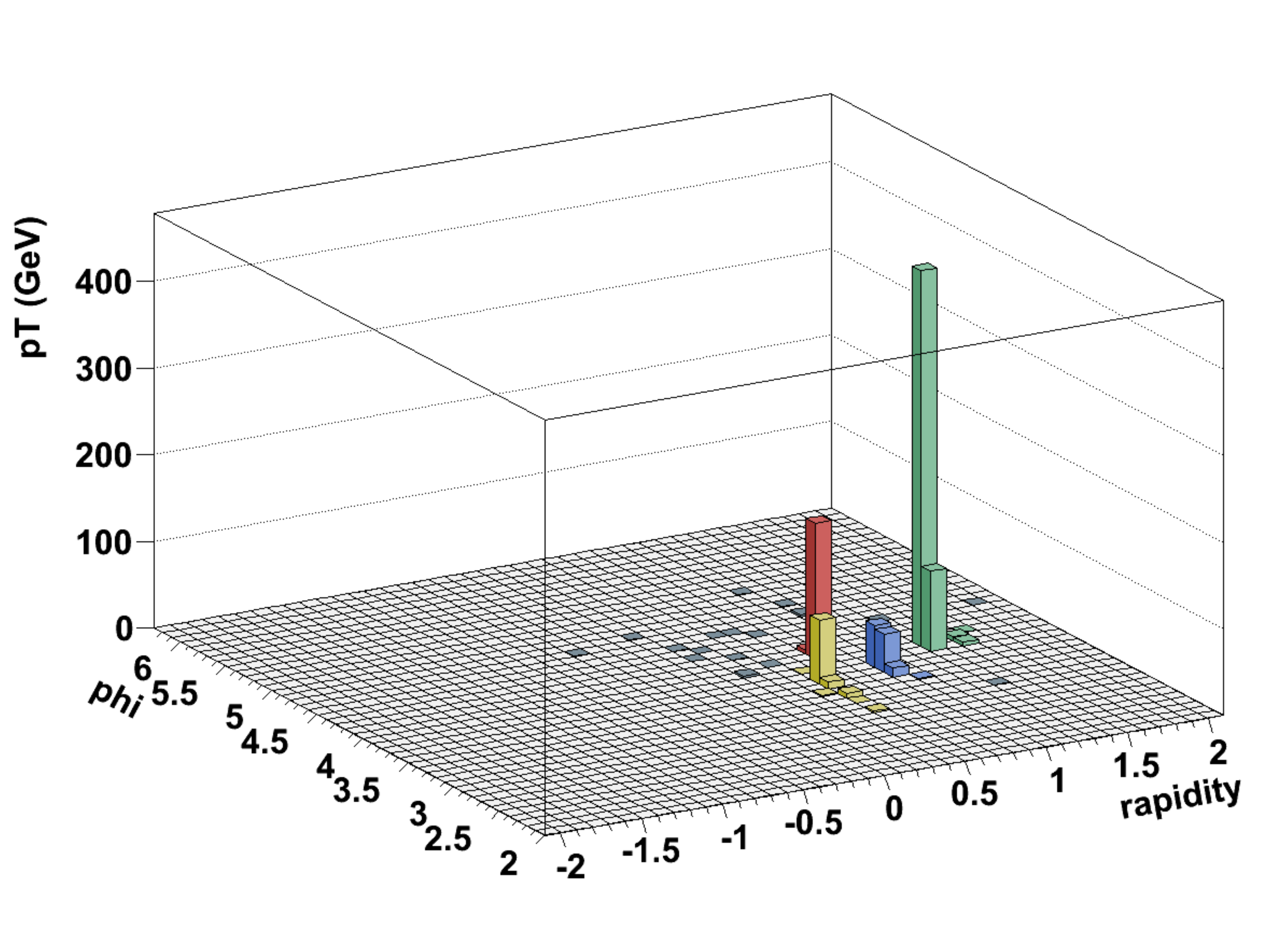}
\caption{Example of a fat (and very massive) QCD jet with $p_T=910$ GeV, $m=360$ GeV and with its two $\nkt$ subjets (left) and
its four $\nca$ subjets (right) indicated. Note that this jet has not been trimmed to better illustrate the different treatment
of soft radiation in $\nca$ and $\nkt$ (the dark gray cells on the right do not belong to any identified subjets).\label{Fig: exampleNs}}
\end{figure}

\begin{figure}
\includegraphics[width = 0.484\linewidth]{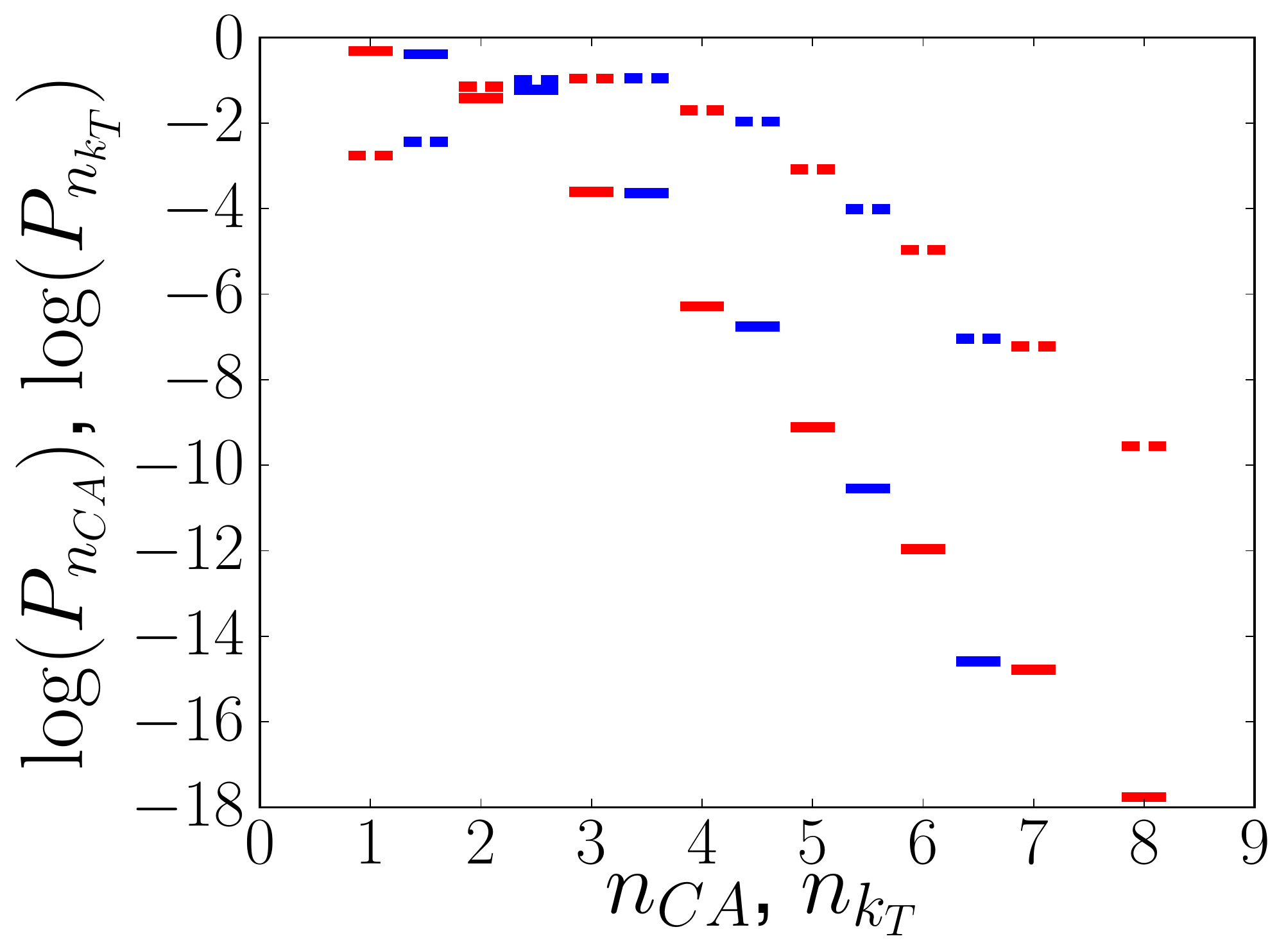}
   \qquad
\includegraphics[width = 0.46\linewidth]{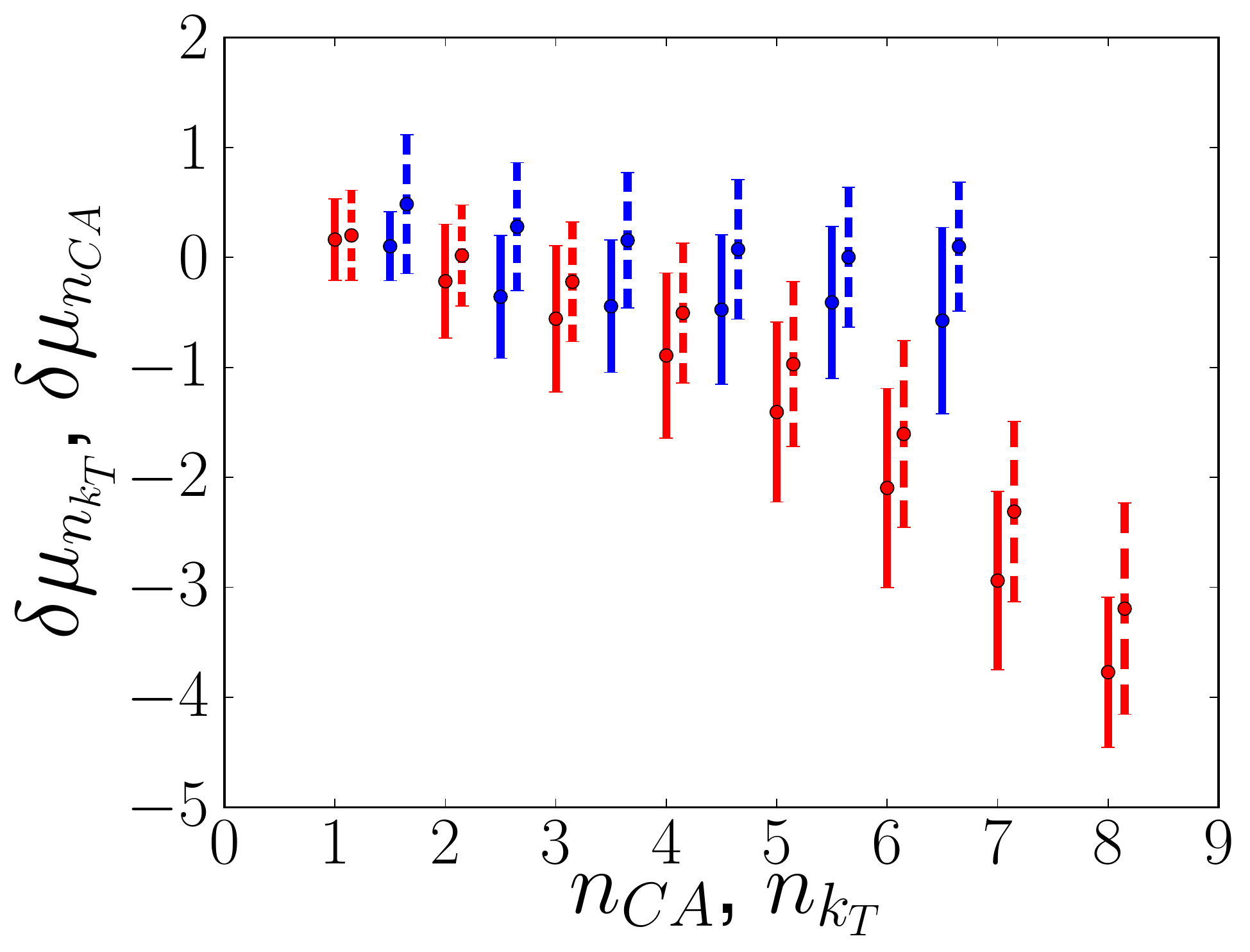}
    \caption{{\it Left:} The subjet distributions $\log_{10} P_\nkt(\nkt)$ and $\log_{10} P_\nca(\nca)$ in blue and red, respectively.  {\it Right:} The blue and red distributions show $\delta\mu_\nca(\nkt)$ and $\delta\mu_\nkt(\nca)$, respectively, which are defined in Eq.~\ref{Eq: CorrelationMean2}. The error bars on 
  $\delta \mu_\nca(\nkt)$ and $\delta\mu_\nkt(\nca)$ correspond to the standard deviations $\sigma_\nca$ and $\sigma_\nkt$.
 For both panels the solid lines correspond to a sample of leading QCD jets generated by {\tt MadGraph} and with $p_T\ge 100\GeV$, while dashed lines
 correspond to a signal-like sample described in the text.\label{Fig: Correlation}}
\end{figure}

On a jet-by-jet basis, there are strong correlations between $\nkt$ and $\nca$, with $\nca$ 
typically yielding more subjets than $\nkt$.  Fig.~\ref{Fig: exampleNs} illustrates a pronounced example of the tendency for $\nca$ to identify more subjets.   
For a given ensemble of jets, it is useful to define the normalized distribution $P(n)$, 
which is the fraction of jets with $n$ subjets. The left panel of Fig.~\ref{Fig: Correlation} shows $P(n)$ for both algorithms for a 
sample of leading QCD jets generated by {\tt MadGraph} and with $p_T\ge 100\GeV$.  Also illustrated is $P(n)$ for a sample\footnote{The {\tt MadGraph} and signal-like samples, which are used throughout this section, are described in more detail in \secref{QCDMG} and \ref{Sec: SimplifiedModels}, respectively.  The {\tt MadGraph} samples are used because of the superior statistics available.} of leading jets drawn from 
signal events where pair produced gluinos decay via $\tilde{g}\rightarrow t\bar{t}\chi$ ($m_{\tilde{g}}=600\GeV$ and $m_\chi=60\GeV$).  

To study how $\nkt$ and $\nca$ are correlated, it is useful to introduce the joint distribution
\begin{eqnarray}
\label{Eq: CorrelationProb}
P( \nkt, \nca) \qquad \text{ with} \qquad \sum_{\nkt, \nca} P(\nkt,\nca) =1
\end{eqnarray}
From the joint distribution one can define the mean of $\nca$ as a function $\nkt$ as well as the mean of $\nkt$ as a function of $\nca$:
\begin{eqnarray}
\label{Eq: CorrelationMean}
\mu_\nca(\nkt) = \sum_\nca  \nca {P}(\nkt,\nca) 
\quad \text{ and }\quad
\mu_\nkt(\nca)= \sum_\nkt  \nkt {P}(\nkt, \nca)
\end{eqnarray}
From this one can then define the quantities
\begin{eqnarray}
\label{Eq: CorrelationMean2}
\delta \mu_\nca(\nkt)= \mu_\nca - \nkt \qquad \text{ and } \qquad 
\delta \mu_\nkt(\nca) = \mu_\nkt - \nca
\end{eqnarray}
which are shown in the right panel of Fig.~\ref{Fig: Correlation}.  It can be seen that that $\nkt$ and $\nca$ track one another
pretty closely for small $n$, but that for larger numbers of subjets $\nca$ tends to pull further and further ahead of $\nkt$.  The correlation
between $\nkt$ and $\nca$ is somewhat tighter in the signal-like sample than in the QCD sample.

For fixed $\nkt$ the distribution $P(\nkt, \nca)$ is a function of $\nca$ with standard deviation $\sigma_\nca(\nkt)$. The standard deviation $\sigma_\nca(\nkt)$
is a steadily rising function of $\nkt$, as illustrated by the error bars in the right panel of Fig.~\ref{Fig: Correlation}. For the QCD sample it grows from 
$\sigma_\nca(1) \simeq 0.3$  to $\sigma_\nca(6)\simeq 1.0$, indicating that for small $n$ the two algorithms identify very similar numbers of subjets, 
but that as the amount of substructure grows there is less agreement between the two algorithms.
Similarly the standard deviation $\sigma_\nkt(\nca)$ grows approximately linearly from $\sigma_\nkt(1)\simeq 0.4$ to $\sigma_\nkt(8)\simeq 0.9$. 
Note that this dispersion is logically distinct from the divergence in the mean seen in $\delta\mu_\nca(\nkt)$.

Note that both $\nca$ and $\nkt$ are peaked at 3 for the leading jet of the signal (see Fig.~\ref{Fig: Correlation}).
This is as expected, since this signal contains up to 12 final state quarks, with the result that, if the leading four fat jets
capture all of the decay products, an average of 3 subjets per fat jet are expected.


Interestingly, the subjet distributions for QCD fat jets are not governed by approximate ``staircase scaling,'' as one might have expected. 
This is a scaling pattern defined by the condition that, if we define the ratio
\begin{eqnarray}
w(n)\equiv\frac{P(n+1)}{P(n)}
\end{eqnarray}
then $w(n)$ is a constant independent of $n$.
Instead, a significantly steeper distribution is seen.    The variable
\begin{eqnarray}
r(n) \equiv \frac{w(n+1)}{w(n)}= \frac{P(n) P(n+2)}{P(n+1)^2}  
\end{eqnarray}
\begin{figure}[t]
    \includegraphics[width = 0.45\linewidth]{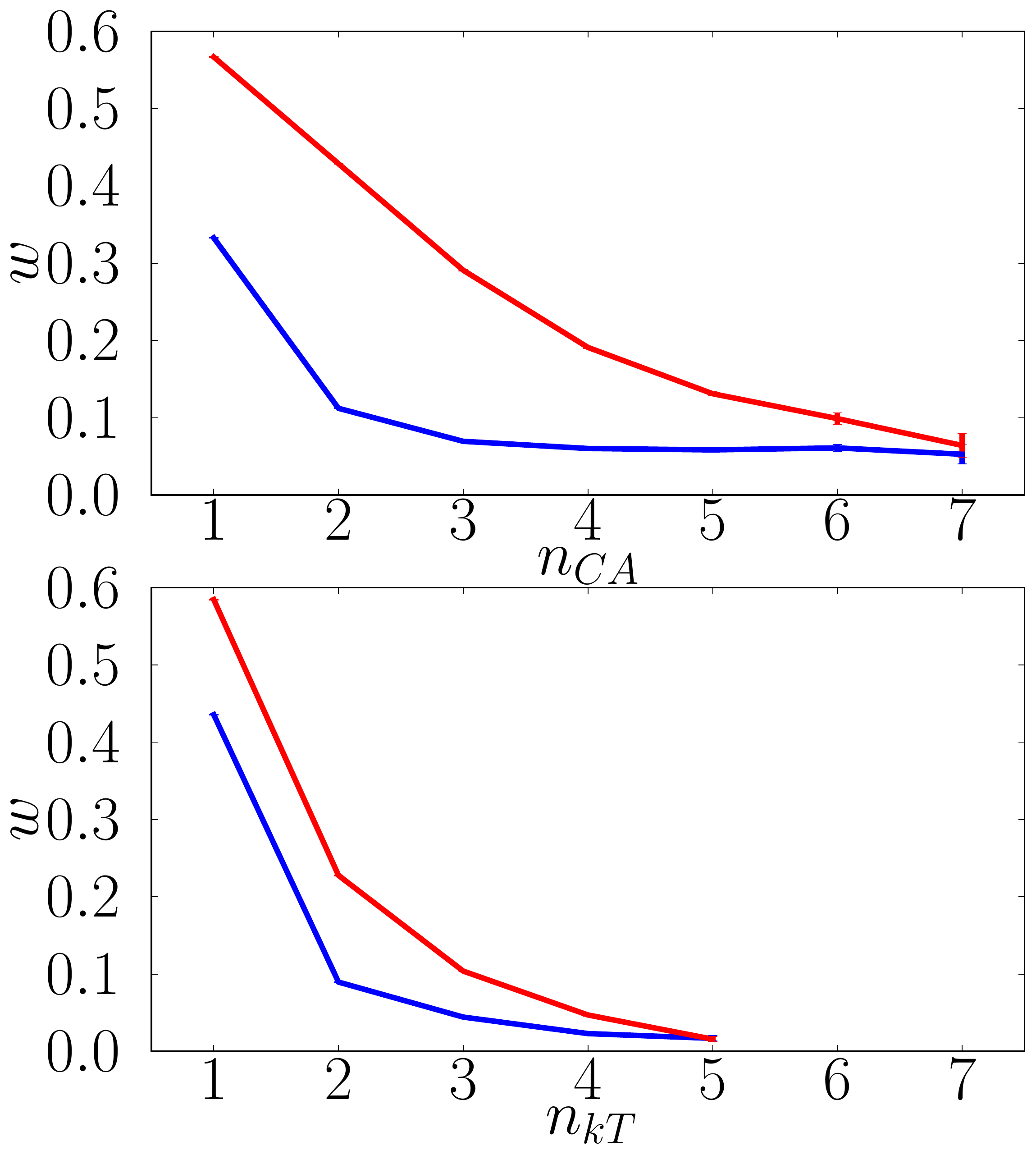}\qquad
    \includegraphics[width = 0.45\linewidth]{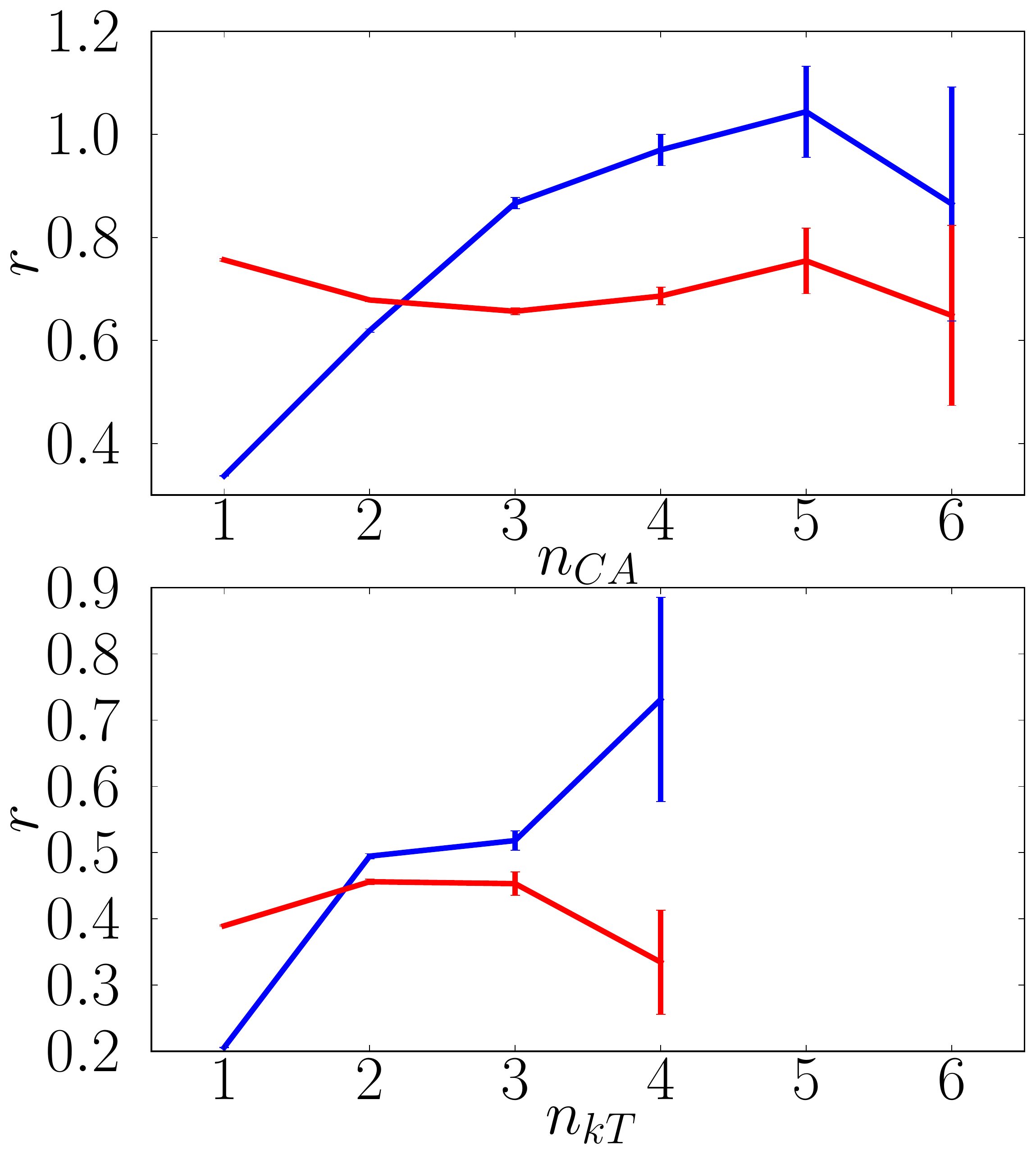}
    \caption{ Scaling patterns for the subjet distributions of leading QCD fat jets generated by {\tt MadGraph}. {\it Left}: The ratio $w(n)$ for $\nkt$ (bottom) and $\nca$ (top) for 
     jets with $p_T\ge 100\GeV$ (blue) and $p_T\ge 500\GeV$ (red).  
    ``Staircase scaling'' corresponds to a flat $w(n)$.  {\it Right}: The ratio $r(n)$ for $\nkt$ (bottom) and $\nca$ (top) for jets with 
    $p_T\ge 100\GeV$ (blue) and $p_T\ge 500\GeV$ (red).    ``Double staircase scaling'' corresponds to a flat $r(n)$ and
     appears to be emerging in the high $p_T$ regime.\label{Fig: Spectrum}}
\end{figure}
is useful for characterizing deviations from constant $w$, with staircase scaling corresponding to $r=1$.
For the case of QCD fat jets the ratios $w(n)$ and $r(n)$ for both $\nkt$ and $\nca$ are illustrated in Fig.~\ref{Fig: Spectrum}.
Since there are intrinsic energy scales in both $\nkt$ and $\nca$, low $p_T$ jets  are sensitive to these scales.  
For $p_T\ge 100\GeV$, the $w(n)$ distribution drops rapidly
before asymptoting to a constant of   $w_\nca \simeq 0.05$ and $w_\nkt\simeq 0.03$, indicating staircase scaling with a very hard spectrum.  
Requiring higher $p_T$ jets makes the influence of the intrinsic scales in the counting algorithms less relevant.  If the minimum $p_T$ of the jet is raised to $500\GeV$, the scales inside the subjet counting algorithm will play a much smaller role.  
Even for these high $p_T$ jets staircase scaling is {\it not} observed; instead $w(n)$ appears to have a staircase behavior and $r(n)$ is approximately
constant with $r_\nca\simeq 0.5$ and $r_\nkt\simeq 0.4$.  This can be called ``double staircase scaling'' and indicates a distribution of subjets that scales like
$$P(n) \sim r^{n^2/2}.$$
This is in contrast to the more traditional staircase scaling that gives
$$P(n)\sim w^{n}$$
demonstrating that subjet counting is not related in a straightforward manner to jet counting.  Deviations from staircase scaling is well known and have even been explained in \cite{Gerwick:2012fw}; however, the deviations from staircase scaling noted here do not appear to be ``Poisson" which predicts that $r$ should asymptote to unity which does not appear to happen.

Finally, it is worth noting that some of the quantitative differences between $\nca$ and $\nkt$ are being driven by the $p_T$-ordering of the
parton shower algorithms in {\tt PYTHIA} and {\tt SHERPA}.  This is important because the definition of $\nca$ is motivated by (the unclustering
of) angular-ordered emissions, while $\nkt$ is motivated by $p_T$-ordered emissions. As we will see in \secref{Searches}, from the point of view
of discriminating QCD backgrounds from high multiplicity signals, what is most important is the heaviness of QCD tails (i.e.~the fraction of fat QCD
jets with $\nkt, \nca \gtrsim 3$).
Preliminary studies indicate that a $p_T$-ordered (angular-ordered) parton shower tends to lead to fatter QCD tails for $\nca$ ($\nkt$). Consequently,
the results in this paper, which rely on $p_T$-ordered parton showers, probably place $\nkt$ at a relative advantage over $\nca$ as far as QCD
discrimination is concerned.  
Given the relatively better agreement between LHC jet substructure data and {\tt HERWIG++} as compared to {\tt PYTHIA 6.4}
(at least as far as leading order Monte Carlo is concerned; see e.g.~ref.~\cite{Aad:2013gja}), we suspect that $\nca$ is likely
to outperform $\nkt$ on real data.
Closer scrutiny of parton shower dependence would be an important part of any future experimental study.


\section{Monte Carlo Calculations}
\label{Sec: MC}

This article studies the use of the total number of subjets in an event as a way to separate new physics scenarios that produce many final state quarks and gluons from QCD and electroweak-scale backgrounds.  The ultimate goal is to reduce the reliance upon missing transverse energy ($\MET$) and lepton
requirements so as not to veto on signals that have neither, while still obtaining relatively low background search regions. Since $\MET$ and leptons are the standard handles used to reduce QCD backgrounds, it is particularly important to have reliable estimates of multijet production rates and differential distributions. Of course, the same holds true for the non-QCD backgrounds. Consequently this section and the next play an important role in everything that follows.  Throughout, the calculations are performed at a center of mass energy of $\sqrt{s} = 8 \TeV$.

The rest of this section is organized as follows.
\secref{EWBGs} describes the calculation of non-QCD backgrounds such as $V$+jets and $t\bar{t}+$jets.  
\secref{QCD} describes the two approaches used to generate QCD Monte Carlo events.   
\secref{detector} describes how detector effects and jet clustering are implemented.
These latter two subsections are complemented by the discussion in \secref{DD}, which focuses on data driven methods.

\subsection{Non-QCD Backgrounds}
\label{Sec: EWBGs}

The dominant Standard Model backgrounds are QCD, $V$+jets and $t\bar{t}$+jets.  Since, however, any backgrounds where 
there is an intrinsic mass scale are potentially important, the leading subdominant backgrounds were also computed for completeness, see Table~\ref{Tab: BGs}.  
The non-QCD backgrounds used for our analyses were generated using {\tt MadGraph 4.5.1} \cite{Steltzer:1994aa,Maltoni:2003aa,Alwall:2007st} and showered and hadronized using {\tt PYTHIA 6.4} \cite{Sjostrand:2006za} . The five-flavor MLM matching scheme with a shower-$k_\perp$ scheme was used to account for the extra radiation \cite{mlm}.

\begin{table}[t]
\begin{tabular}{|c||c|c|c|c|c|}
\hline
Process&Order &$p_T$ range (GeV) & Cross Section (fb)&Events&Event Weight\\
\hline\hline
  $V+ n_3 j$ &$\OO(\alpha_w\alpha_s^3)$&0  - 100&8,586,000&$2.5\times10^6$& 103.0 \\
    &$\OO(\alpha_w\alpha_s^3)$&100- 200&949,000&$2.5\times10^6$& 11.4 \\
  &$\OO(\alpha_w\alpha_s^3)$&200- 300&72,400&$2.5\times10^6$& 0.87 \\
   &$\OO(\alpha_w\alpha_s^3)$&300+&15,200&$2.5\times10^6$& 0.18 \\
\hline
$t\bar{t}+ n_2 j $&$\OO(\alpha_s^4), \OO(\alpha_w^2\alpha_s^2)$&0-150&94,300 &$3.5\times10^5$&8.08 \\
&$\OO(\alpha_s^4), \OO(\alpha_w^2\alpha_s^2)$&150-300&33,900&$3.5\times10^5$&2.90 \\
&$\OO(\alpha_s^4), \OO(\alpha_w^2\alpha_s^2)$&300+&4,440&$3.5\times10^6$&0.38 \\
\hline
$VV'+n_2 j\;*$&$\OO(\alpha_w^2\alpha_s^2)$&0+&51,500&$5.4\times10^5$&2.86\\
\hline
$Vt+n_2j\;*$&$\OO(\alpha_w\alpha_s^3)$&0-200&13,870&$1.8\times10^5$&2.31\\
&$\OO(\alpha_w\alpha_s^3)$&200-300&984&$1.8\times10^5$&0.16\\
&$\OO(\alpha_w\alpha_s^3)$&300+&319&$1.8\times10^5$&0.053\\
\hline
$t+n_3 j$&$\OO(\alpha_w^2\alpha_s^2)$&0-200&11,800&$2.6\times10^5$&1.32\\
&$\OO(\alpha_w^2\alpha_s^2)$&200-300&1,430&$2.6\times10^5$&0.16\\
&$\OO(\alpha_w^2\alpha_s^2)$&300+&355&$2.6\times10^5$&0.040\\
\hline
$VH$&$\OO(\alpha_w^2)$&0+&975&$3.0\times 10^4$&0.98\\
\hline
$t \bar{t} V+n_1 j$&$\OO(\alpha_w\alpha_s^3)$&0+&310&$3.0\times 10^4$&0.31\\
\hline
$t\bar{t} H+ n_1j$&$\OO(\alpha_w\alpha_s^3)$&0+&130&$3.0\times 10^4$&0.13\\
\hline
$t\bar{t}t\bar{t}$&$\OO(\alpha_s^4)$&0+&0.8&$1.0\times 10^4$&0.00024\\
\hline
\end{tabular}
\caption{The non-QCD backgrounds used in this analysis.   The subscript $i$ in $n_{i}$ indicates the highest jet multiplicity considered in the matched sample.   Thus $t\bar{t}+n_2j$ is $t\bar{t}+0j$, $t\bar{t}+1j$, $t\bar{t}+2^+j$, where the last jet multiplicity is an inclusive process that
can include higher jet multiplicities generated through the parton shower.  The $p_T$ slicing is with respect to the leading massive object. The two samples marked with a $*$ denote that resonant top production is excluded to avoid double counting. The last column indicates the ratio between the expected
number of events at $30\ifb$ and the number of Monte Carlo events generated.} \label{Tab: BGs}
\end{table}
 
It is the high $p_T$ tails of these backgrounds that will make the dominant contribution to signal regions.  Since some of these backgrounds have quite large cross sections, it is not feasible to generate $30\ifb$ worth of Monte Carlo.  Instead, the backgrounds are generated in different $p_T$ bins of the heavy particle, $p_{T\;\text{heavy}}$, where a heavy particle is any one of $t$, $W^\pm$, $Z^0$, or $H^0$.\footnote{When there are two or more heavy particles in the event, e.g.~$t\bar{t}$, $p_{T\; \text{heavy}}$ denotes the larger $p_T$.}  This ensures that the regions of phase space that 
result in the largest contamination of the signal region are computed with sizeable significance.
Another important consideration is that events with small $p_{T\text{ heavy}}$ can still have large $H_T$ 
as a consequence of high jet activity.  Since these backgrounds are potentially important for regions of phase space with sizeable $\MET$, it is important
for them to be computed with sufficient significance.  However, if $p_{T\text{ heavy}}$ is small, then so is the intrinsic $\MET$, with the consequence that
the high $H_T$/low $p_{T\text{ heavy}}$ region of phase space is not important for the signal region because it is subdominant to the QCD contribution. In practice, the large event weights of the low $p_{T\text{ heavy}}$ regions do not  limit the statistical accuracy of the background estimate.

The resulting backgrounds are shown in Table~\ref{Tab: BGs}, where the subscript $i$ in $n_{i}$ indicates the highest jet 
multiplicity considered in the matched sample.  The different $p_{T\;\text{heavy}}$ bins are listed in the third column. The five flavor matching scheme introduces one additional complication for diboson and single top production.  This arises because there can be resonant top production in these two channels that has already been included in ordinary $t\bar{t}$ production.  In order to exclude this double counting, a requirement of $|m_t -m_{bW}|> 15 \Gamma_t$ is imposed. 

The two most important non-QCD backgrounds for our search are $V+$jets and $t\bar t+$jets.   These backgrounds not only have large cross sections but are also jet rich and result in a reasonable amount of $\MET$. Backgrounds like $t\bar tt\bar t$, $t\bar t V$ and $t\bar t H$, which have jet multiplicities and $\MET$ comparable to that of some of the benchmark signals we study, have small cross sections and make a negligible contribution to the total background (although we include them for completeness).

\subsection{QCD}
\label{Sec: QCD}

 Several techniques are used to calculate the QCD contribution to signal regions. 
  \secref{QCDMG} describes a calculation of the QCD background using an MLM-matching scheme implemented in {\tt MadGraph} and {\tt PYTHIA} 
  using unweighted events with up to four partons matched.   This is a relatively low multiplicity method of generating backgrounds and relies heavily on the parton shower to generate high multiplicities; nevertheless, it is a standard calculational method that makes it easy to to get good Monte Carlo statistics over the entire signal region.  \secref{QCDSh} describes a calculation of the QCD background using a CKKW-matching scheme implemented in {\tt SHERPA} 
using weighted events with up to six partons matched. This method allows significantly higher multiplicities to be generated and samples the high energy
and high multiplicity tails with weighted events. The use of weighted events tends to hurt convergence and gives relatively poor Monte Carlo statistics.  

\subsubsection{{\tt MadGraph}+{\tt PYTHIA}}
\label{Sec: QCDMG}

One set of QCD backgrounds was generated with {\tt MadGraph 4.5.1} \cite{Steltzer:1994aa,Maltoni:2003aa,Alwall:2007st} at $\OO(\alpha_s^4)$ and showered with {\tt PYTHIA 6.4} \cite{Sjostrand:2006za} using MLM matching \cite{mlm}.  The events were generated in multiple exclusive samples varying both the matrix-element parton multiplicity from $n_j = 2$ to $n_j=4$ and the scalar transverse energy $H_T$ using the 5-flavor matching scheme \cite{Alekhin:2009ni}.  It is not practical to generate multiplicities higher than $n_j=4$ in {\tt MadGraph} due to computational limitations. Since the event selection for our search
strategy requires $n_j\ge4$, all jet substructure will be modeled by the parton shower.  This is clearly insufficient for
gaining much confidence in the resulting background estimates.  Nevertheless, the {\tt MadGraph} events will serve as a useful crosscheck in
what follows.
In addition, the high statistics available from {\tt MadGraph} will be very useful in creating missing energy templates in \secref{QCDMET}.

\subsubsection{\tt{SHERPA}}
\label{Sec: QCDSh}

The second event generator used to generate QCD backgrounds is {\tt SHERPA 1.4.0} \cite{Gleisberg:2008ta,Krauss:2001iv,Schumann:2008aa,Gleisberg:2008bb,Hoeche:2009aa}.  {\tt SHERPA} uses CKKW matching \cite{Catani:2001cc} and is capable of generating up to $n_j=6$ at the matrix element level.  {\tt SHERPA} can therefore generate more hard substructure using matrix elements without relying on the parton shower. Consequently we will primarily be relying on {\tt SHERPA} for our QCD background estimates.
This sample was generated and fully described in \cite{Cohen:2012yc}\footnote{We thank Tim Cohen, Mariangela Lisanti, Eder Izaguirre, and Tim Lou for letting us use this Monte Carlo sample.}. 

One of the drawbacks of {\tt SHERPA} is that it is not straightforward to generate separate samples binned by $H_T$. Instead weighted Monte
Carlo events can be generated.  The main problem with relying on weighted Monte Carlo is that it becomes more difficult to obtain
convergence in the signal region.  One frequently observes that a single (or handful of) high weight event(s) can make large
contributions to the tails of distributions, with the consequence that statistical uncertainties in the tails become large.

Two separate weighting methods were used in generating the QCD backgrounds.  The first is the default weighting procedure in {\tt SHERPA}.  
The second skews the weights towards higher multiplicities, with
\begin{eqnarray}
w(n_j) = 4^{n_j-2}
\end{eqnarray}
for $2\le n_j \le 6$.  Thus, 256 times as many $n_j=6$ events are generated as compared to $n_j=2$ events. 
This allowed for the generation of relatively more high multiplicity events so as to better fill out the tails of the distributions. 
Together these two weighting methods resulted in a total of $4.8\times10^6$ events passing our basic fat jet requirements (see \secref{fatjets}).

\subsubsection{Comparison of {\tt MadGraph} and {\tt SHERPA}}

In Figs.~\ref{Fig: SherpaPythiaJetComparison} and \ref{Fig: SherpaPythiaEventComparison} the {\tt SHERPA} results are compared to {\tt MadGraph+PYTHIA}.  We see that there is generally good agreement between the two.  Even the subjet count $N_{\rm{CA}}$ shows good agreement up to
$N_{\rm{CA}}=8$, a regime in which both generators (especially {\tt MadGraph}) are relying on the parton shower to generate substructure.  
The biggest differences appear in the tails of the distributions.
As discussed above, the presence of high weight events in {\tt SHERPA} can lead to poor convergence. Whenever there is a large disagreement
between the two generators in these figures, it seems to be largely driven by this effect (c.f.~the large statistical uncertainties in the regions
of largest disagreement).

The disagreement in the tails of the distributions between {\tt SHERPA} and {\tt MadGraph + PYTHIA} deserves further comment.    While the naive expectation is that {\tt SHERPA} should provide a better description in the high $H_T$, $M_J$ and $N_J$ regime, there are significant statistical uncertainties in the {\tt SHERPA} sample arising from slow convergence of the weighted Monte Carlo calculations.  The features of the distributions also appear to be slightly pathological in their behavior, appearing as inflection points in regions that one would expect to be relatively smooth.   In the case of $N_{\text{CA}}$, some of the bins are not even monotonically decreasing.  
This article will rely on the {\tt SHERPA}  calculation for its background estimates; fortunately the design of the search regions in \secref{Searches} will not be heavily influenced by these features. 
In \secref{DD}, a data driven approach to calculating these backgrounds is presented, and the disagreement between the data driven approach and the straight Monte Carlo calculation will be, at least in part, related to these same features.

\begin{figure}
\includegraphics[width=2.12in]{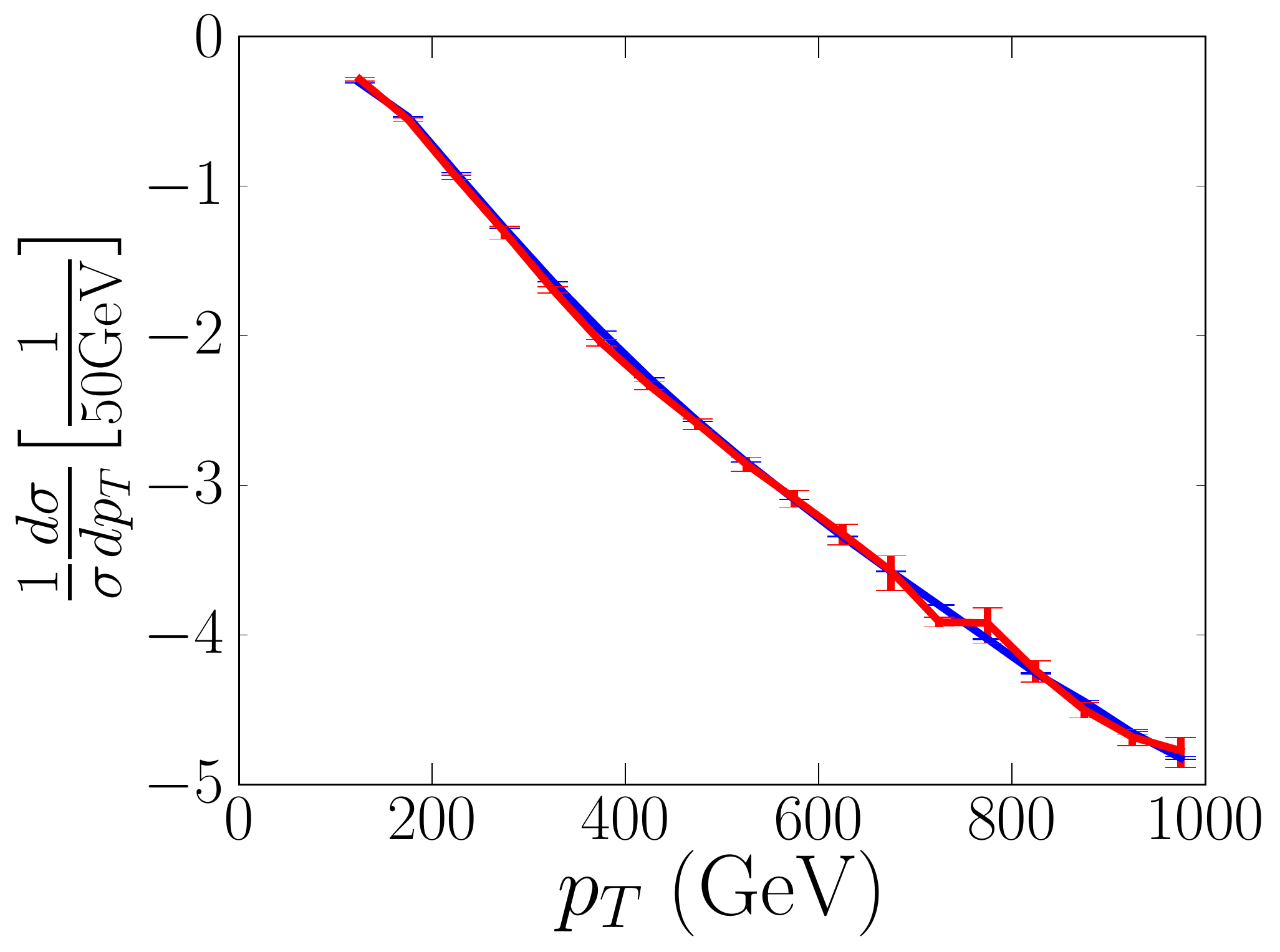}
\includegraphics[width=2.12in]{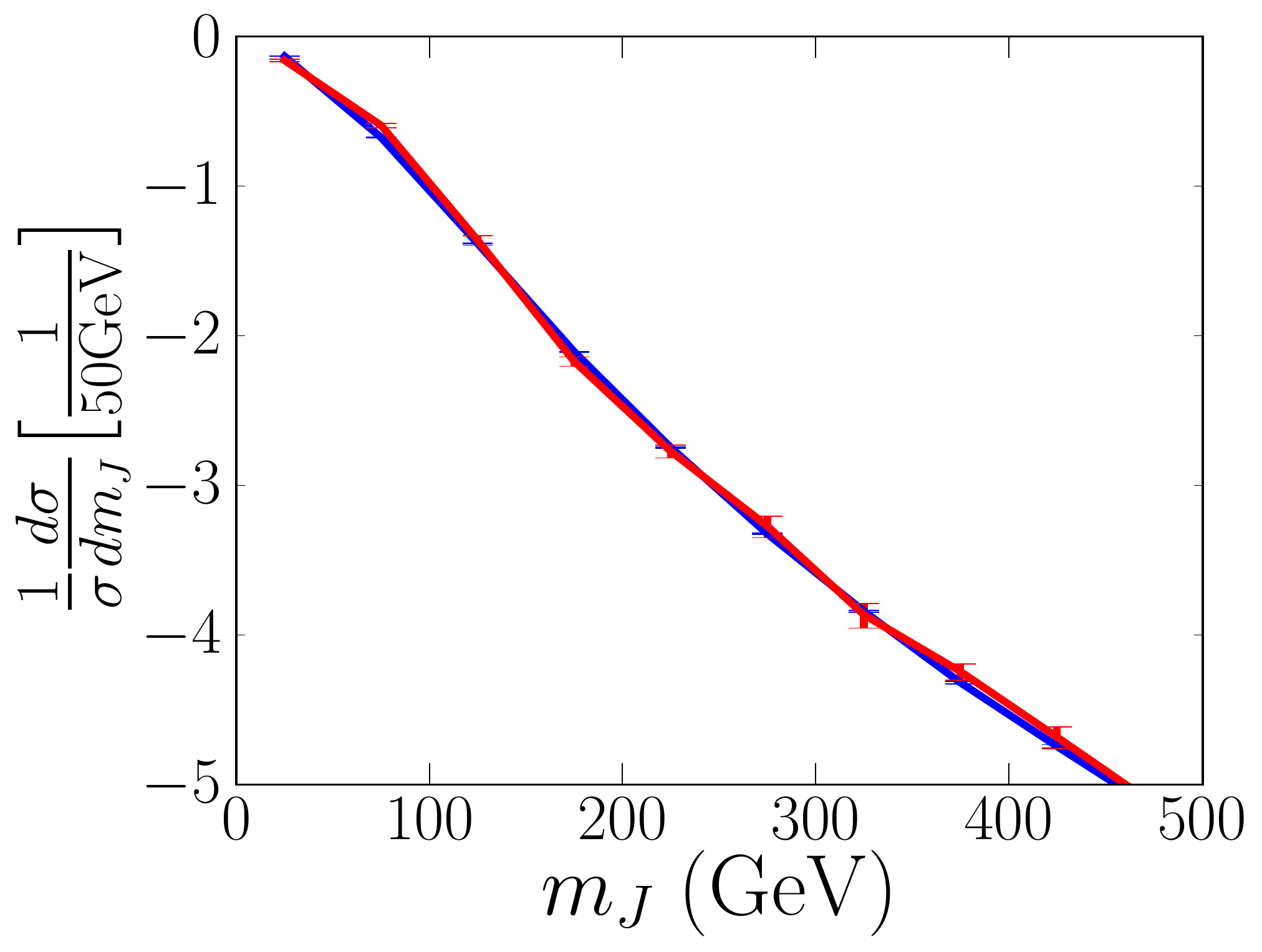}
\includegraphics[width=2.07in]{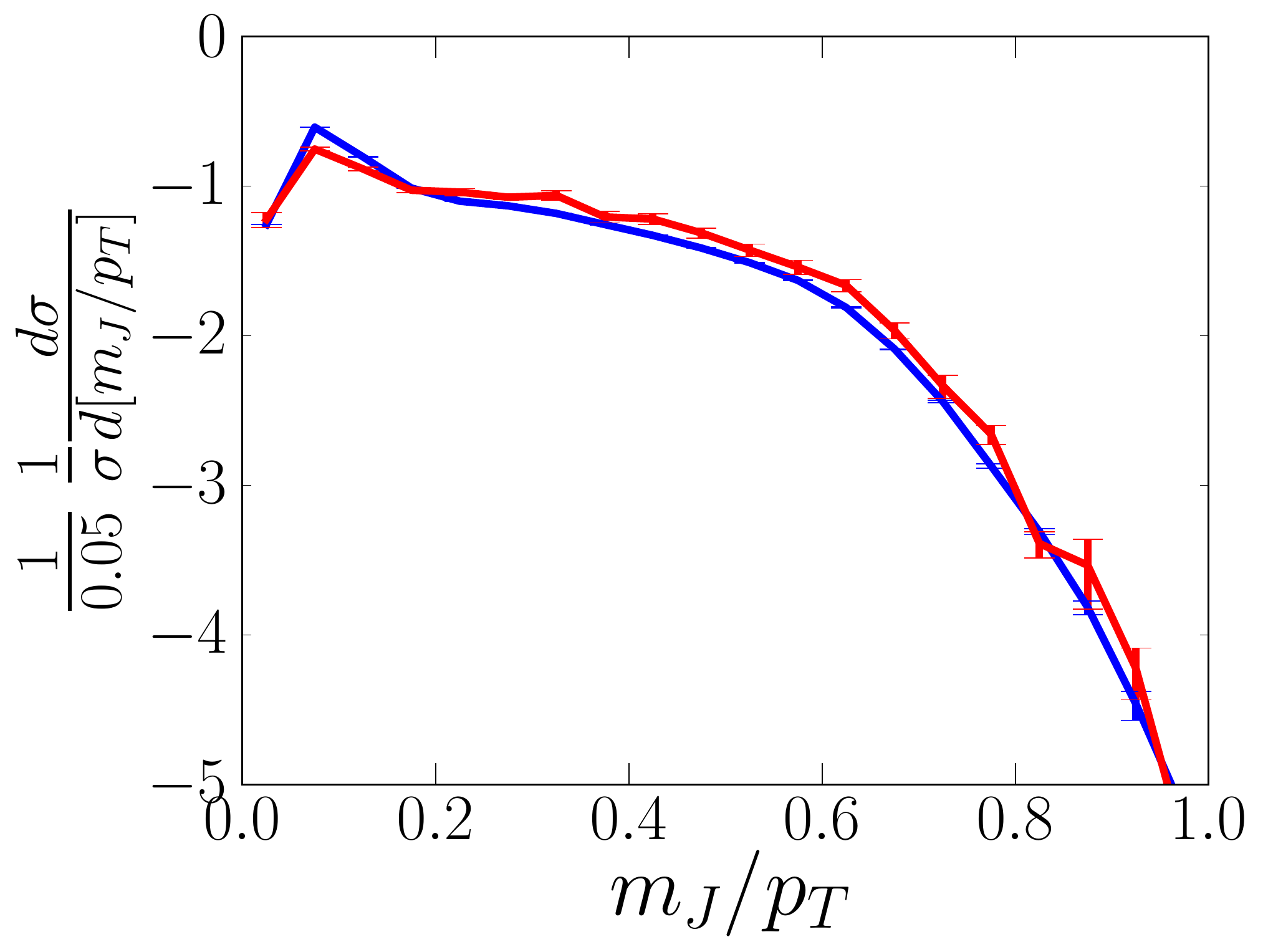}
\caption{Comparison between {\tt SHERPA} (red) and {\tt MadGraph+PYTHIA} (blue) for three relevant kinematic variables of the leading jet.} \label{Fig: SherpaPythiaJetComparison}
\end{figure}

\begin{figure}
\includegraphics[width=0.42\linewidth]{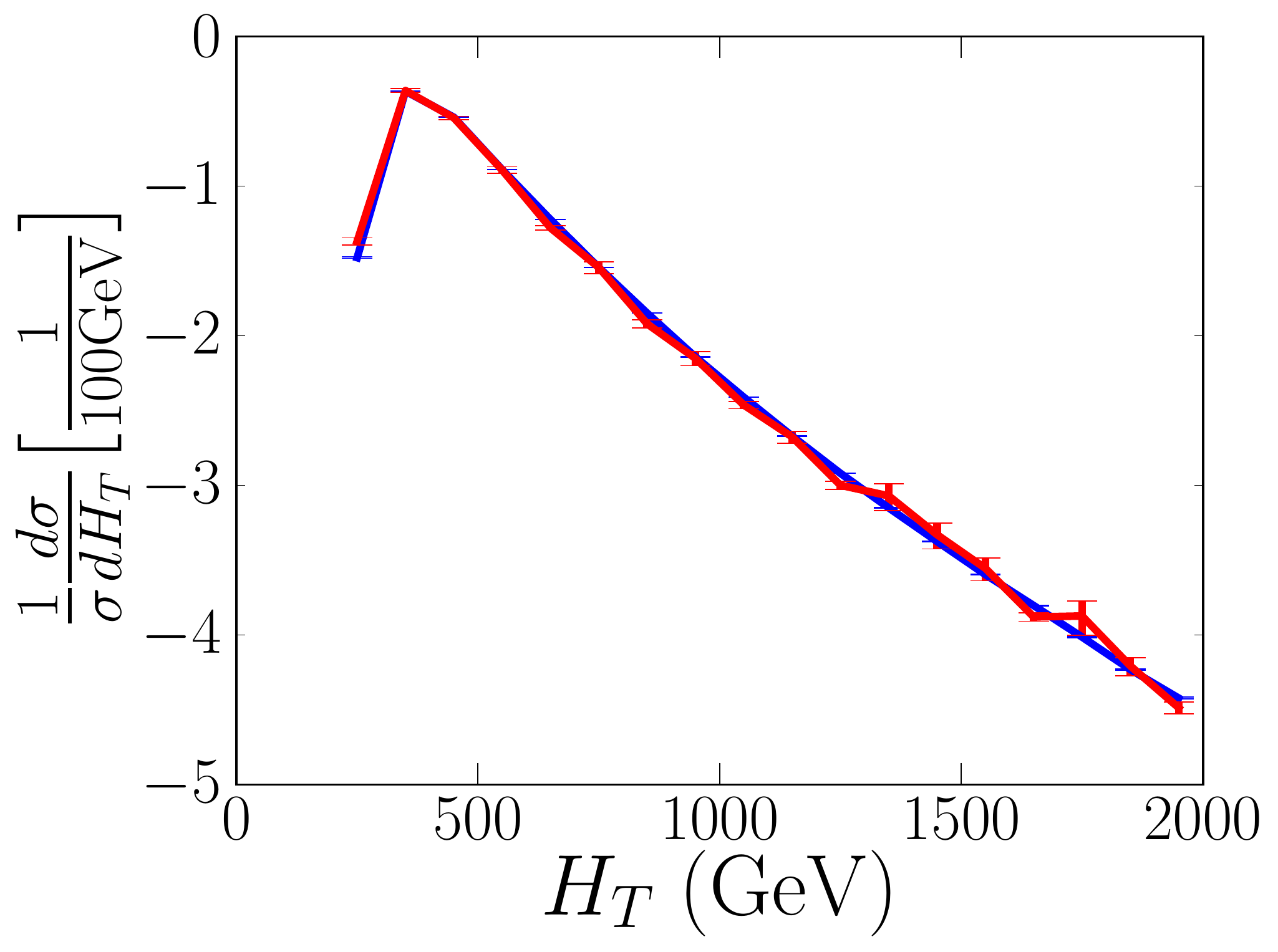}
\includegraphics[width=0.42\linewidth]{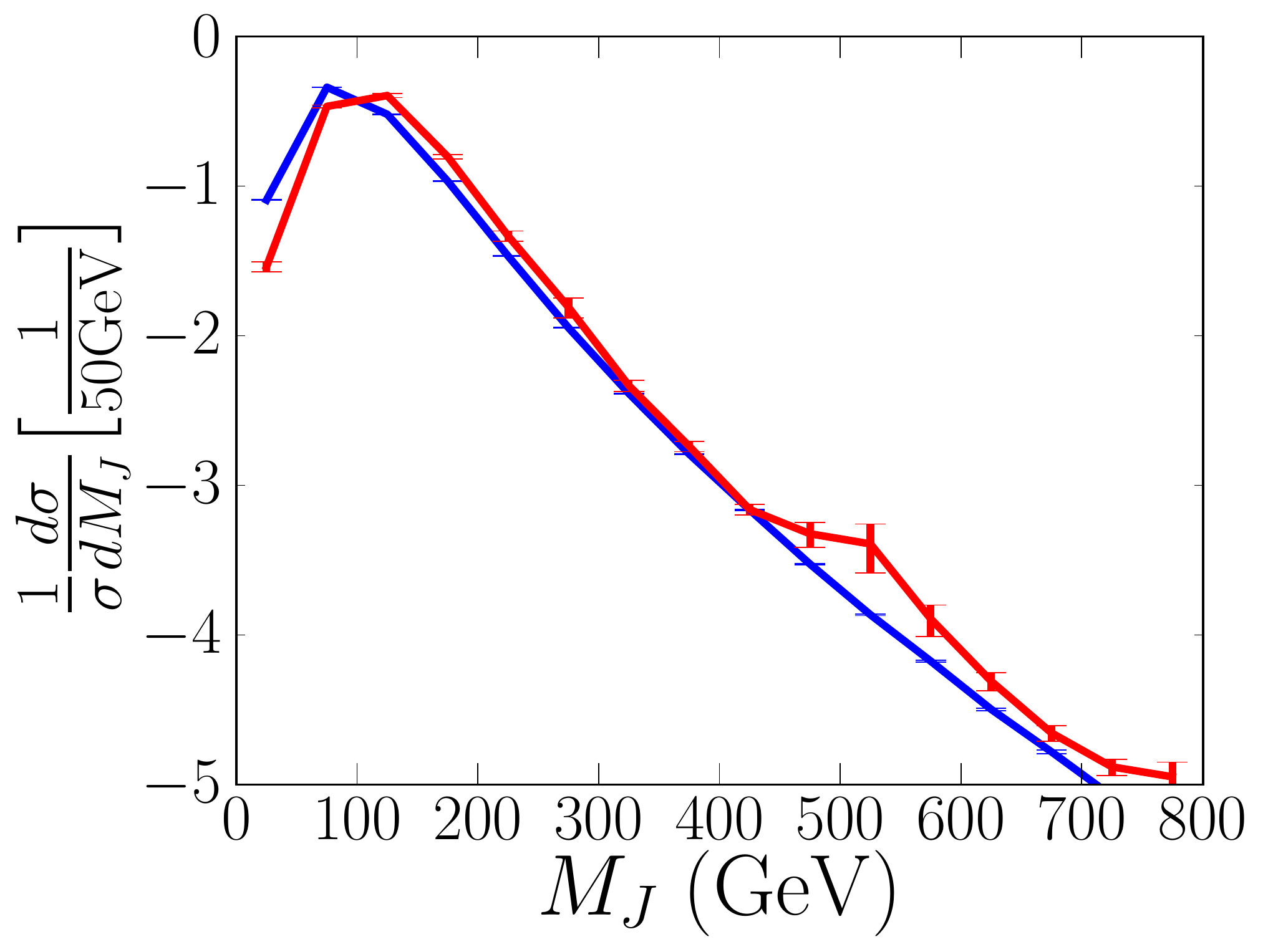}
\includegraphics[width=0.42\linewidth]{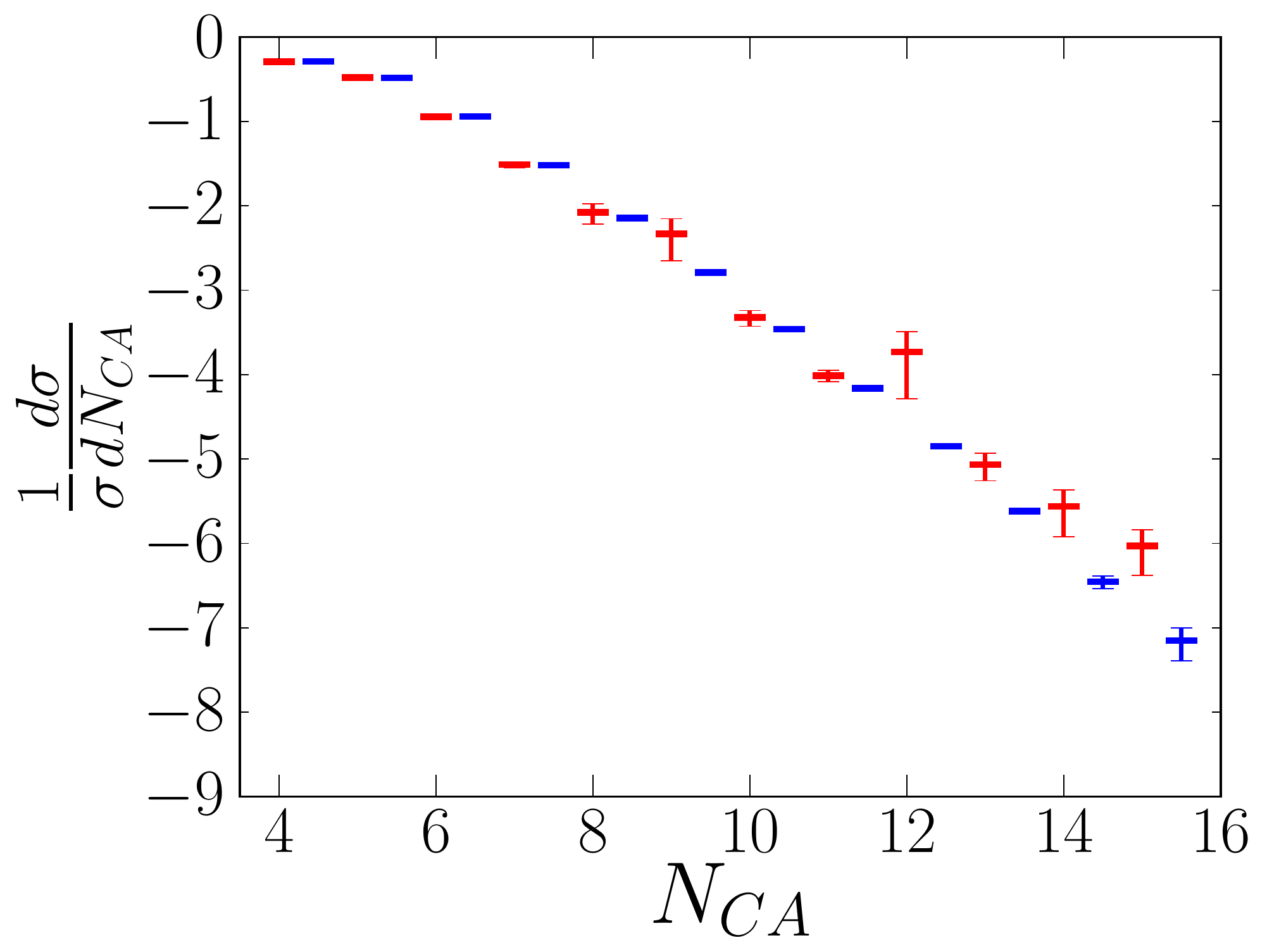}
\includegraphics[width=0.42\linewidth]{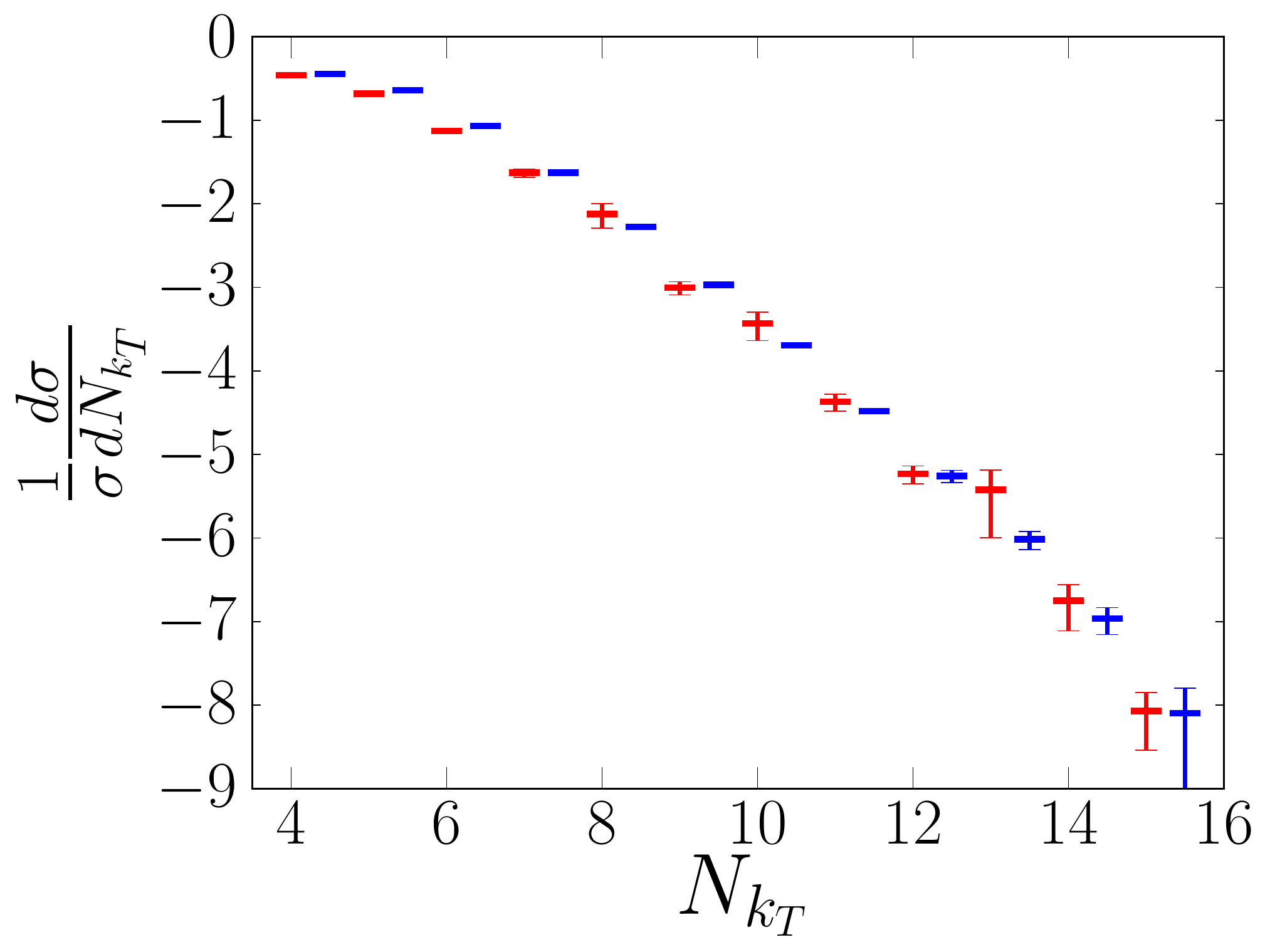}
\caption{Comparison between {\tt SHERPA} (red) and {\tt MadGraph+PYTHIA} (blue) for $H_T$ (upper left), $M_J$ (upper right), $N_{\rm{CA}}$ (lower left) and $N_{k_T}$ (lower right).  
For the definition of the latter two observables see \secref{Subjets}. For the bottom two distributions, the {\tt MadGraph+PYTHIA} points have been shifted by half a unit to the right in order to facilitate the comparison.} \label{Fig: SherpaPythiaEventComparison}
\end{figure}

\subsection{Detector mockup and jet clustering}
\label{Sec: detector}
After showering, all hadron-level events are passed to the {\tt PGS 4 }\cite{Conway:2009aa} detector simulation, which parameterizes the detector response. The detector parameters used are those of the default ATLAS {\tt PGS} card. The {\tt PGS} output is clustered into $0.1 \times 0.1$ cells in $\eta\!-\!\phi$ 
space, and then each cell is represented as a massless four-vector pseudoparticle.  Finally, those pseudoparticles with rapidities $|y|<2.5$ 
are fed into {\tt FastJet} 3, which we use for jet clustering \cite{Cacciari:2011ma}.  

The primary purpose of {\tt PGS} is to give estimates of the missing energy that arises from imperfect detectors.  As we will see in \secref{QCDMET}, for the QCD backgrounds
this is best accomplished by parametrizing the {\tt PGS} QCD missing energy spectrum in terms of template functions. In Sec.~\ref{Sec: validation}, our Monte Carlo background calculations are compared against published ATLAS data.  There we will see that the QCD missing energy templates will need to be rescaled to obtain a better fit to the data.  
{\tt PGS} is also useful for simulating lepton identification efficiencies.  However, for the primary proposal of this paper, no lepton requirements or vetoes are made, and
so lepton identification efficiencies will not play a large role.  

\subsection {Treatment of leptons}
The goal of this paper is to develop a search strategy that can dramatically reduce Standard Model backgrounds while making the least number
of assumptions about the characteristics of the final state apart from its having a high multiplicity.  Consequently, while
it may be advantageous to require or veto on something like b-tagged jets or isolated leptons for a particular signal model, we do not
do so here.  For the broad class of
signals we are interested in probing, the high final state multiplicity may be exclusively hadronic in origin or it may include some number of leptons.
For models with multiple possible cascade topologies (or indeed {\it any} with top quarks or electroweak gauge bosons in the final state) both the hadronic and
semi-hadronic modes may be simultaneously present, and signal discovery may require sensitivity to both channels.
Consequently throughout this study we treat leptonic energy as hadronic energy.  That is to say that in both fat jet clustering and 
subjet counting, hadronic and leptonic energy are treated democratically. 
This helps to ensure that signal efficiencies are not unnecessarily degraded.  It is interesting to ask whether alternative treatments
of the leptons might lead to effective search strategies without having to sacrifice the relative inclusiveness of the present search. Doing
so, however, lies outside the scope of this paper.  
\section{Data Driven Backgrounds}
\label{Sec: DD}

High multiplicity searches push into regions of phase space that are challenging to model, 
particularly in the case of the pure QCD backgrounds. 
For this reason it is important to have as many handles on the backgrounds as possible. In particular a data driven extrapolation 
of the background from a  control region to the signal region would be especially valuable for corroborating background estimates available from Monte Carlo.  In this section 
we explore how such a data driven estimate might be made.  In \secref{QCDMET} we specialize to the particular case of missing energy. The resulting
missing energy templates allow us to achieve significantly improved statistics in our Monte Carlo estimates of QCD missing energy acceptances.
In \secref{validation} we compare our Monte Carlo background estimates to published ATLAS data.
Finally in \secref{DataDriven} we discuss the possibility of extending the template method to take into account $M_J$ and $N_J$. These latter results are preliminary and
are not used for the background estimates that enter in the expected limits in \secref{Searches}.

\subsection{QCD missing energy templates}
\label{Sec: QCDMET}

One of the purposes of this work is to reduce the dependence of new physics searches on missing energy requirements, which are particularly
effective in reducing QCD backgrounds.  Thus it is important that we model $\MET$, and in particular QCD $\MET$, as accurately as possible.

QCD missing energy typically arises from two distinct sources at the LHC.  The first is from neutrinos lost in semi-leptonic decays of bottom and charm quarks.  This irreducible form of missing energy gives a long non-Gaussian tail to missing energy distributions but can be estimated through Monte Carlo calculations. The second form of missing energy arises from detector effects that result in particles being lost or otherwise mismeasured.  This form of missing energy is usually parameterized as a response function on the jet-by-jet level and is typically Gaussian for several standard deviations.  The 
typical amount of missing energy scales as the square root of the jet energy, 
although there is a small linear term that takes over at large jet energies. For QCD events it is
this latter form of missing energy, which arises from detector effects, that is dominant.

This article uses the approach of creating a probability distribution function for the missing energy of QCD events 
as a function of $H_T$ (c.f.~\secref{DataDriven}). This allows for a huge reduction in the number of Monte Carlo events 
necessary for accurate background estimates in the presence of missing energy requirements.
Because the detector response is orthogonal to other jet properties that will be used 
(and is in any case being parametrized by {\tt PGS}) this approach should faithfully reproduce 
the results of a much larger Monte Carlo calculation. 

\begin{figure}[t]
\includegraphics[width=3.2in]{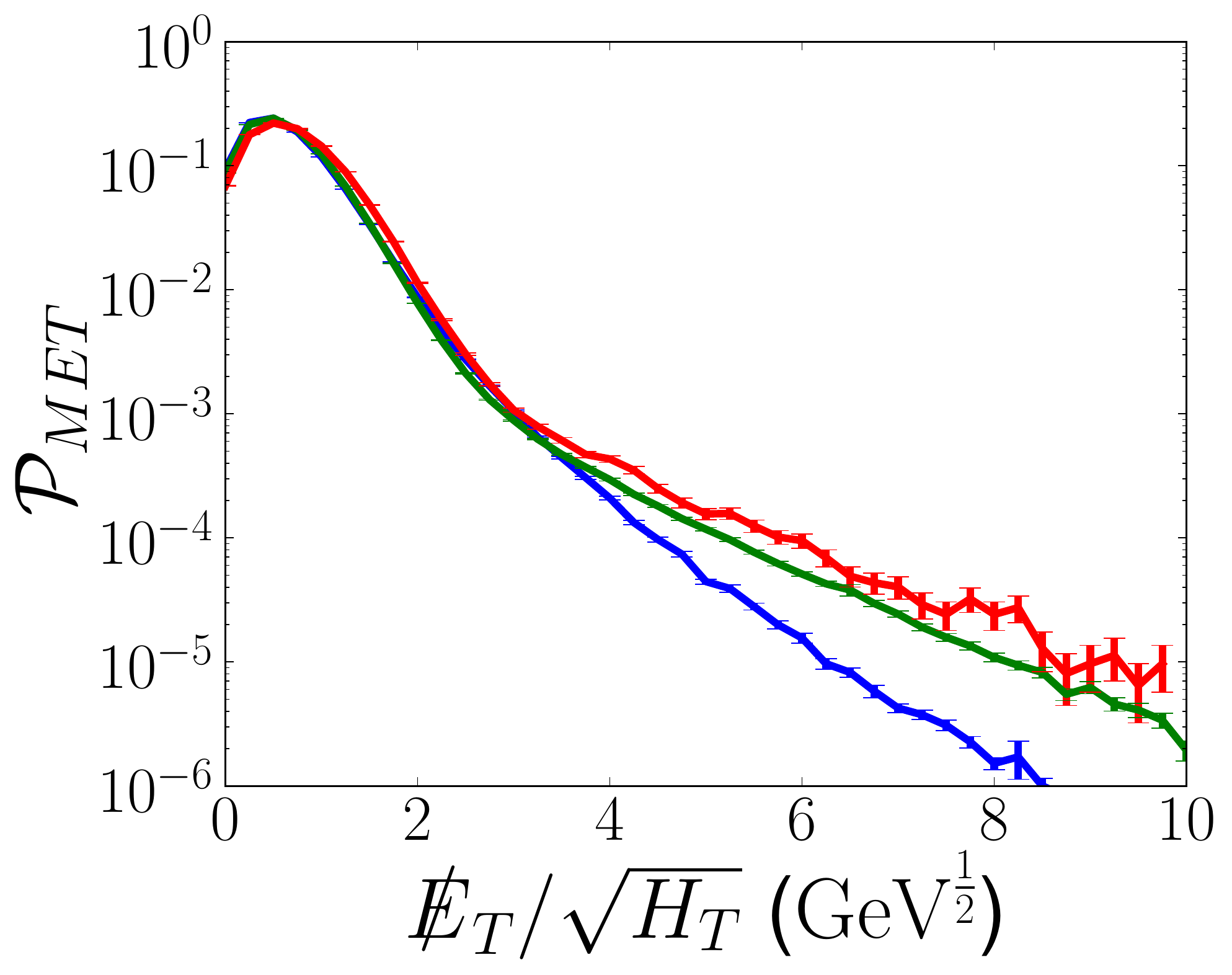}
\caption{The missing energy significance, $y=  \MET/\sqrt{H_T} \;(\GeV^\half)$ in three different $H_T$ bins: [300, 600]~GeV (blue), [900, 1200]~GeV (green), and [1500, 1800]~GeV (red).\label{Fig: METtemplates}}
\end{figure}

Specifically, a probability distribution function for missing energy significance, $$y\equiv\MET/\sqrt{H_T},$$ as a function of $H_T$ 
is constructed from the unweighted {\tt MadGraph} QCD sample:  
\begin{eqnarray}
    \mathcal{P}_{\rm{MET}}\left(y; H_T\right)\qquad\text{with}\qquad \int_0^\infty \!dy\; \mathcal{P}_{\rm{MET}}(y;H_T) =1
\end{eqnarray}
Thus for the rest of this study (where {\tt SHERPA} will be used for all our QCD background estimates), although the $H_T$ distribution will be modeled by 
${\tt SHERPA}$ the mapping $H_T \to \MET$ will be modeled by a combination of {\tt PGS} and {\tt MadGraph}.  We use {\tt MadGraph}
for this purpose because doing so yields much better statistics (the ability to target specific $H_T$ bins is one advantage).  
This PDF can be used event-by-event to determine the probability that an event, $e_i$, passes a specific $\MET$ requirement:
\begin{eqnarray}
P(e_i) = \int_{\MET{}_{\text{cut}}}^{\infty}\! dy \; \mathcal{P}_{\rm{MET}}(y ; H_{T\;e_i})
\end{eqnarray}
Fig.~\ref{Fig: METtemplates} shows the missing energy significance distributions for various $H_T$ windows. Note that these PDFs are just another way of
parametrizing {\tt PGS}'s detector response.  We will see below, in \secref{validation}, that it will be necessary to scale these PDFs to ensure a better
fit to published ATLAS data.

Looking forward to \secref{DataDriven}, the factorization we have introduced is given succinctly by 
\begin{eqnarray}
d\sigma(\MET, H_T) =  \mathcal{P}_{\rm{MET}}(\MET/\sqrt{H_T}; H_T) d\sigma(H_T)
\end{eqnarray}
which can be compared to the analogous expression in Eq.~\ref{probfactorization} below.


\subsubsection{Background validation}
\label{Sec: validation}

\begin{table}[t]
\begin{tabular}{|ccc||cc|cc|cc|cc|cc|}
\hline
\multicolumn{3}{|c||}{Signal Region}&\multicolumn{2}{|c|}{$Z^0/\gamma^*$+jets}&\multicolumn{2}{|c|}{$W^\pm$+jets}&\multicolumn{2}{|c|}{$VV$+jets}&\multicolumn{2}{|c|}{$t\bar{t}$}&\multicolumn{2}{|c|}{QCD, $\alpha=0.8$}
\\
\hline
$n_j$&$m_{\text{eff}}$&$\MET/m_{\text{eff}}$& {\tt MG4}& ATLAS& {\tt MG4}& ATLAS& {\tt MG4}& ATLAS&{\tt MG4}& ATLAS& {\tt SHERPA}& ATLAS\\
\hline\hline
2&1000&0.40&{\scriptsize 66.51}&{\scriptsize $61.0\pm 5.3$}&{\scriptsize 41.38}&{\scriptsize $31.0\pm 24.1$}&{\scriptsize 4.84}&{\scriptsize $9.14\pm3.97$}&{\scriptsize 18.59}&{\scriptsize $11.6\pm 2.8$}&{\scriptsize 0.14}&{\scriptsize $0.10\pm0.10$}\\
\hline
2& 1900&0.30&{\scriptsize 1.99}&{\scriptsize $1.17\pm 0.38$}&{\scriptsize 1.35}&{\scriptsize $0.52\pm 0.68$}&{\scriptsize 0.19}&{\scriptsize $0.57\pm 0.52$}&{\scriptsize 0.53}&{\scriptsize $0.13\pm 0.13$}&{\scriptsize 0.00}&{\scriptsize ---}\\
\hline
3&1300&0.30&{\scriptsize 9.25}&{\scriptsize $10.2\pm 1.0$}&{\scriptsize 5.31}&{\scriptsize $5.5\pm 5.5$}&{\scriptsize 0.99}&{\scriptsize $1.9\pm 0.9$}&{\scriptsize 3.23}&{\scriptsize $2.4\pm 0.9$}&{\scriptsize 0.01}&{\scriptsize $0.03\pm0.03$}\\
\hline
4&1000&0.30&{\scriptsize 9.93}&{\scriptsize $11.6\pm 1.0$}&{\scriptsize 7.49}&{\scriptsize $6.9\pm 6.9$}&{\scriptsize 1.25}&{\scriptsize $1.0\pm 0.7$}&{\scriptsize 9.22}&{\scriptsize $6.7\pm1.2$}&{\scriptsize 0.05}&{\scriptsize ---}\\
\hline
5& 1700&0.15&{\scriptsize 0.37}&{\scriptsize $0.43\pm 0.19$}&{\scriptsize 0.42}&{\scriptsize ---}&{\scriptsize 0.06}&{\scriptsize $0.14\pm 0.07$}&{\scriptsize 0.84}&{\scriptsize $0.44\pm 0.28$}&{\scriptsize 0.18}&{\scriptsize $0.07\pm0.09$}\\
\hline
6&1300&0.25&{\scriptsize 0.18}&{\scriptsize $0.10\pm 0.09$}&{\scriptsize 0.08}&{\scriptsize $0.16\pm0.19$}&{\scriptsize 0.03}&{\scriptsize ---}&{\scriptsize 0.54}&{\scriptsize $0.34\pm 0.24$}&{\scriptsize 0.00}&{\scriptsize ---}\\
\hline
6&1000&0.30&{\scriptsize 0.18}&{\scriptsize $0.34\pm0.17$}&{\scriptsize 0.13}&{\scriptsize $0.14\pm0.22$}&{\scriptsize 0.05}&{\scriptsize ---}&{\scriptsize 0.64}&{\scriptsize $0.43\pm0.16$}&{\scriptsize 0.00}&{\scriptsize ---}\\
\hline
\end{tabular}
\caption{Comparison between backgrounds and ATLAS data driven estimates \cite{ATLAS:2012ona}.  The QCD backgrounds
are generated with {\tt SHERPA} as described in Sec.~\ref{Sec: QCDSh}, whereas the others are generated with {\tt MadGraph + Pythia}
as described in \secref{EWBGs}.}
\label{Tab:validation}
\end{table}

We validated our backgrounds against the signal regions used by an ATLAS search for squarks and gluinos \cite{ATLAS:2012ona}.  This
lead us to rescale our $\MET$ PDFs (see \secref{QCDMET} and below),
since the ATLAS {\tt PGS} parameters were resulting in too large QCD $\MET$ acceptances.
The results after the rescaling are shown in Table~\ref{Tab:validation}.  Overall, our Monte Carlo background estimates agree with the data driven
estimates given in the ATLAS study.  In those cases where there is disagreement, our Monte Carlo
always overestimates the ATLAS estimates so that the expected sensitivities presented in Sec.~\ref{Sec: Searches} should be
reasonably conservative.

Each of the ATLAS signal regions includes a lepton veto.   
Because ATLAS lepton identification is only approximately reproduced by {\tt PGS}, we expect larger differences between
Monte Carlo estimates and ATLAS estimates for those backgrounds with leptons.  This is indeed what we find in Table~\ref{Tab:validation},
where backgrounds with no leptons (like $Z^0/\gamma^*$+jets) match the data well.  For events with $\MET$ arising from the decay of $W^\pm$ bosons,  such as $W^\pm$+jets and $t\bar{t}$+jets, the comparisons are systematically off because {\tt PGS} is not identifying isolated leptons with sufficiently
high efficiency.  This is not a major issue in the validation because the searches described in this article neither require nor veto on leptons.

As mentioned above, the {\tt PGS} treatment of angular and energy smearing does not faithfully reproduce the response of the ATLAS detector.  In particular, in the presence of $\MET$ cuts the ATLAS {\tt PGS} parameters result in an overestimate of QCD backgrounds by an order of magnitude solely due to the
treatment of $\MET$.  In order to better reproduce the QCD backgrounds, the $\MET$ templates are rescaled by
making the replacement
\begin{eqnarray}
  \mathcal{P}_{\rm{MET}}\left(\frac{\MET}{\sqrt{H_T}};H_T\right)\to  \mathcal{P}_{\rm{MET}}\left(\frac{1}{\alpha}\frac{\MET}{\sqrt{H_T}};H_T\right)
\end{eqnarray}
The best fit to ATLAS estimates for QCD backgrounds in $\MET$-rich regions is with $\alpha=0.8$.  This rescaled $\MET$ template leads to significantly improved agreement between the Monte Carlo and ATLAS  estimates of the QCD background.


\subsection{Fat jet templates}
\label{Sec: DataDriven}

The main obstacle to modeling 4-jet QCD production is the large dimensionality of the space of observables under consideration. The quantity we would like to understand is the 
9-dimensional 4-(fat)jet differential cross section $d\sigma_{\rm{4J}}(\MET,  m_i, n_i)$. 
Here $\MET$ is the missing energy of the 4-jet event, the $m_i$ are the masses of the
four jets, and the $n_i$ are the four subjet counts (using e.g.~the $n_{\rm{k_T}}$ algorithm defined in \secref{Subjets}).  With $d\sigma_{\rm{4J}}$
in hand the chosen cuts on $\MET$,
\begin{eqnarray}
 M_J\equiv \sum_i m_i \qquad\text{and}\qquad N_J \equiv \sum_i n_i
 \end{eqnarray}
 can be imposed, thus yielding the expected 
QCD background in the signal region. 

To make progress it is useful to reduce the dimensionality of the problem. This can be done by making the {\it assumption} that each of the four jets is
governed by a universal probability distribution 
\begin{eqnarray}
\rho_J(x, n; p_T)  
\end{eqnarray}
with
$$
x\equiv m/p_T
$$
which describes the probability of a fat jet having $n$ subjets and a particular value of $m/p_T$, as a function of the fat jet $p_T$. The mass of a fat jet 
is correlated with the number of subjets it contains, since it is impossible to get multiple subjets without having a sizable mass; consequently $\rho_J$ does not factorize and it is necessary to construct a two dimensional PDF.  

The assumption of universality is not completely valid, for one reason because quark-initiated jets and gluon-initiated jets will have different distributions.  The hope is that ensembles of jets will have similar ratios of quark- versus gluon-initiated jets and that the distribution functions will not be radically different.  In practice, this is the case since it is challenging to distinguish quark-initiated jets from gluon-initiated jets and since it is difficult to construct selection criteria that isolate one from the other.  The assumption of universality is even more aggressive, however, since it implies that these distribution functions are independent of their environment.  This assumption is known to be violated to some degree, particularly as jets come closer together, but the pull of the environment on properties of fat jets tends to be less than $\OO(10\%)$ in magnitude.\footnote{See e.g.~Figure 2 in ref.~\cite{Hook:2012fd}.}
In this section we will
be satisfied to check these assumptions empirically with Monte Carlo calculations, leaving a more detailed study to future work. Note that we will not be applying the resulting
background estimates when calculating the estimated sensitivity of our search strategy: the results presented in Sec.~\ref{Sec: Searches} will use the {\tt SHERPA} Monte Carlo calculations described in Sec.~\ref{Sec: QCDSh}. The results in this section are a first attempt at studying more aggressive uses of data driven approaches to QCD backgrounds.  

The assumption underlying the form of this jet template is that
a jet's substructure (e.g.~its mass) is determined by its $p_T$ and is independent of other jets in the event.
The full 4-jet distribution is then obtained via the product:
\begin{eqnarray}
d\sigma_{\rm{4J}}(\MET,  m_i, n_i;  p_{Ti}) = d\sigma_{\rm{4J}}(\MET,  p_{Ti}) \prod_{i=1}^4 \rho_J(x_i, n_i; p_{T\,i}) 
\end{eqnarray}
Here the $p_{Ti}$ are the transverse momenta of the four jets. Thus the 9-dimensional distribution $d\sigma_{\rm{4J}}(\MET,  m_i, n_i)$ has been
re-expressed as a function of the 5-dimensional distribution $d\sigma_{\rm{4J}}(\MET,  p_{Ti})$ and the 3-dimensional jet template $\rho_J(x, n; p_T)$. 
A further reduction can be made by assuming that $\MET$ only depends on the quantity
 $H_T \equiv \sum p_{Ti}$.  With this assumption we can
introduce the $\MET$ template $\mathcal{P}_{\rm{MET}}(y; H_T)$, with
$$
y\equiv \MET/H_T^{\half}
$$
 thus ending up with the factorization
\begin{eqnarray}
\label{probfactorization}
d\sigma_{\rm{4J}}(\MET,  m_i, n_i; H_T, p_{Ti}) = d\sigma_{\rm{4J}}(p_{Ti})
 \mathcal{P}_{\rm{MET}}(y; H_T) \prod_{i=1}^4 \rho_J(x_i, n_i; p_{T\,i})
\end{eqnarray}
Ultimately it is an experimental question whether such a factorization holds. At some level we certainly expect correlations between the 
four jets and deviations from the form of Eq.~\ref{probfactorization}.  For example, we would expect correlations to arise from color (re)connections as
well as out-of-jet radiation. The presence of significant pile-up (so long as it remains unsubtracted) would also tend to result in (positive) 
correlations between the jets.

In the case that the correlations are large it may be necessary to systematically include corrections to
Eq.~\ref{probfactorization}.  We anticipate that some kind of principal component analysis or form of tensor decomposition would
be applicable. We leave this interesting question to future work. For the remainder of this section we would like to explore the degree
to which the universality assumptions underlying Eq.~\ref{probfactorization} are valid in the only data sample available to us, namely the 
4.8 million {\tt SHERPA} events described in Sec.~\ref{Sec: QCDSh}.  

In a realistic experimental study one would presumably want to measure $d\sigma_{\rm{4J}}(p_{Ti})$ and $\rho_J(x, n; p_T)$
from independent samples.  Given our somewhat limited statistics, however, we will instead `measure' 
$d\sigma_{\rm{4J}}(p_{Ti})$ and $\rho_J(x, n; p_T)$ from the same 4-jet sample and use 
Eq.~\ref{probfactorization} to construct an estimate of the full 9-dimensional distribution. This will allow us to estimate acceptances after imposing
 $\MET$, $M_J$ and $N_J$ cuts. The degree to which
this procedure reproduces cut acceptances in the raw event sample will reflect the viability of the jet template and 
$\MET$ template ans\"{a}tze.

Given our somewhat limited statistics, it is difficult to judge whether deviations between the raw and
template cut acceptances (see Fig.~\ref{Fig: templateansatz}) are an indication of deviations from Eq.~\ref{probfactorization} or just statistical
fluctuations.  Nevertheless, the approximate agreement over 7 orders of magnitude of cut acceptance in Fig.~\ref{Fig: templateansatz} is 
a promising result, although more sophisticated statistical methods would likely be required for a robust experimental analysis.
\begin{figure}[t]
    \includegraphics[width=0.49\linewidth]{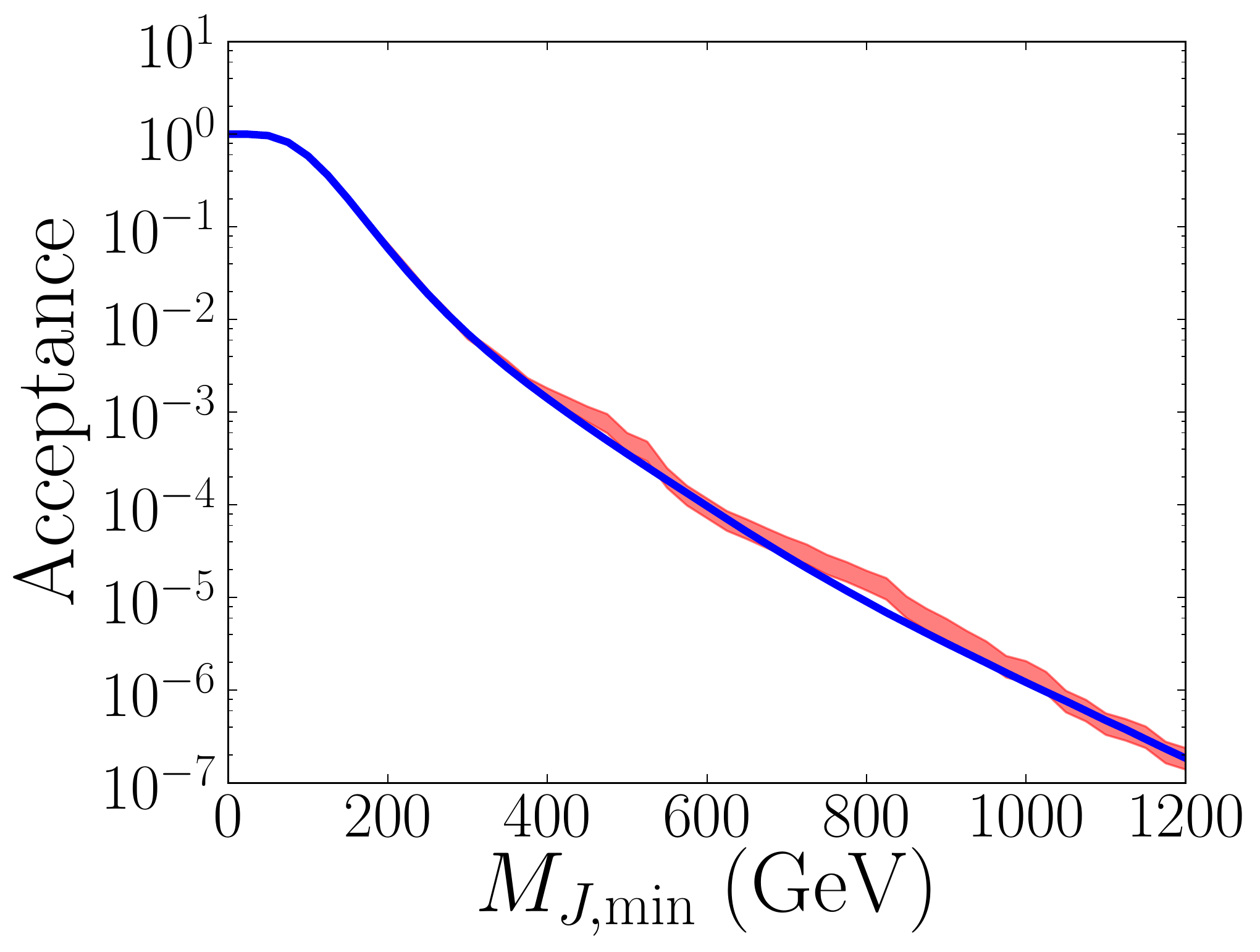}\hspace{-1mm}
      \includegraphics[width=0.45\linewidth]{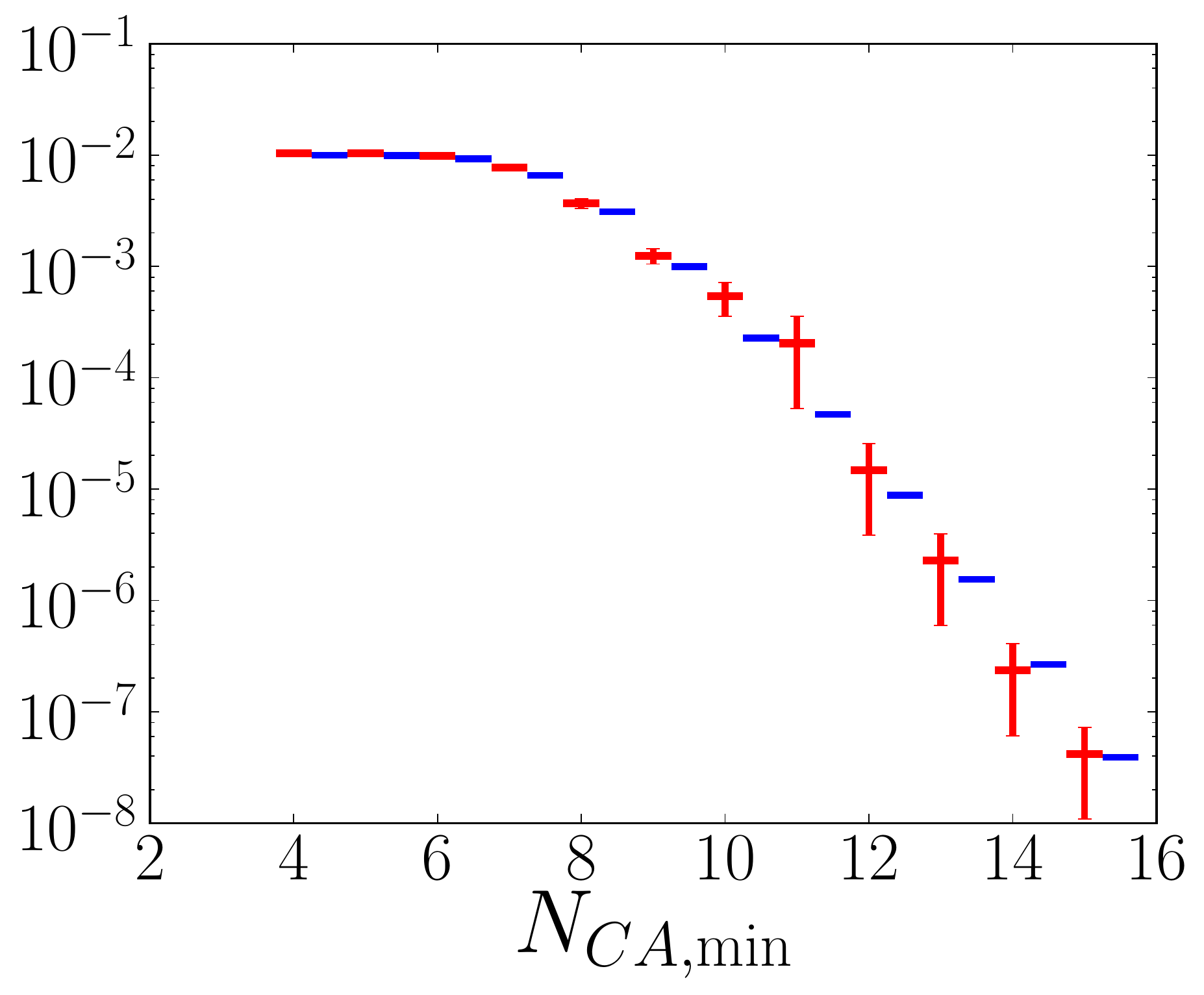}
    \caption{Testing the jet template ansatz.  The figure on the left compares the raw $M_J$ cut acceptance (red) to
    the template estimate (blue).  The figure on the right is analogous, with a sliding $N_{\rm{CA}}$ cut and a fixed cut $M_J > 280$ GeV.  The
    red uncertainties are statistical (the statistical uncertainties for the template estimate are not shown, since by construction they are
    parametrically smaller than the raw uncertainties).}
        \label{Fig: templateansatz}
\end{figure}

When the jet template and $\MET$ template ans\"{a}tze are appropriate, they have the advantage of reducing the statistical uncertainties in 
$d\sigma_{\rm{4J}}(\MET,  m_i, n_i)$.
This follows directly from the reduced dimensionality of the problem.  This reduction is especially significant in the tails of the distribution,
where statistical uncertainties are parametrically reduced by virtue of the fact that---due to the convolution---they receive sizeable contributions 
from statistically rich parts of the component probability distributions. For example, a rare 4-jet event with $N_J=12$ will be dominated by 
contributions from four jets with three subjets each.  The probability of a single fat jet having three subjets at a given $m/p_T$ and $p_T$ 
can be measured (or calculated) readily. 

Note that in the above we have discussed template methods in the context of LHC data.  Another possibility is to use template methods
to extend Monte Carlo calculations---indeed this is precisely what we did in the previous section for the specific case of missing energy.  In the context of
Monte Carlo template methods have the obvious advantage of reducing statistical uncertainties in the tails of distributions. They also offer
the possibility of extending lower multiplicity calculations to a higher multiplicity regime in the following sense.  Take for example {\tt SHERPA},
which can generate up to $n_j=6$ jets in the final state at the matrix element level.  Thus a {\tt SHERPA} dijet sample will include jets with up to five partons
generated through matrix elements.  When a fat jet template obtained from such a dijet sample is combined with a 4-jet distribution $d\sigma_{\rm{4J}}(p_{Ti})$, the resulting 
distributions will extend the nominal reach of {\tt SHERPA} beyond $n_j=6$.  Of course, the theoretical validity of such a method is a delicate
matter.  Given the importance of having reliable Monte Carlo estimates in the tails of distributions, however, such an approach deserves further
study.  For example, it would be important to investigate correlations between fat jets.  Since {\tt SHERPA} extends to $n_j=6$, one would
be able to, among other possibilities, study 3-jet events and observe what happens as fat jets come closer together.

\section{Results}
\label{Sec: Searches}

This section investigates the benefit of incorporating a subjet counting observable, namely $N_J$, into high multiplicity searches based off the summed 
jet mass observable, $M_J$.
\secref{SimplifiedModels} discusses the models used to quantify the improvement in searches that results from incorporating $N_J$.  These are
supersymmetric models whose phenomenology involves the pair production of gluinos that subsequently decay into the lightest supersymmetric particle.
Both R-parity conserving and R-parity violating models are considered.   We choose these benchmark signals because they
are well known in the literature and are easy to implement in Monte Carlo event generators.  
\secref{SigBGDist} describes the signal and background distributions
in signal-like regions.  \secref{Efficacy} describes the criterion that was used to create optimized search regions for the benchmark signals.
\secref{ExLim} describes the expected sensitivity of the optimized search regions to the benchmark signals.  
Finally, \secref{Comparison} compares the optimized search regions to previous searches.

\subsection{Benchmark signals}
\label{Sec: SimplifiedModels} 

The goal of this work is to gain access to a large class of signals without specifically targeting any one signal.  Nevertheless, it is useful to have some benchmark models to consider.  While these benchmark models are plausible extensions of the Standard Model, more than anything else they are meant to exhibit features of theories that produce high multiplicity final states.
For any single theory, there are numerous handles beyond large multiplicity that could allow for additional discrimination between signal and background. 
For instance, many of our benchmark signals 
have leptons and b-jets.  These are powerful handles that can be used in conjunction with the methods in this article, but they are not generic to all signals that produce high multiplicity final states.  Therefore, these additional handles will not be used.

The eight benchmark models considered in this article arise from the pair production of gluinos. These benchmarks provide relatively straightforward ways of toggling the multiplicity of final state partons within a class of models that is easily implemented in all the standard Monte Carlo calculation packages.  Each
benchmark model  is generated using {\tt MadGraph} 4.5.1~\cite{Steltzer:1994aa,Maltoni:2003aa,Alwall:2007st} and with up to two additional jets:
\begin{eqnarray}
    pp\rightarrow \tilde{g} \tilde{g} + n_{\tilde{g}} j\qquad \rm{with}\qquad n_{\tilde{g}}\le 2
\end{eqnarray}
The MLM matching scheme with a shower-$k_{\perp}$ scheme was used to account for the extra radiation.  The events were showered and hadronized using {\tt Pythia}~6.4~\cite{Sjostrand:2006za}. K factors for the signals were calculated using Prospino 2.1~\cite{Prospino}.

A collection of signals with diverse phenomenology is considered in order to better explore/delineate the efficacy of the subjet techniques used in this paper. 
This diversity arises through the variety of gluino decay topologies that are possible.  The gluino can decay to light or heavy quarks; it can decay
directly to the LSP or instead through a cascade. The theory can be R-parity conserving or R-parity violating. Decay topologies with cascade decays or decays involving top quarks as well as certain RPV topologies will lead to very high multiplicity events with 12 or more
final state partons. Indeed one benchmark signal we consider ($\GG_7$) has a spectacular  26 final state partons.
The expected sensitivities to all the signals presented here will be given in \secref{ExLim}.  

\begin{center}
\begin{table}[t]
\begin{tabular}{|c||c|c||c|c||c|c||c|}
\hline
\text{Model}& \multicolumn{2}{|c||}{\text{Gluino Decay}}&\multicolumn{2}{|c||}{\text{ Electroweakino}}&\multicolumn{2}{|c||}{\text{LSP Decay}}&\text{Final State Partons}
\\
& $q\bar{q}\chi (+4)$& $t\bar{t}\chi_i (+12)$ & $\;\;\chi_0\;\;$& $\chi_2 (+8)$ & $\rm{Stable}(+0)$& $cbs(+6)$&\\
\hline\hline
$\GG_0$&$\checkmark$&&$\checkmark$ &&$\checkmark$&&4\\
$\GG_1$&&$\checkmark$&$\checkmark$ &&$\checkmark$&&12\\
$\GG_2$&$\checkmark$&&&$\checkmark$ &$\checkmark$&&12\\
$\GG_3$&&$\checkmark$& &$\checkmark$&$\checkmark$&&20\\
$\GG_4$&$\checkmark$&&$\checkmark$ &&&$\checkmark$&10\\
$\GG_5$&&$\checkmark$&$\checkmark$ &&&$\checkmark$&18\\
$\GG_6$&$\checkmark$&& &$\checkmark$&&$\checkmark$&18\\
$\GG_7$&&$\checkmark$& &$\checkmark$&&$\checkmark$&26\\
\hline
\end{tabular}
\caption{\label{Table: Benchmarks} The eight benchmark signals used in this paper.  The numbers in parentheses indicate the number of final state
partons added by choosing that particular branch of the decay topology.} 
\end{table}
\end{center}
%

In more detail, all the processes outlined here start with a gluino decaying to either a light quark or a top quark pair and a neutralino:
\begin{eqnarray}
\tilde{g} \rightarrow q \bar{q} \chi \qquad \text{ or } \qquad \tilde{g} \rightarrow t\bar{t} \chi
\end{eqnarray}
The neutralino $\chi$ may be the LSP $\chi_0$ or one of the heavier electroweakinos.  For simplicity, only decays to the LSP and to the NNLSP $\chi_2$ are considered,
where the latter decay chain results in a 2-step cascade:
\begin{eqnarray}
\chi_2 \rightarrow  V \chi_1 \rightarrow V V' \chi_0
\end{eqnarray}
Finally the LSP may or may not decay into jets.  The constraints on R-parity violation are much weaker for decays into heavy flavor. 
If the LSP is lighter than $200\GeV$, then the decays will be dominantly with the $\lambda_{ijk} U^c_iD^c_j D^c_k$ flavor structure $(ijk)=(2,3,2)$.  The resulting decay topology is
\begin{eqnarray}
\chi_0 \rightarrow qqq =  c\,b\,s .
\end{eqnarray}
If the mass of the LSP $\chi_0$ is above $200\GeV$, then it is possible for the dominant R-parity violating decay mode to be $tbs$, resulting in four more final state partons over the $cbs$ decay mode. To keep the number of benchmark signals to a manageable number, we do not include this decay mode in any of our benchmarks.
All of these possibilities taken together result in eight different gluino decay topologies that span a range of 
final state parton multiplicities, see Table \ref{Table: Benchmarks}.\footnote{Here we use the term parton in a loose sense that includes the leptons from gauge boson decays.}

For all signals, we choose the LSP mass using the formula
\bea
m_{\chi_0} = m_{\tilde{g}} / 10
\eea
For the signal models involving heavier electroweakinos we choose  the intermediate masses as follows:
\bea
m_{\chi_2} = (m_{\tilde{g}} + m_{\chi_0} )/2 \qquad
m_{\chi_1} = (m_{\chi_2} + m_{\chi_0} )/2 
\eea

Note that if the gluinos decay to light quarks and the LSP, the final state will have between $4$ (RP conserving) and $10$ (RPV) partons, which will make these topologies hard to discriminate against the $t\bar t$ background, especially in the former case. In the case of cascade decays or decays involving top quarks, however, there will be at least $12$ partons in the final state and the method outlined in this article should prove more effective. Moreover, if R-parity is violated, each LSP will decay to three quarks, thus adding \mbox{6 jets} to the final state. Cuts on the total number of subjets could then provide a
competitive replacement for MET cuts for these kinds of signals.  

These signals are simply meant as benchmark models to test the sensitivity of our search to high multiplicity final states. The search presented here should prove effective for any signal implying the existence of final states with $ 8$ or more final state jets.

\begin{figure}[t]
  \includegraphics[width=0.45\linewidth]{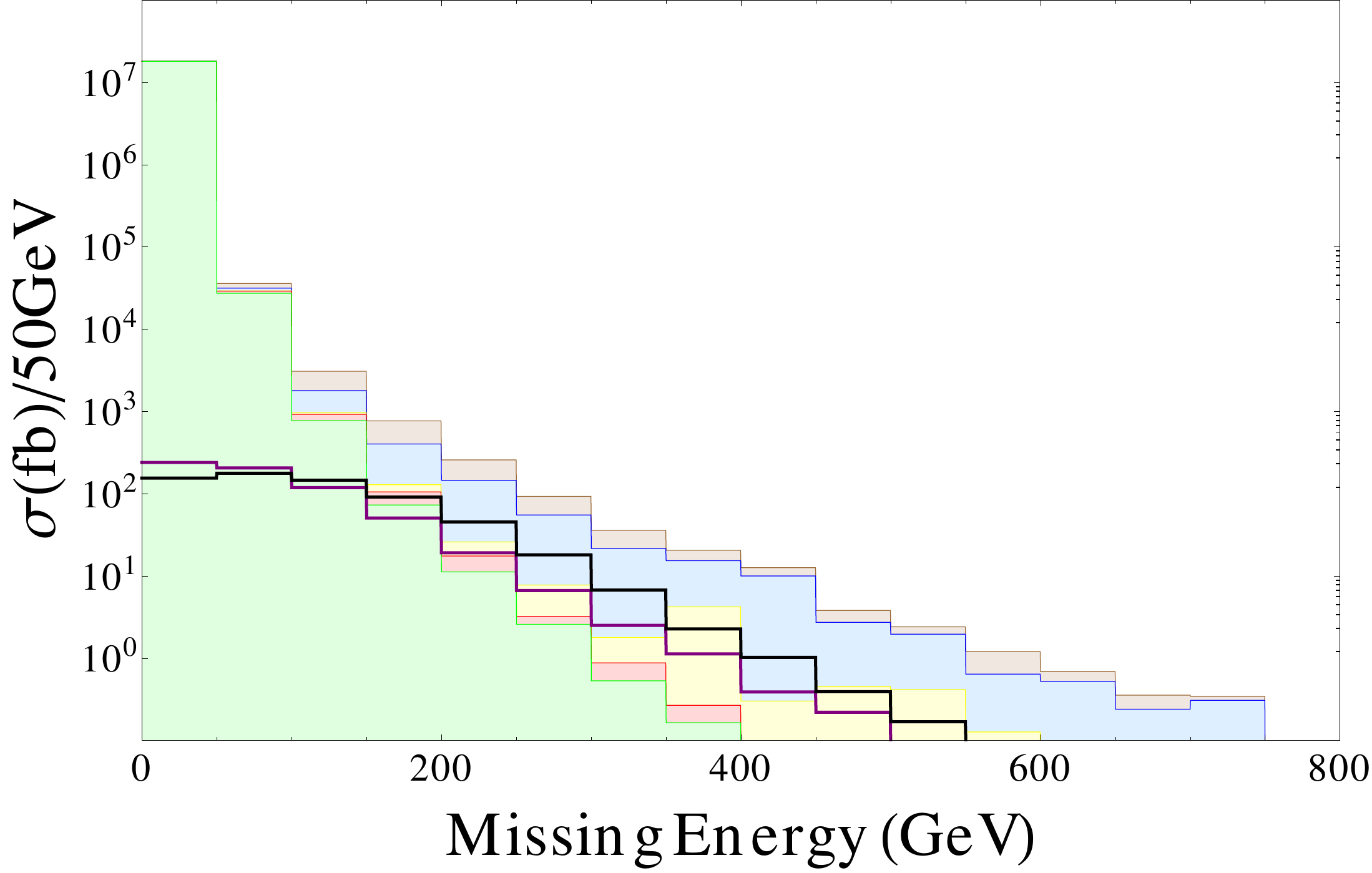}
    \includegraphics[width=0.45\linewidth]{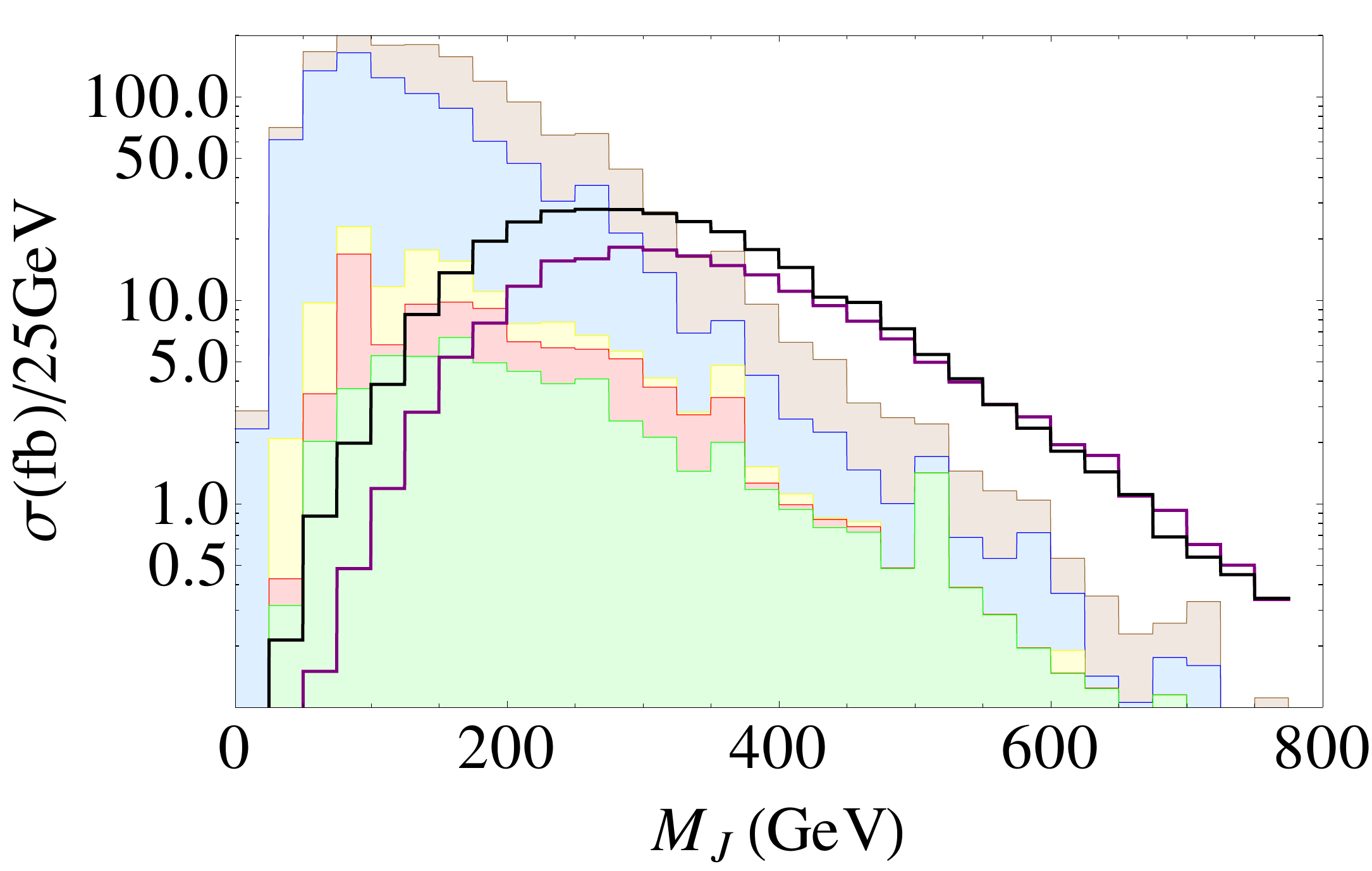}
      \includegraphics[width=0.46\linewidth]{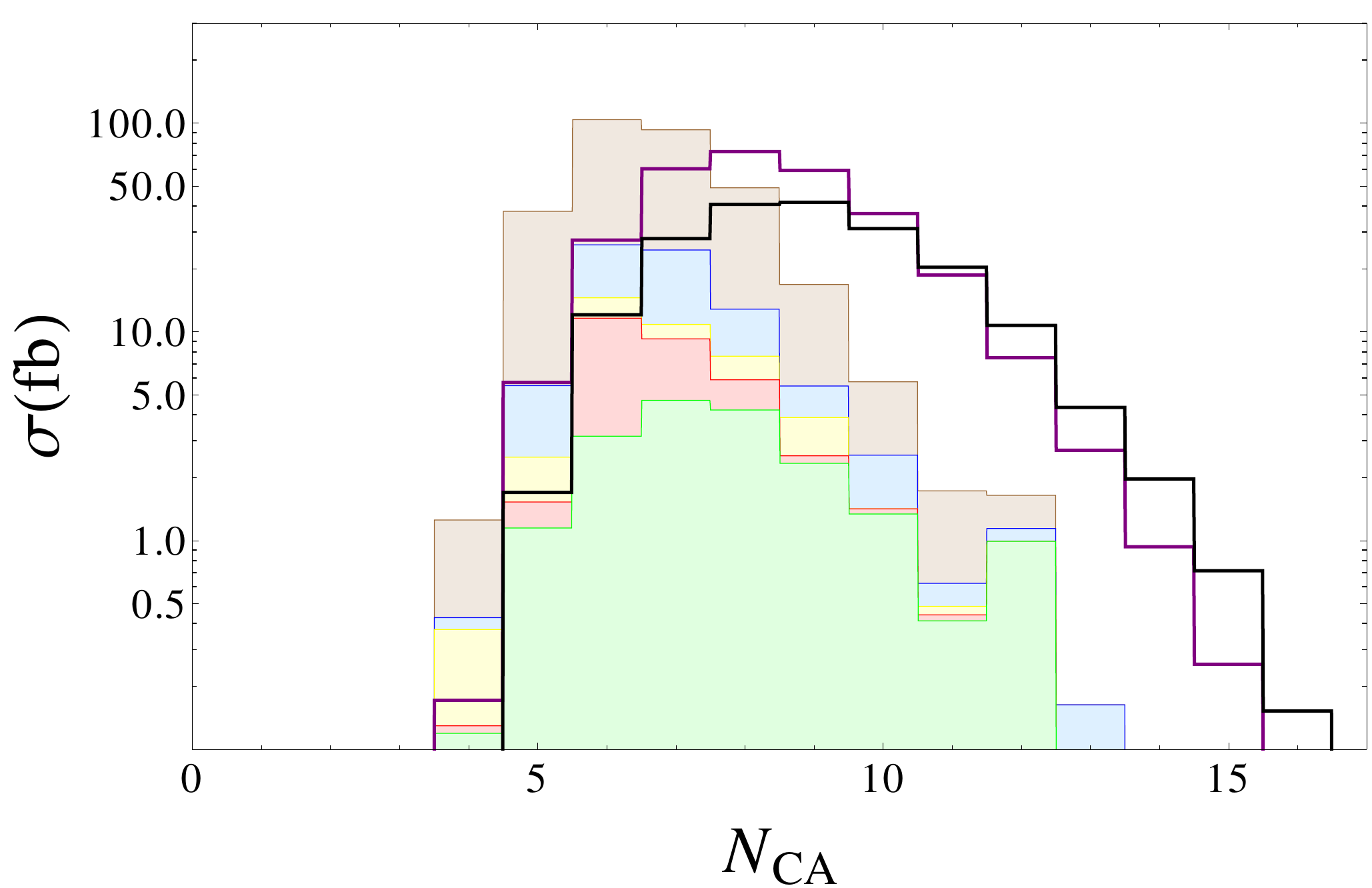}
       \includegraphics[width=0.46\linewidth]{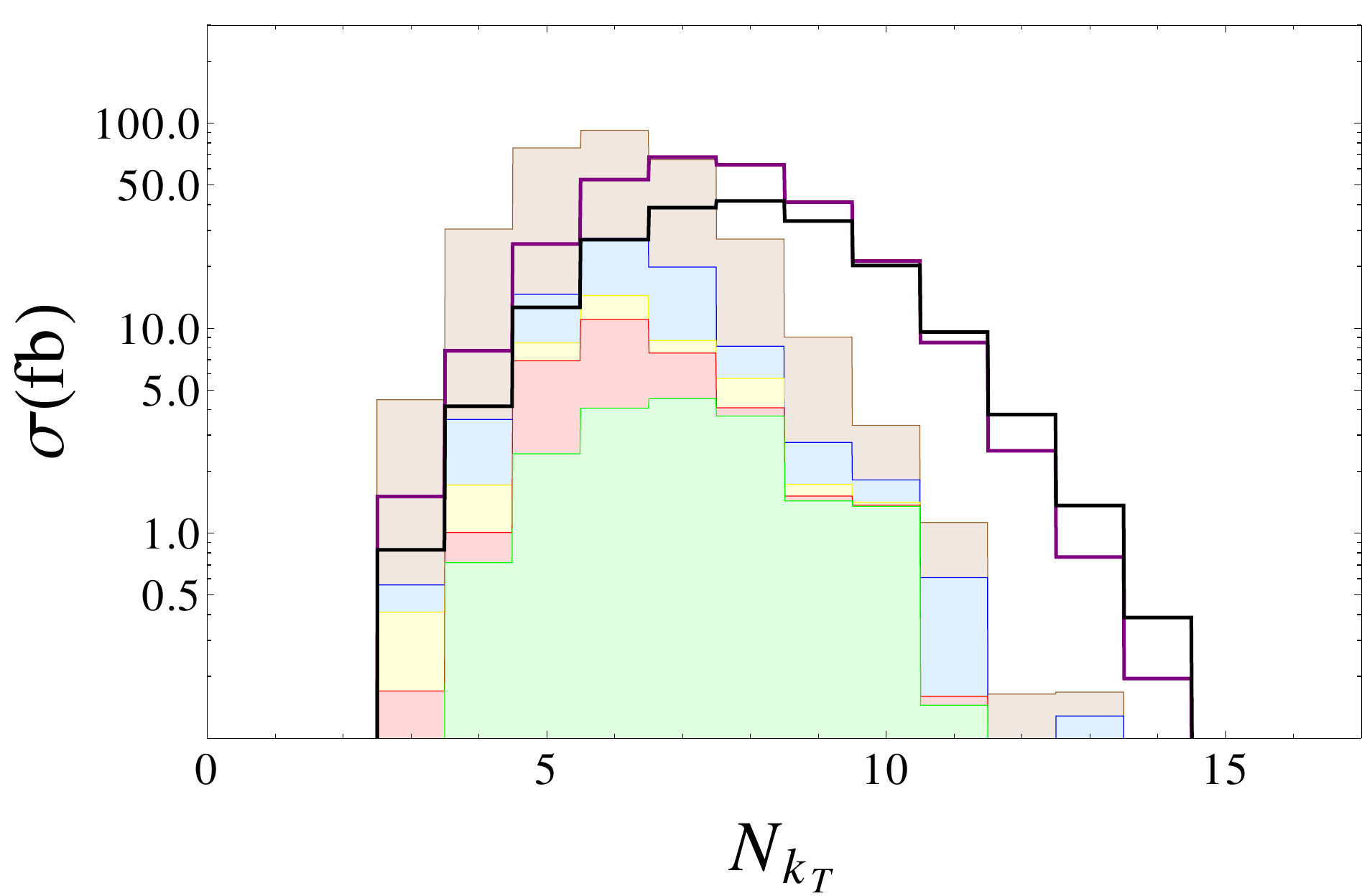}
    \caption{
    {\it Top Left}: $\MET$ distributions for signals and backgrounds after requiring four or more fat jets.
    {\it Top Right}: $M_J$ distributions, after requiring four or more fat jets and $\MET > 150$ GeV.
    {\it Bottom Left}: $N_{\rm{CA}}$ distribution after requiring four or more fat jets, $M_J>280$ GeV and $\MET > 150$ GeV.  
    {\it Bottom Right}: $N_{\rm{kT}}$ distribution after requiring four or more fat jets, $M_J>280$ GeV and $\MET > 150$ GeV.  
    Stacked histograms show the SM backgrounds, which include top and single top (light brown), $V + n j$
(light blue), diboson (light yellow), QCD (light green), and the remaining non-QCD backgrounds mentioned in Sec~\ref{Sec: EWBGs} (light red). 
The distributions for a 600 GeV gluino in the $\mathcal{G}_1$ and $\mathcal{G}_3$ topologies are shown in purple and black, respectively.
Note that the $N_{\rm{CA}}$ and $N_{\rm{kT}}$ distributions for $\mathcal{G}_3$ (with 20 final state partons) are not substantially different from 
$\mathcal{G}_1$ (with 12).}\label{Fig:Njets}
\end{figure}

\subsection{Distributions for signals and backgrounds}
\label{Sec: SigBGDist}
One of the goals of this paper is to investigate the degree to which $\MET$ cuts for the signals of interest
can be substituted (more realistically, loosened) by 
requiring particular jet substructure. That this is challenging can be inferred from Fig.~\ref{Fig:Njets}, which shows 
the $\MET$ distributions of the various backgrounds with the $\MET$ distributions of two benchmark
signals superimposed.  Although the dominant QCD background is significantly reduced by a $\MET$ cut of order 
$150\GeV$, any loosening of this cut dramatically increases the number of QCD events in the search region.

The remaining backgrounds, most of which have intrinsic $\MET$, require additional cuts to be suppressed.  As shown in Fig.~\ref{Fig:Njets},
a cut on the sum of the jet masses of order $300\GeV$ is effective. That a cut on $M_J$ does not exhaust the discrimination available
from jet substructure can be seen in Fig.~\ref{Fig:NvsMJ}, which illustrates how even at large values of $M_J$ the $N_{\rm{CA}}$ and $N_{\rm{kT}}$ distributions of a typical high multiplicity signal are well separated from that of the QCD background. The observation of similar behavior for the N-subjettiness ratio 
$\tau_3/\tau_2$ \cite{ATLAS:2012kla} suggests that this separation should hold for real QCD data as well.  

Thus $N_J$ cuts should be complementary to $M_J$ cuts, allowing for the possibility that $\MET$ cuts could be loosened.
This complementarity is made more explicit in the bottom row of Fig.~\ref{Fig:Njets}, which shows the distributions of $N_{\rm{CA}}$ and $N_{\rm{kT}}$ for the various backgrounds and two benchmark signals after imposing $\MET$ and $M_J$ cuts.  
Once these cuts have been imposed, the dominant remaining background comes from $t\bar{t}+$jets (and to a lesser extent $V+$jets),
as it has the largest number of final state partons.  In order for $t\bar{t}$+jets to pass the $\MET$ cut, the $W^\pm$ bosons have to decay semi-leptonically; 
while to pass the $M_J$ cut, the final state partons must be distributed in phase space such that they form massive jets upon fat jet clustering.
The combination of all these cuts strongly suppresses the various backgrounds.

\begin{figure}
    \includegraphics[width=2.5in]{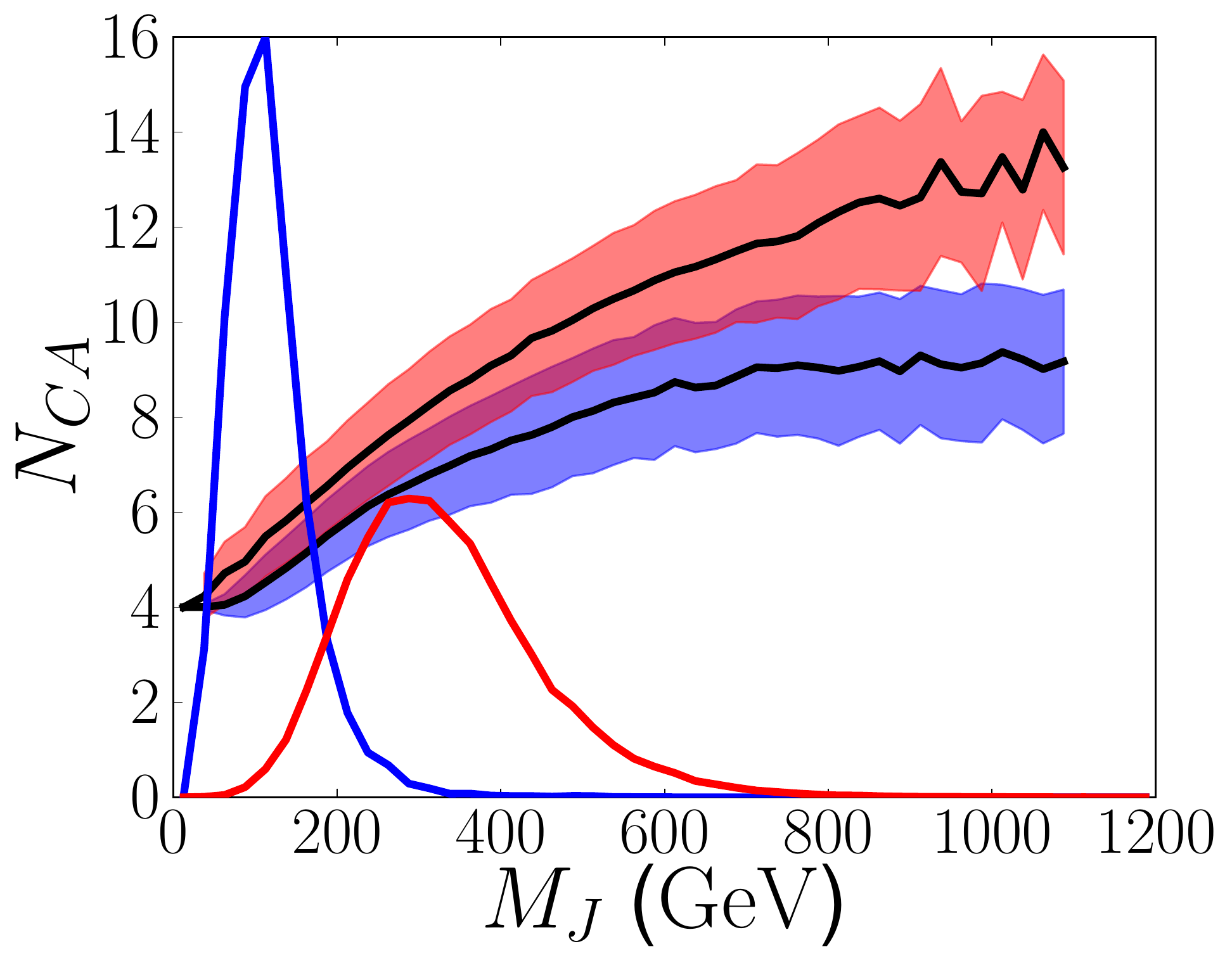}
    \includegraphics[width=2.5in]{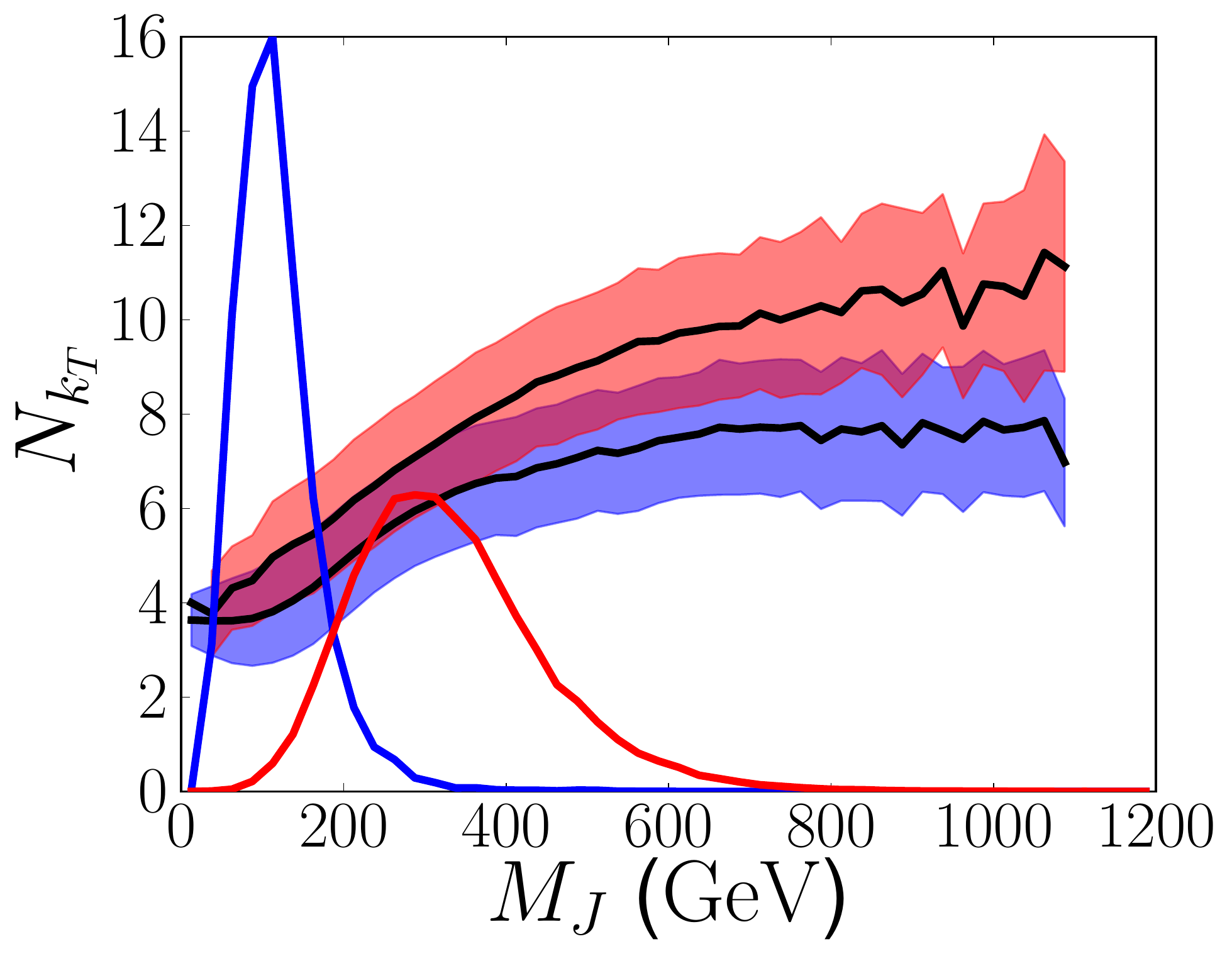}
    \caption{\label{Fig:NvsMJ}$N_{\rm{CA}}$ (left) and $N_{k_T}$ (right) versus $M_J$ for signal topology $\mathcal{G}_4$ with $m_{\tilde{g}} = 600\GeV$ (red band) and QCD background (blue band). Each band illustrates $\pm 1 \sigma$ standard deviation about the mean, which is denoted by a black line. 
    The superimposed $M_J$ distributions are for the signal (red) and the QCD background (blue). As defined in Sec~\ref{Sec: Subjets}, $N_{CA}$ and $N_{k_T}$ are the sum of the number of subjets of the four leading jets of each event. The events considered are required to have at least four jets with $p_T>50\GeV$ and the $p_T$ of their leading jet must be greater than $100\GeV$.\label{Fig: nsub}}
\end{figure}

\subsection{Optimizing search strategies}
\label{Sec: Efficacy}

The simplified models introduced in \secref{SimplifiedModels} can be used to develop broad search strategies that cover the model space. 
This section describes the method that was used to construct the minimal number of signal regions necessary to cover the 
entire space of simplified models.  The method used was introduced in ref.~\cite{Alves:2011sq}, developed further further in ref.~\cite{Essig:2011qg},
and is based off the variable ``efficacy,'' which is defined below.

In order to demonstrate the usefulness of $N_{J}$ cuts, we present two separately optimized search strategies. 
The first uses only $M_J$ and $\MET$ 
cuts, while the second uses $N_J$, $M_J$ and $\MET$ cuts.  Since the first set of searches is a subset of the second, the second will always do better.  The degree to which the more complex search strategy can be judged superior (if at all) will depend on the resulting sensitivities, the number of 
search regions required, and the sorts of cuts favored by the introduction of $N_J$.

The two search strategies are defined as
\be\label{eq:cuts}
\hat{\CC} =\{ ( \MET{}_{\text{ min}}, M_{J\text{ min}})\}\qquad \rm { and } \qquad
\CC =\{(N_{\text{subjet min}}, \MET{}_{\text{ min}}, M_{J\text{ min}})\}.
\ee
where the values of each of the cut requirements are taken from the following sets:
\begin{eqnarray*}
N_{J\text{ min}} &\in& \{0, \ldots, 16\}\\
\MET{}_{\text{ min}} &\in& \{ 0, 25,\ldots, 600\}\GeV\\
M_{J\text{ min}} &\in& \{ 0,25, \ldots, 1600\}\GeV\,.
\end{eqnarray*}
This results in 1625 search regions for $\hat{\CC}$ and $27,\hspace{-.2mm}625$ search regions for $\CC$.  The optimized search strategies will make
use of only a small subset of these search regions.

A given signal region or set of cuts, $C_i$, will yield an expected limit on the cross section times branching ratio, $\sigma\times\mathcal{B}$, for a given simplified model at the 95\% C.L. given by
\be
(\sigma\times\mathcal{B})_i &=& \frac{\Delta(B)_{i}}{\mathcal{L} \times\epsilon(M)_i}
\ee
Here $\epsilon(M)_i$ is the efficiency of $C_i$ for the model $M$ and $\mathcal{L}$ is the integrated luminosity, while $\Delta(B)_i$ 
is the maximum number of allowed signal events at the 95\% C.L. if $B$ background events are expected after the cuts and in fact fit the data. We take 
\be
\Delta(B) &=& 2 \times \sqrt{\text{Stat}(B)^2 + (\epsilon_{\text{syst}}B)^2},
\ee
where Stat$(B)$ is the Poisson limit on $B$ and $\epsilon_{\text{syst}}$ is the systematic uncertainty.  Throughout we will take $\epsilon_{\text{syst}}=30\%$.
The Monte Carlo statistical uncertainties in $B$ and $\epsilon(M)_i$, i.e.~$\delta B$ and $\delta \epsilon(M)_i$, are taken into account by making the
replacements
\begin{align}
   B \rightarrow B + \delta B \qquad \rm{ and } \qquad \epsilon(M)_i \rightarrow \epsilon(M)_i- \delta \epsilon(M)_i
\end{align}
which result in conservative limits.

The optimal limit on a model $M$ is then given by
\be
(\sigma\times\mathcal{B})_{\text{opt}} &=& \{\text{min}((\sigma\times\mathcal{B})_i): i\in\{1,N_{\text{cuts}}\}\}\,, 
\ee
where the number of search regions is $N_{\text{cuts}}=1625\text{ or } 27,\hspace{-.2mm}625$ depending on whether $N_J$ cuts are being used.  
It is natural to quantify the ``goodness'' of a cut $C_i$ by how close it is to optimal.  For this purpose, we introduce the efficacy of a cut
\be
\EE(C_i) &=&\frac{(\sigma\times\mathcal{B})_i}{(\sigma\times\mathcal{B})_{\text{opt}}}. 
\ee
This is  the ratio of the expected limit on the production cross section using a particular cut $C_i$ divided by the expected limit on the cross section using the optimal set of cuts.  An efficacy of $1.0$ is ideal.  Thus the best search strategy for covering a collection of model points $\{M_j\}$ will be a combination of cuts $\{C_i\}$ such that $\EE$ is close to one for each $M_j$ for at least one of the $C_i$.  This article will use $\EE \le \EE_{\text{crit}}= 1.5$ as the criterion for optimizing the number of search regions.  That is, each model point $M$ will be
covered by at least one cut that yields a limit on $\sigma\times\mathcal{B}$ that is within a factor of $1.5$ of the optimal limit. The efficacy approach has several advantages:
\begin{itemize}
\item it ensures near optimal coverage over the range of signals;
\item it allows for a fair comparison between different sets of observables;  
\item it allows for a reasonable comparison to the ATLAS high multiplicity search, which makes use of 6 search regions; and 
\item  each signal is grouped with like signals  on the basis of which search region it is covered by.
\end{itemize}

Finding a search strategy that covers all models with a desired efficacy is computationally challenging 
because the configuration space is enormous, with $2^{N_{\text{cuts}}}$ possible search strategies. 
Since a brute force search is not feasible, we use a genetic algorithm to construct the minimal set of search regions needed to cover 
the entire space of models.  This algorithm, which we find to be quite effective for the task at hand, is described in App.~\ref{app:genetic} and is
based off a genetic algorithm described in detail in \cite{Essig:2011qg}.

\subsection{Expected sensitivity}
\label{Sec: ExLim}

\begin{table}[t]
\begin{tabular}{|c||c|c||c|c||c|c|c|c||c|}
\hline
 \multicolumn{3}{|c||}{   Search Region } & \multicolumn{2}{|c||}{Models Covered}& \multicolumn{5}{|c|}{Background (for $30 \ifb$)}\\
 \hline
&   $M_J$ & $\MET$ &Class& $m_{\tilde{g}}$&QCD&  $t\bar t$ &V+jets& Other& Total\\ 
\hline
\hline
1&1000 & 0 &  $\mathcal{G}_{4}$ &{\scriptsize $m_{\tilde{g}}\lsim 1.0 \TeV$ }
& {\footnotesize $495\pm 61.5$} 
&{\footnotesize $2.38\pm 0.69$} 
& {\footnotesize$6.93\pm 2.73$} 
& {\footnotesize$0.13\pm 0.10$} 
& {\footnotesize$ 505\pm 62$}\\
    \hline
    2&1350 & 0 &  $\mathcal{G}_{4}$ & {\scriptsize$m_{\tilde{g}}\gsim 1.0 \TeV$ }
& {\footnotesize$13.7\pm 1.5$} 
& {\footnotesize $\lsim 0.1$} 
& {\footnotesize$0.54\pm 0.54$} 
&{\footnotesize $\lsim 0.1$} 
& {\footnotesize$14.3\pm 1.6$}\\
    \hline
\multirow{2}{*}{3}&\multirow{2}{*}{400} & \multirow{2}{*}{400} &   $\mathcal{G}_{0}$ & {\scriptsize$m_{\tilde{g}}\lsim 1.2 \TeV$}
&\multirow{2}{*}{{\footnotesize$0.38\pm 0.04$}}
&\multirow{2}{*}{{\footnotesize$16.63\pm 1.81$}}
&\multirow{2}{*}{{\footnotesize$14.30\pm 2.62$}}
&\multirow{2}{*}{{\footnotesize$4.40\pm 1.52$}}
&\multirow{2}{*}{{\footnotesize$35.71\pm 3.53$}} \\ 
&&&  $\mathcal{G}_{1}$ & {\scriptsize$0.8\TeV\gsim m_{\tilde{g}}\gsim 1.1\TeV$ }&&&&&
\\
 \hline
 \multirow{2}{*}{4}&\multirow{2}{*}{500} & \multirow{2}{*}{200} &  $\mathcal{G}_{1}$ &{\scriptsize $m_{\tilde{g}}\lsim 0.8 \TeV$}
 &\multirow{2}{*}{{\footnotesize$23.9\pm 4.9$}}
&\multirow{2}{*}{{\footnotesize$54.6\pm 3.3$}}
&\multirow{2}{*}{{\footnotesize$28.0\pm 5.6$}}
&\multirow{2}{*}{{\footnotesize$6.26\pm 1.52$}}
&\multirow{2}{*}{{\footnotesize$112.8\pm 8.2$}} \\
&&&  $\mathcal{G}_{2,3}$ & {\scriptsize$m_{\tilde{g}}\lsim 0.9\TeV$}&&&&&
\\
 \hline
  \multirow{3}{*}{5}&\multirow{3}{*}{625} & \multirow{3}{*}{425} &  $\mathcal{G}_{0}$ & {\scriptsize$m_{\tilde{g}}\gsim 1.2 \TeV$ }
 &\multirow{3}{*}{{\footnotesize$0.09\pm 0.02$}}
&\multirow{3}{*}{{\footnotesize$0.59\pm 0.34$}}
&\multirow{3}{*}{{\footnotesize$0.73\pm 0.73$}}
&\multirow{3}{*}{{\footnotesize$0.47\pm 0.29$}}
&\multirow{3}{*}{{\footnotesize$1.89\pm 0.86$}}\\
&&&  $\mathcal{G}_{1}$ & {\scriptsize$m_{\tilde{g}}\gsim 1.1\TeV$ }&&&&&\\
&&&  $\mathcal{G}_{2,3}$ &{\scriptsize $m_{\tilde{g}}\gsim 1.3\TeV$}&&&&&
\\
 \hline
   \multirow{2}{*}{6}&\multirow{2}{*}{725} & \multirow{2}{*}{175} &  $\mathcal{G}_{2,3}$ & {\scriptsize$0.9\TeV\lsim m_{\tilde{g}}\lsim 1.3 \TeV$}
&\multirow{2}{*}{{\footnotesize$5.28\pm 0.72$}}
&\multirow{2}{*}{{\footnotesize$5.34\pm 1.03$}}
&\multirow{2}{*}{{\footnotesize$2.85\pm 1.08$}}
&\multirow{2}{*}{{\footnotesize$0.41\pm 0.18$}}
&\multirow{2}{*}{{\footnotesize$13.87\pm 1.67$}} \\
&&&  $\mathcal{G}_{5,6,7}$ &all&&&&&
\\
 \hline
\end{tabular}
\caption{Search regions for the $M_J$ + $\MET$ search with cuts in GeV and assuming 30\% systematic uncertainties. For each search region $C_i$ the column `Models Covered' 
lists the benchmark models that are optimally covered by $C_i$. The search regions are chosen using the efficacy criterion $\EE < 1.5$. The background uncertainties shown are statistical.}\label{Table:mjmet}
\end{table}

\begin{table}[t]
\begin{tabular}{|c||c|c|c||c|c||c|c|c|c||c|}
\hline
\multicolumn{4}{|c||}{    Search Region } &\multicolumn{2}{|c|}{ Models Covered }& \multicolumn{5}{|c|}{Background (for $30 \ifb$)} \\
\hline
  &$M_J$ & $\MET$ & $N_{\rm{CA}}$ & Class & $m_{\tilde{g}}$&QCD& $t\bar t$ &V+jets& Other& Total \\ 
\hline \hline
1&450 & 450 & 0 &  $\mathcal{G}_{0}$ & all 
&{\footnotesize$0.18\pm 0.26$} 
& {\footnotesize$8.31\pm 1.28$} 
& {\footnotesize$2.05\pm 1.08$} 
&{\footnotesize $0.64\pm 0.26$} 
& {\footnotesize$11.18\pm 1.70$}\\
 \hline
2& 1050 & 0 & 13 &  $\mathcal{G}_{4}$& all 
&{\footnotesize $21.60\pm 3.03$} 
&{\footnotesize $\lsim 0.1$} 
& {\footnotesize$\lsim 0.1$} 
& {\footnotesize$0.03\pm 0.01$} 
& {\footnotesize$21.63\pm 3.03$}\\
 \hline
 \multirow{3}{*}{3} & \multirow{3}{*}{475} &\multirow{3}{*} {275} & \multirow{3}{*}{11} & $\mathcal{G}_{1}$ &all  &\multirow{3}{*}{{\footnotesize$0.96\pm 0.46$}}
&\multirow{3}{*}{{\footnotesize$4.16\pm 0.91$}}
&\multirow{3}{*}{{\footnotesize$0.78\pm 0.59$}}
&\multirow{3}{*}{{\footnotesize$0.03\pm 0.01$}}
&\multirow{3}{*}{{\footnotesize$5.90\pm 1.18$}} \\
&&  &&$\mathcal{G}_{2}$& {\scriptsize $m_{\tilde{g}} \gsim 0.8\TeV$}&&&&&  \\
&&  &&$\mathcal{G}_{3}$&{\scriptsize  $m_{\tilde{g}} \gsim 0.9\TeV$ }&&&&&
\\
\hline
 \multirow{3}{*}{4} & \multirow{3}{*}{525} &\multirow{3}{*} {125} & \multirow{3}{*}{12} & $\mathcal{G}_{2}$ &{\scriptsize$m_{\tilde{g}} \lsim 0.8\TeV$}
&\multirow{3}{*}{{\footnotesize$7.86\pm 1.92$}}
&\multirow{3}{*}{{\footnotesize$7.72\pm 1.24$}}
&\multirow{3}{*}{{\footnotesize$6.71\pm 4.58$}}
&\multirow{3}{*}{{\footnotesize$0.33\pm 0.19$}}
&\multirow{3}{*}{{\footnotesize$22.65\pm 5.11$}}  \\
&&  &&$\mathcal{G}_{3}$&  {\scriptsize$m_{\tilde{g}} \lsim 0.9\TeV$} &&&&& \\
&&  &&$\mathcal{G}_{5,6}$& {\scriptsize $m_{\tilde{g}} \gsim 0.9\TeV$} &&&&&
\\
\hline
 \multirow{2}{*}{5} & \multirow{2}{*}{425} &\multirow{2}{*} {125} & \multirow{2}{*}{14} & $\mathcal{G}_{5,6}$ &{\scriptsize$m_{\tilde{g}} \lsim 0.9\TeV$} 
&\multirow{2}{*}{{\footnotesize$1.08\pm 0.32$}}
&\multirow{2}{*}{{\footnotesize$1.19\pm 0.49$}}
&\multirow{2}{*}{{\footnotesize$\lsim 0.1$}}
&\multirow{2}{*}{{\footnotesize$0.01\pm 0.01$}}
&\multirow{2}{*}{{\footnotesize$2.26\pm 0.58$}} \\
&&  &&$\mathcal{G}_{7}$&  all &&&&&
\\
\hline
\end{tabular}
\caption{Search regions for the $M_J$ + $\MET$ + $N_{\rm{CA}}$ search with $M_J$ and  $\MET$ cuts in GeV and assuming 30\% systematic uncertainties. For each search region $C_i$ the column `Models Covered' 
lists the benchmark models that are optimally covered by $C_i$. The search regions are chosen using the efficacy criterion $\EE < 1.5$. The background uncertainties shown are statistical.}\label{Table:mjmetnca}
\end{table}

Expected sensitivities to the various benchmark signals at $\sqrt{s} =8$ TeV are depicted in Figures \ref{exclG0}-\ref{exclusionSys30}.  
These are presented as expected 95\% exclusion 
limits on $\sigma \times\mathcal{B}$ (the production cross section times the branching ratio into that particular gluino decay topology) as a function
of the gluino mass and for an integrated luminosity of 30 $\rm{fb}^{-1}$.
As expected, the performance of the $M_J$ + $\MET$ +$N_J$ search depends strongly on the final state multiplicity as well as the
intrinsic $\MET$ of the signal. 

The results of the subjet counting search are best revealed by comparing the optimal search regions for the
$M_J$ + $\MET$ search to those of the $M_J$ + $\MET$ +$N_J$ search.  For the case of $N_{\rm{CA}}$ 
the former has 6 search regions, while the latter has 5 search regions, see Tables~\ref{Table:mjmet} and \ref{Table:mjmetnca}.  
Interestingly, the efficacy criterion groups the signals into roughly similar signal classes, with the difference that some of the $M_J$ 
and $\MET$ cuts move around once $N_{\rm{CA}}$ cuts are introduced. 

The first class of signals consists of $\GG_0$ alone, has intrinsic $\MET$ from
the stable (non-RPV) LSP and only 4 final state partons. Consequently the cuts (and expected limits, see Fig.~\ref{exclG0}) 
do not change substantially after the introduction of a (trivial) $N_{\rm{CA}}$ cut.  

The second class of signals consists of $\GG_4$ alone, which differs from $\GG_0$ in
that the LSP undergoes the RPV decay $\chi \rightarrow cbs$. Consequently there is no intrinsic $\MET$, and both search strategies cover
$\GG_4$ with search regions that have trivial $\MET$ cuts.  Since, however, $\GG_4$ is a high multiplicity signal
with 10 final state partons, the corresponding $N_{\rm{CA}}$ search region imposes a significant cut $N_{\rm{CA}} \ge 13$ 
with a loosened $M_J$ cut. This results in an expected limit on $\sigma \times\mathcal{B}$
that is better by a factor 2 to 4 compared to the optimized $M_J$ + $\MET$ search (see Fig.~\ref{exclG4}) 
but that is nevertheless weaker than what would be needed to exclude the benchmark gluino cross section.

The third class of signals consists of $\GG_5$, $\GG_6$ and $\GG_7$.  These signals have intrinsic $\MET$ from top quarks or electroweak
gauge bosons produced in the gluino decay chain.  They also have especially large final state multiplicities, 
since the LSPs at the end of the decay chain end in the RPV decay $\chi \rightarrow cbs$.  
The inclusion of a cut $N_{\rm{CA}}\ge12-14$ improves the expected limit by a factor of $ 2-5$ 
depending on the specific signal and gluino mass.  The $\MET$ cut is loosened by $50\GeV$,
while the $M_J$ cut is lowered by $200-300 \GeV$.  This represents a modest success in trading our reliance on 
$\MET$ cuts for a more refined use of jet substructure observables. 

The fourth and final class of signals consists of $\GG_1$, $\GG_2$ and $\GG_3$.  These signals have large intrinsic $\MET$ because the
LSPs at the end of the gluino decay chain are stable and because top quarks and/or electroweak gauge bosons are produced in the decay chain.
The top quarks and electroweak gauge bosons also ensure that the final state multiplicity is high. The inclusion of a $N_{\rm{CA}}$ cut 
improves the expected limit on $\sigma\times \mathcal{B}$ by a factor $ 2-4$ for low and intermediate gluino masses, with little or 
no improvement at large gluino masses. The inclusion of a $N_{\rm{CA}}$ cut also loosens (in most places) the requirements on $M_J$ and $\MET$ 
by $ 50-200 \GeV$ and $ 50-150\GeV$, respectively.  This demonstrates that for these signals with significant 
$\MET$ there is more room for loosening $\MET$ requirements in favor of $N_J$ requirements.

\begin{figure}
\includegraphics[width=0.6\linewidth]{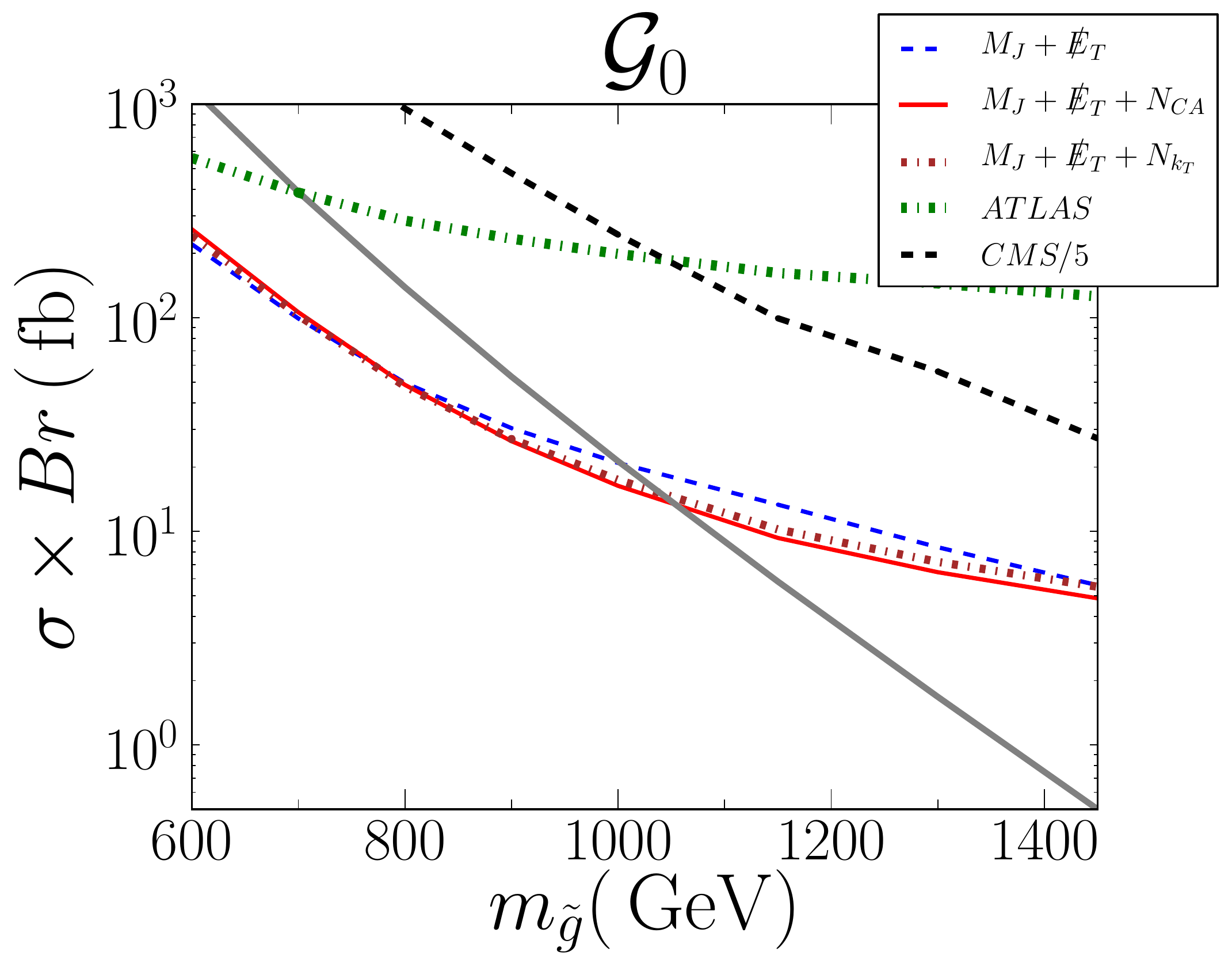}
\caption{95\% exclusion limits on $\sigma\times\mathcal{B}$ for the $M_J$ + $\MET$ search (dashed blue), the $M_J$+$\MET$+$N_{\rm{CA}}$ search (solid red),
and the $M_J$+$\MET$+$N_{\rm{k_T}}$ search (dash-dotted brown) for 
signal $\GG_0$, which has only 4 final state partons. The exclusion limits given by the ATLAS high multiplicity search \cite{ATLAS:2012una} (dash-dotted green line) and the CMS black hole search \cite{Chatrchyan:2012taa} (dashed black) as well as the NLO gluino production cross section (grey solid) are also shown. The systematic uncertainty on the background is assumed to be $30$\%.  Note that the CMS limit is rescaled by a factor of 5.}\label{exclG0}
\end{figure}

\begin{figure}
\includegraphics[width=0.6\linewidth]{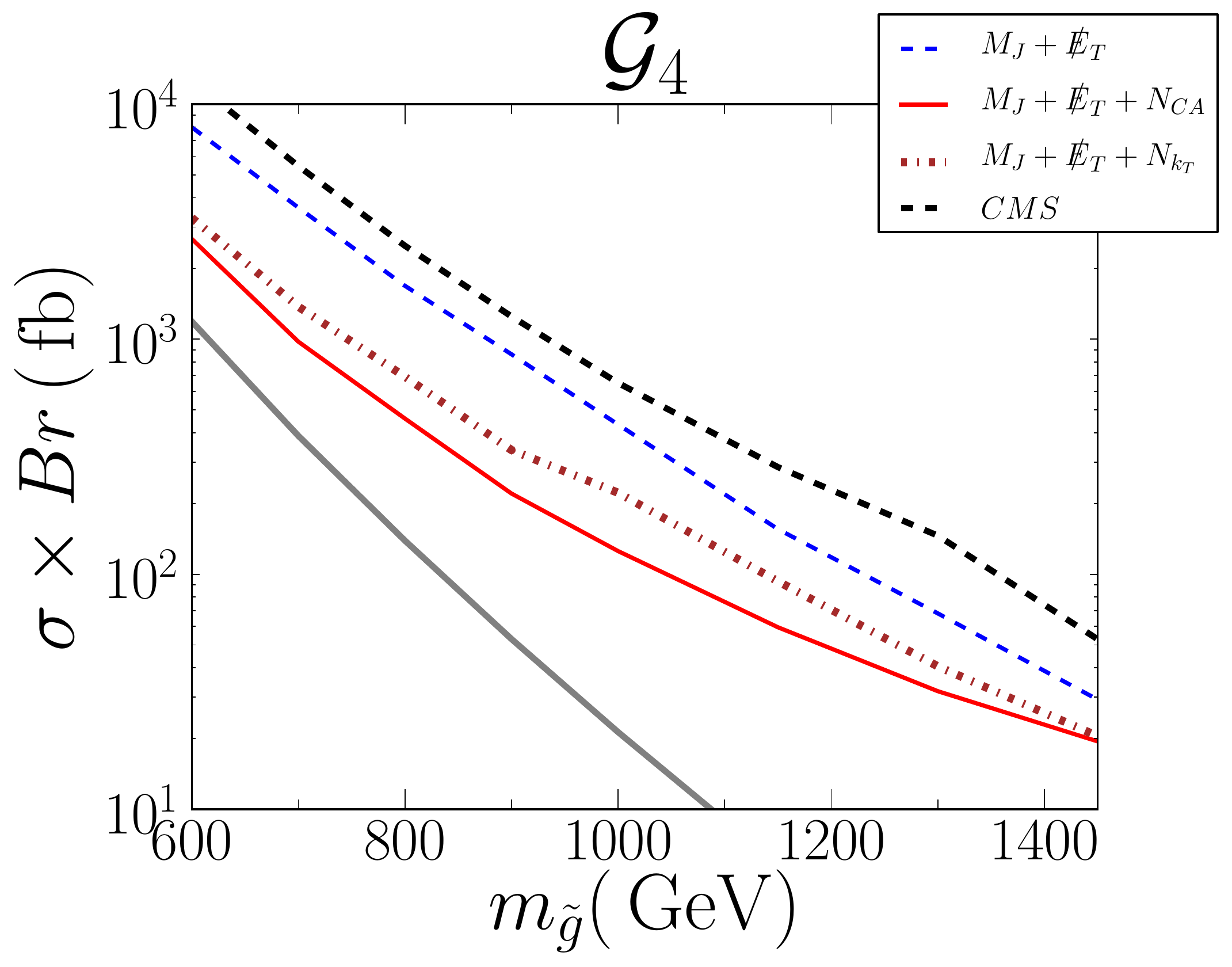}
\caption{95\% exclusion limits on $\sigma\times\mathcal{B}$ for the $M_J$ + $\MET$ search (dashed blue), the $M_J$+$\MET$+$N_{\rm{CA}}$ search (solid red),
and for the $M_J$+$\MET$+$N_{\rm{k_T}}$ search (dash-dotted brown) 
for signal $\GG_4$, which has 10 final state partons and no intrinsic $\MET$. The exclusion limit given by the CMS black hole search \cite{Chatrchyan:2012taa} (dashed black) as well as the NLO gluino production cross section (grey solid line) are also shown. The systematic uncertainty on the background is assumed to be $30$\%.
The ATLAS limit is not shown because it is orders of magnitude worse than the others due to its strict $\MET$ requirement.}\label{exclG4}
\end{figure}

\begin{figure}
\includegraphics[width=0.49\linewidth]{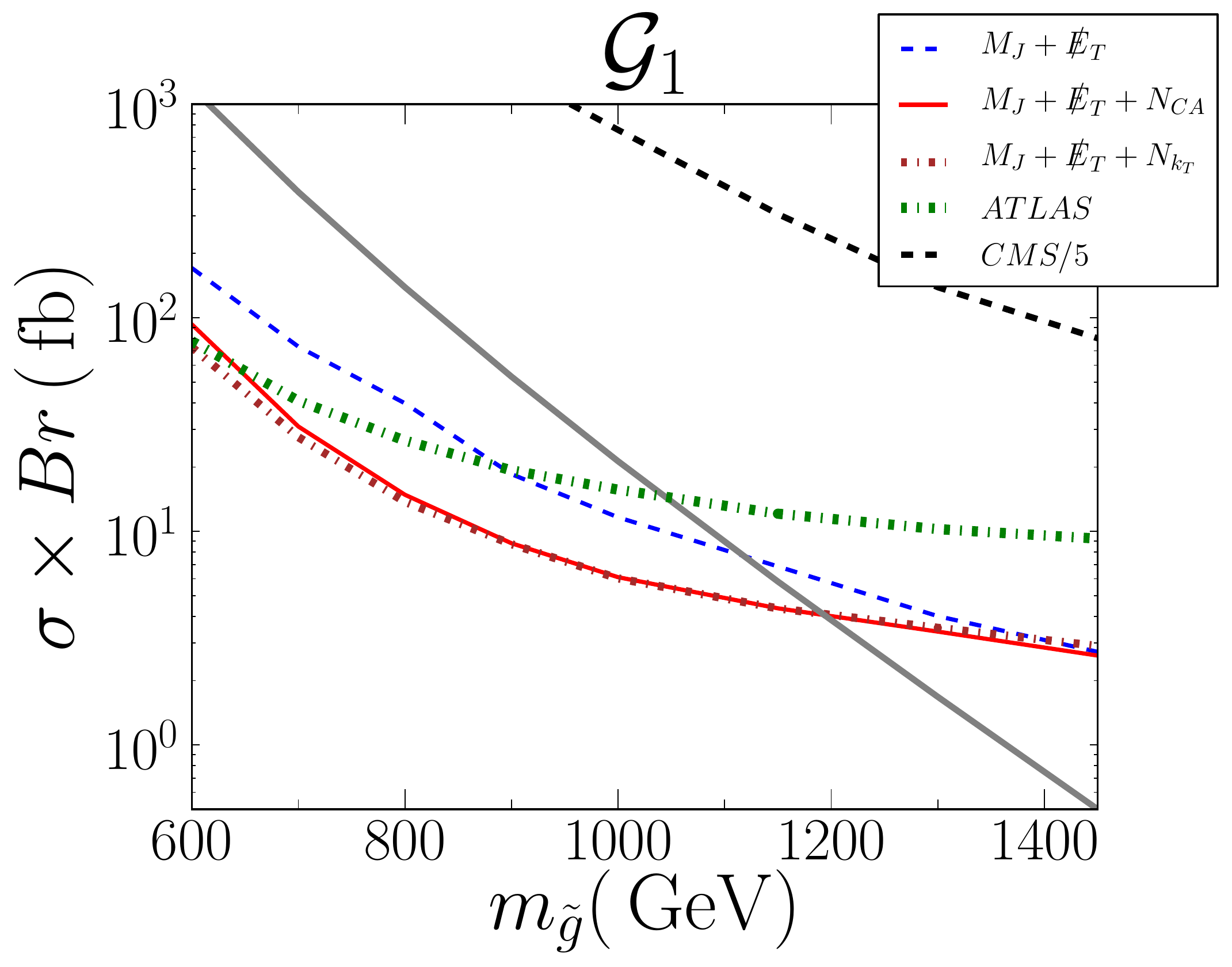}
\includegraphics[width=0.49\linewidth]{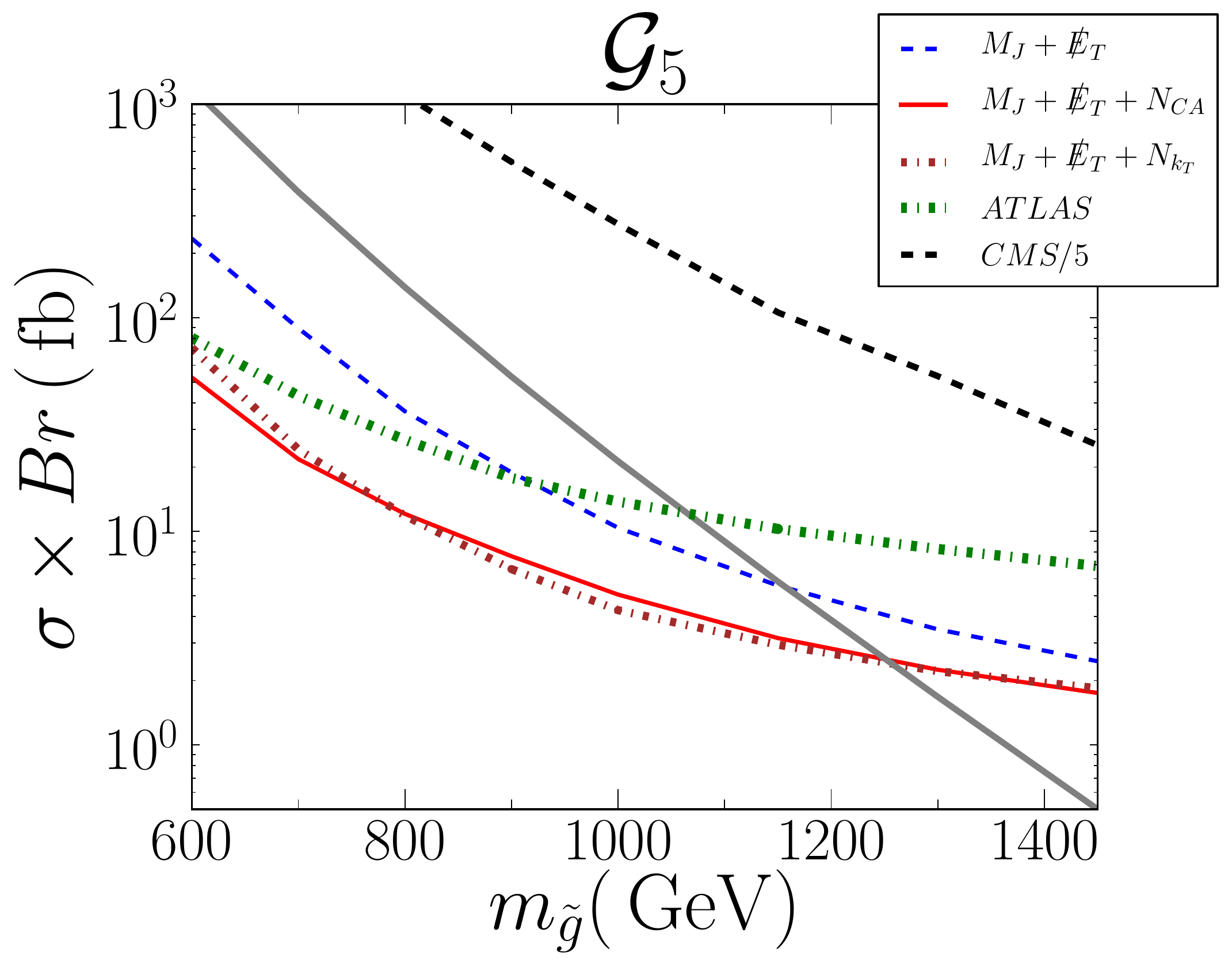}
\includegraphics[width=0.49\linewidth]{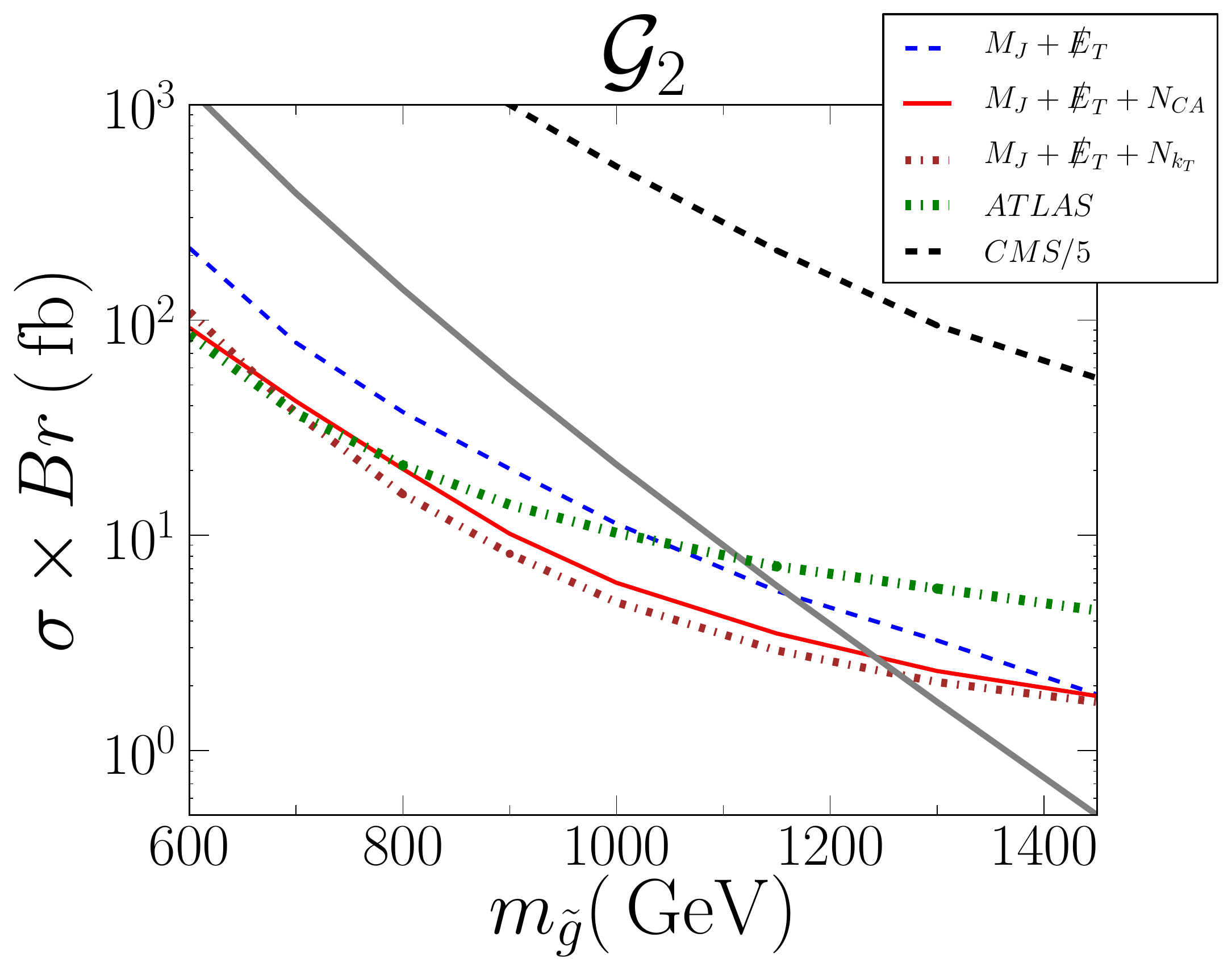}
\includegraphics[width=0.49\linewidth]{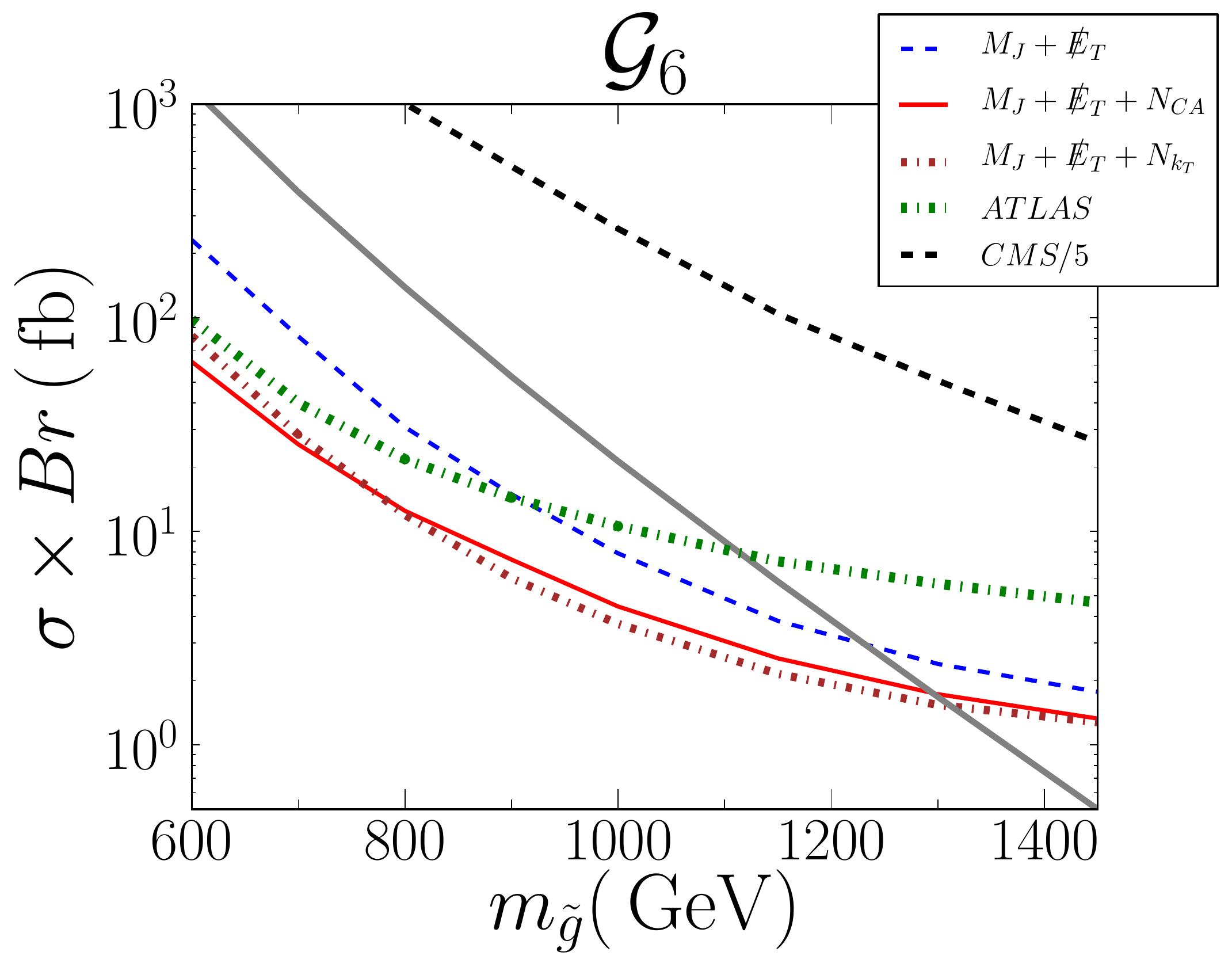}
\includegraphics[width=0.49\linewidth]{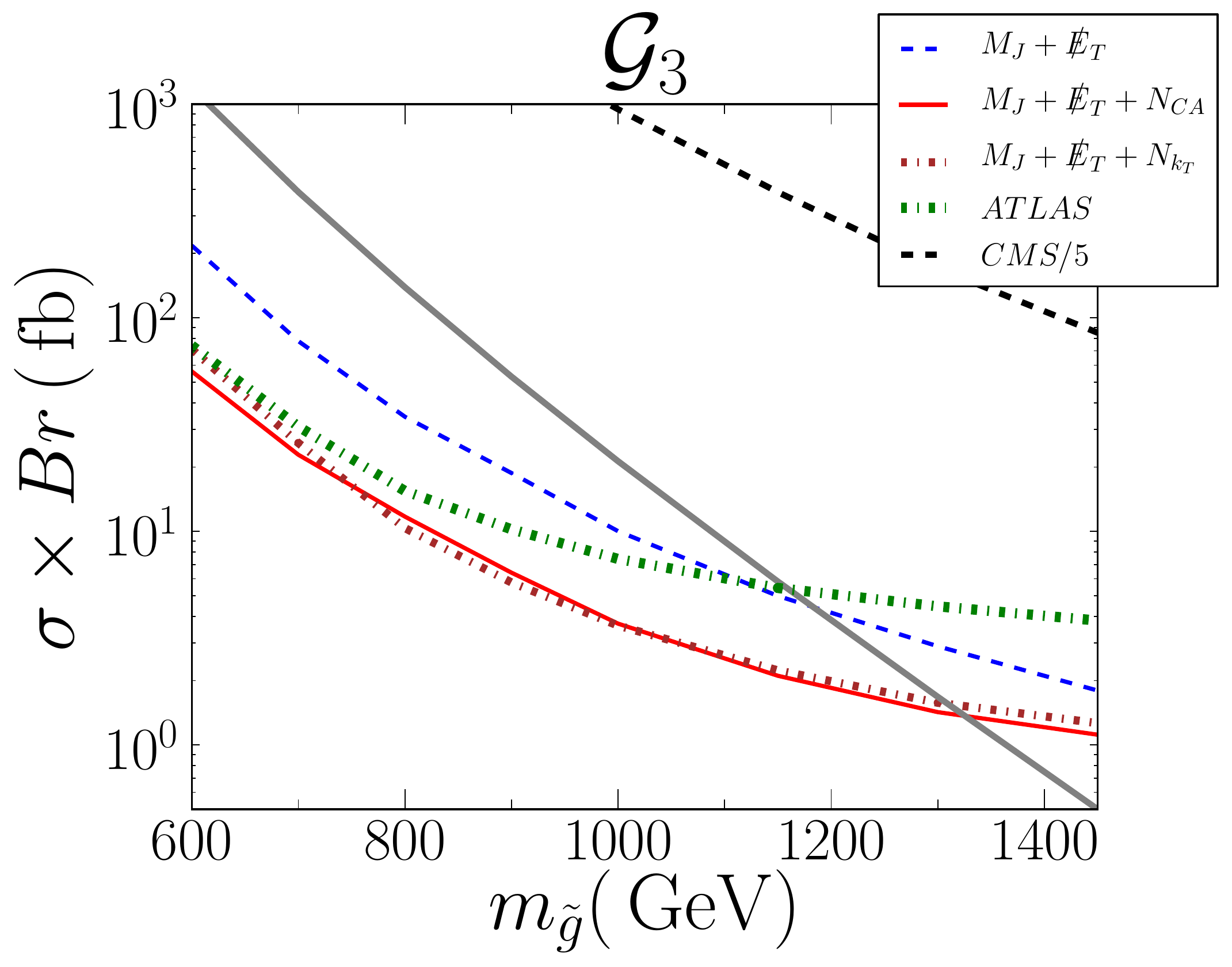}
\includegraphics[width=0.49\linewidth]{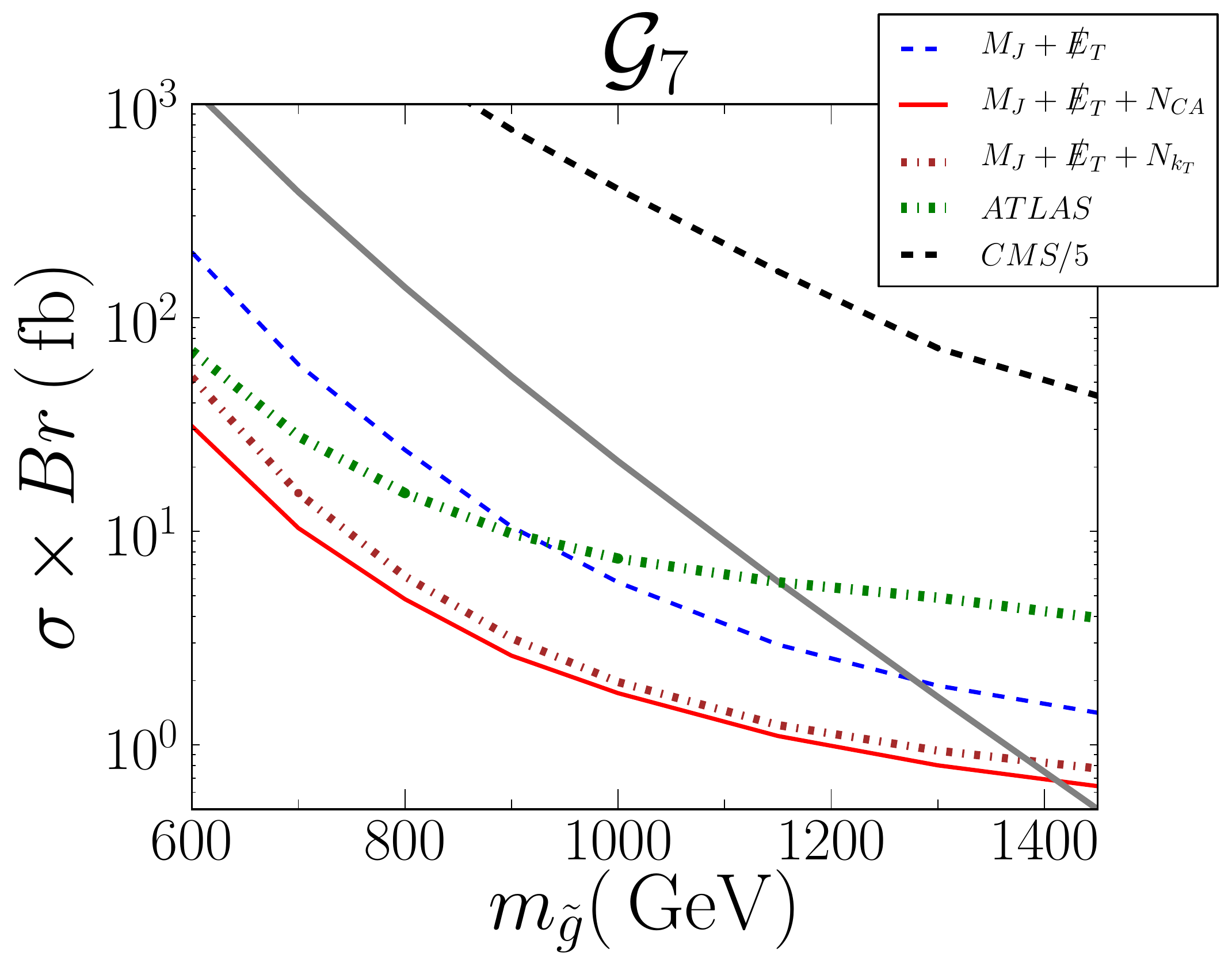}
\caption{95\% exclusion limits on $\sigma\times\mathcal{B}$ for the $M_J$ + $\MET$ search (dashed blue), the $M_J$+$\MET$+$N_{\rm{CA}}$ search (solid red), 
and the $M_J$+$\MET$+$N_{\rm{k_T}}$ search (dash-dotted brown) for the R-parity conserving topologies $\mathcal{G}_1$, $\mathcal{G}_2$, and $\mathcal{G}_3$ (left, top to bottom) and the corresponding RPV ones, $\mathcal{G}_5$, $\mathcal{G}_6$, and $\mathcal{G}_7$ (right, top to bottom). The exclusion limits given by the ATLAS high multiplicity search \cite{ATLAS:2012una} (dash-dotted green) and the CMS black hole search \cite{Chatrchyan:2012taa} (dashed black) as well as the NLO gluino production cross section (grey solid line) are also shown. The systematic uncertainty on the background is assumed to be $30$\%. Note that the CMS limit is rescaled by a factor of 5.}\label{exclusionSys30}
\end{figure}

\subsection{Comparison with previous searches}
\label{Sec: Comparison}

This section presents a comparison of the techniques proposed in this article to previous searches.  This is not meant to be a complete survey of all the searches that have been performed at the LHC and that are sensitive to high multiplicity signals.   Two searches are considered.  The first, presented in \secref{ATLASSearch}, is an ATLAS search that requires up to 9 $R=0.4$ jets with missing energy.  The second search, presented in \secref{CMSSearch}, is a search for ``black holes'' at CMS.  These two searches are different attempts at gaining access to high multiplicity final states. We find that the methods presented in this article are competitive.

\subsubsection{ATLAS High Multiplicity Search}
\label{Sec: ATLASSearch}
A comparison is made with ATLAS's most up-to-date high multiplicity search, which makes use of $5.8 \ifb$ at 8 TeV \cite{ATLAS:2012una}.  This search clusters
events into $R=0.4$ anti-$k_T$ jets and looks at 6 search regions:
\begin{itemize}
\item $n_j\ge 7$ with $p_T\ge 55\GeV$
\item $n_j\ge 8$ with $p_T \ge 55\GeV$ 
\item $n_j  \ge 9$ with $p_T\ge 55\GeV$
\item $n_j\ge 6$ with  $p_T\ge 80\GeV$ 
\item $n_j \ge 7$  with $p_T\ge 80\GeV$ 
\item $n_j \ge 8$ with $p_T \ge 80\GeV$ 
 \end{itemize}
It is further required that events contain no isolated leptons and that $\MET/\sqrt{H_T} > 4 \GeV^{\half}$.  Note that for an event
with $H_T = 1000 \GeV \,(3000 \GeV)$ the latter cut corresponds to a $\MET$ cut of $126 \GeV \,(219 \GeV)$. 

In order to compare the performance of the ATLAS search to the optimized search strategies in Tables~\ref{Table:mjmet} and \ref{Table:mjmetnca}, 
we assumed that the ATLAS search could be scaled up to $30\ifb$ while keeping the cuts fixed.  The re-estimated expected limits 
have been computed by linearly rescaling the expected number of background events and the corresponding uncertainties (as given by ATLAS)
to the new luminosity and computing the exclusion
limit as outlined in Sec~\ref{Sec: Efficacy}. Such a linear rescaling, 
which assumes that systematic uncertainties do not come to dominate the limits, is probably overly optimistic.

We find that for those benchmark signals with large final state multiplicities and intrinsic $\MET$ ($\mathcal{G}_{1,2,3}$ and 
$\mathcal{G}_{5,6,7}$), i.e.~the sorts of signals that the ATLAS search is designed to be sensitive to, the $M_J$+$\MET$+$N_{\rm{CA}}$ 
search generally outperforms the ATLAS search,
particularly at higher gluino masses, where the exclusion limit on $\sigma \times\mathcal{B}$ is improved by a factor $2-5$.  For these
benchmark signals this corresponds to an extended gluino mass reach of order $100\GeV$ to $250 \GeV$.  While it is difficult to know the extent to 
which this promising result can be realized at ATLAS or CMS, it is worth emphasizing that whatever the final search sensitivity should turn
out to be, it is already valuable to have a search strategy that is governed by different systematic uncertainties. 

\subsubsection{CMS Black Hole Search}
\label{Sec: CMSSearch}
The CMS black hole search \cite{Chatrchyan:2012taa}, which makes use of $4.7 \ifb$ at 7 TeV, is also sensitive to high multiplicity final states.  
This search makes use of 16 search regions,
each of which corresponds to different $S_{T\text{ min}}$ and $N_{\text{min}}$ cuts.  Here $S_{T\text{ min}}\in[1.9,4.1]\TeV$ is a cut on the scalar sum 
over transverse energy and $N_{\text{min}}\in[3,7]$ is a cut on the total number of reconstructed objects with $E_T > 50$ GeV.  See ref. \cite{Chatrchyan:2012taa}
for details.  

In order to compare the expected performance of the CMS search to the optimized search strategies in Tables~\ref{Table:mjmet} and \ref{Table:mjmetnca}, it
is necessary to extrapolate the CMS background estimates from 7 TeV to 8 TeV. 
Because there are no missing energy requirements, the background is completely dominated by
QCD events.  The absence of any intrinsic high energy scale in the background allows us to adopt the following approximate extrapolation. 
For each value of $N_{\text{min}}$, CMS provides the expected number of background events as a function of $S_T$, which we fit to an exponential,
\begin{align}
    N_{\rm QCD}^{7\TeV}(S_T; N_{\text{min}}) = e^{-\alpha(S_T - S_{T}^{(0)})}
\end{align}
where $\alpha$ and $S_{T}^{(0)}$ are fit parameters. The number of background events as a function of $S_T$ at $8\TeV$ is then estimated to be
\begin{align}
    N_{\rm QCD}^{8\TeV}(S_T; N_{\text{min}}) = N_{\rm QCD}^{7\TeV}\left(\frac{7\TeV}{8\TeV}\times S_T; N_{\text{min}}\right)
\end{align}
These background estimates can then rescaled to $30\ifb$ and combined with the efficiencies of the benchmark signals in each of the 16 search regions to
obtain the expected sensitivity of the CMS search at 8 TeV.  While these background estimates are inexact, they are sufficient for demonstrating that the
CMS search results in expected limits that are about two orders of magnitude weaker those obtained by a $M_J$+$\MET$+$N_{\rm{CA}}$ search 
(see Fig.~\ref{exclusionSys30}). This is because for the gluino masses that are accessible at 8 TeV, the $S_{T}$ cuts are highly inefficient, 
and the absence of missing energy requirements results in large QCD backgrounds. 
This demonstrates that high multiplicity searches targeting black holes are not necessarily well suited for other kinds of high multiplicity signals.

\section{Discussion}
\label{Sec: Discussion}

Recent years have seen an impressive amount of research on a large variety of jet substructure techniques.\footnote{For a comprehensive set
of references see the BOOST 2010 \& 2011 proceedings \cite{BOOST}.}  The majority of this work has focused on the development of either
general purpose tools (jet grooming, top tagging, etc.)~or jet substructure analyses tailored to specific search channels (e.g.~the BDRS
boosted Higgs search \cite{Butterworth:2008sd}). One area that has seen less work is the design of search techniques
for topologies that are more complicated or whose structure is not known a priori.
 In this paper we have taken a step in this direction by arguing that jet substructure suggests a different approach to counting jet
multiplicities that results in an effective search strategy that is sensitive to a variety of high multiplicity topologies.

The flexibility inherent in this approach raises the possibility of loosening missing energy cuts in favor of well chosen jet substructure
cuts.  This is of special interest for new physics scenarios in which signals exhibit little or no intrinsic missing energy, such as 
supersymmetric scenarios with baryonic RPV.  In \secref{ExLim} we have seen that for signals with large final state multiplicities
and some (though not necessarily very much) intrinsic $\MET$, the introduction of $N_{\rm{CA}}$ cuts does in fact lead to lower
$\MET$ requirements. While this represents only a modest push towards the regime of (near)-vanishing $\MET$ requirements,
it is nevertheless an encouraging result given how effective $\MET$ requirements are in reducing the huge QCD backgrounds.  In fact,
we find that trading $\MET$ cuts for $N_{\rm{CA}}$ cuts is particularly effective for the QCD background---it is the need to suppress the
$t\bar{t}+$jets background that prevents the $\MET$ cuts from being loosened further in Table \ref{Table:mjmetnca}. We anticipate
that if additional handles were introduced to combat the $t\bar{t}+$jets background (e.g.~vetoing on b-jets), then $\MET$ cuts could be loosened
even further. We have not pursued this interesting direction here, since our goal was to keep the search strategy as inclusive as possible.

One possible concern with high multiplicity searches at the LHC is their potential sensitivity to pile-up, something that becomes more 
pressing as the LHC pushes towards higher and higher luminosities.  In this paper we have advocated the use of jet trimming to 
reduce the present search's sensitivity to pile-up, but something like the technique introduced in ref.~\cite{Soyez:2012hv} might also be
necessary. This whole issue would need to be revisited if the search were to be performed
by ATLAS or CMS.  This is particularly the case because it is impossible for us to thoroughly examine pile-up effects given their sensitivity to detector effects
and the fact that each collaboration has detector specific methods for mitigating pile-up effects.  It is worth pointing out that one
possible advantage of $N_{\rm{CA}}$ is that it includes some built-in jet grooming by virtue of the veto it imposes on asymmetric subjet energy sharing.

In conclusion, we have seen that an effective search strategy can be developed by exploiting missing energy, a sum over fat jet masses,
and a sum over fat jet subjet counts. The two subjet counting algorithms presented, $N_{\rm{CA}}$  and $N_{\rm{kT}}$, yield comparable results, so that
a choice between the two would need to be guided by experimental studies (with a particular focus on inherent systematic uncertainties, performance
under pile-up, etc.). Other subjet counting algorithms are possible,\footnote{We have investigated the possibility of a subjet counting algorithm based on 
N-subjettiness \cite{Thaler:2010tr}, using a boosted decision tree to map $\tau_N$ space to different subjet multiplicities. The resulting algorithm performed similarly
to $N_{\rm{kT}}$, with the difference that it performed no better than a $M_J$+$\MET$ search on the $\GG_4$ topology. In the end we decided not to include this algorithm in the present study because its reliance on a sample of training jets (for learning the parameters of the BDT) raises concerns about its robustness. In
contrast to the small number of parameters that define $\nca$ and $\nkt$ (each of which is readily understandable from a parton shower picture), the many
parameters that define the BDT are not as easily connected to the underlying physics.} but what we would like to stress here is that, as has
been seen in many other jet substructure studies, the flexibility of the fat jet approach is very powerful. In this case the potential for systematic data driven
estimates of the QCD background is of particular importance.  As another example of the flexibility of the fat jet approach, we refer the reader to the related
work in ref.~\cite{Cohen:2012yc}, which focuses on high multiplicity hadronic final states with vanishing missing energy.  

\section*{Note}{\label{sec:note}
An implementation of $\nca$ and $\nkt$ has been made freely available at {\tt http://fastjet.hepforge.org/contrib/}.

\section*{Acknowledgements}{\label{sec:ack}
We thank Timothy Cohen, Eder Izaguirre, Mariangela Lisanti, Jesse Thaler, Gavin Salam,  Stefan H\"oche, Steffen Schumann, Ariel Schwartzman, Ken Van Tilburg, and Xinlu Huang for helpful discussions. MJ would like to thank Michael Spannowsky for interesting conversations. Special thanks to Timothy Cohen for having provided Monte Carlo data for the QCD background. SE and
JW are supported by the US DOE under contract number DE-AC02-76-SF00515. SE is
supported by a Stanford Graduate Fellowship.  AH is supported by the US DOE under contract number DE-FG02-90ER40542.

\appendix
\section{Genetic algorithm for optimizing search regions}
\label{app:genetic}

The genetic algorithm is initialized with 1000 search strategies, where each search strategy is a set of search regions.   Each of these search strategies is formed as follows.
First a random selection of 40 of the $N_{\text{cuts}}$ search regions is chosen. Each of these 40 search regions is assigned a weight proportional
to the number of models it covers with $\EE \le \EE_{\text{crit}}$. Finally, 1000 search strategies are created by sampling (without replacement)
from these 40 search regions. This gives a slight preference in the initialization stage to search regions that are sensitive to more models.
The exact initialization procedure is not critical for rapid convergence of the algorithm 
 
The search strategies are evaluated to see how many models they cover within the desired efficacy, and a ``fitness'' is assigned to them with the formula
\begin{eqnarray} \label{eq:FitnessFunction}
f(C, M) = \frac{1}{M_{\text{max}}^2 - (M^2 - C)}\,,
\end{eqnarray}
where $M$ is the number of models covered, $C$ is the number of search regions in the search strategy, and $M_{\text{max}}$ is the total number of models.  
This fitness function strongly penalizes search strategies that do not cover all models, followed by a penalty 
for having too many search regions.  

After evaluating the fitness of the search strategies, the  least fit  50\% are removed.  Pairs of fit search strategies are then selected and a new search strategy is created by taking a randomly determined fraction of each search strategy's search regions.  For instance, if the two selected search strategies had $N_1$ and $N_2$ search regions, then a uniform random number on the unit line segment, $x$, would determine that $x N_1 $ search regions would be taken from the first search strategy and $(1-x) N_2$ would be taken from the second search strategy. So if $N_1=20$ and $N_2=30$ and $x=0.20$, 4 search regions would be taken from the first search strategy and 24 would be taken from the second.  If duplicate signal regions are selected, the duplicate is removed, reducing the number of search regions.  
After creating a new search strategy, the search is mutated to guarantee that the population of search strategies has sufficient diversity.   Each search region within a search strategy
has a finite probability of being changed to another random search region.  We
use 6\% for this probability known as the ``mutation rate''.  Thus for the 16 search regions in the example, 1 change would be made on average. 

If after ten consecutive generations no progress has been made, i.e.~if no solution has been found that covers the entire model space, then a solution is manually created by forcing every model to be covered by some search region.  This can be done by increasing the number of search regions in the search strategies until full coverage is achieved.   Finally, if every model is covered 
and no further progress is achieved for seven generations, search strategies are scoured to see if any search regions can be removed without reducing coverage.  Either way, the genetic algorithm is restarted.  If no progress in reducing the number of search regions in a search strategy has been made in twenty generations, the program ends.  

Typically, the algorithm converges after 20 to 30 generations, and 10 to 30 distinct optimized search strategies are found each time.  While the termination of the program does not guarantee that the optimal solution has been found, re-running the program multiple times usually results in the same number of required search regions.   The resulting search strategies typically have similar features even if they differ slightly in detail.

\end{document}